\documentclass[traditabstract]{aa}
\usepackage{graphicx}
\usepackage{graphicx}
\usepackage{caption}
\usepackage{amsmath}
\usepackage{natbib}
\usepackage{color}
\usepackage[english]{babel}

\bibpunct{(}{)}{;}{a}{}{,} % to follow the A&A style

\def \aap{A\&A}

\def \apjs{ApJS}

\def \apjl{ApJL}

\begin{document}
\title{Reconstructing the star formation history of the Milky Way disc(s) from chemical abundances}
\titlerunning{Star formation history of the Milky Way from chemical abundances}

\author{O. Snaith\inst{1,2}, M. Haywood\inst{1}, P. Di Matteo\inst{1}, M.~D. Lehnert\inst{3}, F. Combes \inst{4}, D.~Katz\inst{1}, A.~G\'omez\inst{1}}

\authorrunning{Snaith et al.}

\institute{GEPI, Observatoire de Paris, CNRS, Universit$\rm \acute{e}$  Paris Diderot, 5 place Jules Janssen, 92190 Meudon, France\\
\email{owain.n.snaith@ua.edu}
\and
Current Address: Department of Physics \& Astronomy, University of Alabama, Tuscaloosa, Albama, USA
\and
Institut d'Astrophysique de Paris, UMR 7095, CNRS, Universit$\acute{e}$ Pierre et Marie Curie, 98bis Bd Arago, 75014, Paris, France
\and LERMA, Observatoire de Paris, CNRS, 61 Av. de l'Observatoire, 75014, Paris, France
}

\date{\today}

\abstract{We develop a chemical evolution model in order to study the star formation history of the Milky Way. Our model assumes that the Milky Way is formed from a closed box-like system in the inner regions, while the outer parts of the disc experience some accretion. Unlike the usual procedure,  we do not fix the star formation prescription (e.g. Kennicutt law) in order to reproduce the chemical  abundance trends. Instead, we fit the abundance trends  with age in order to recover the star formation history of the Galaxy. Our method enables one to recover with unprecedented accuracy the star formation history  of the Milky Way in the first Gyrs,  in both the inner (R$<$7-8 kpc) and outer (R$>$9-10 kpc) discs as sampled in the solar vicinity. We show that, in the inner disc, half of the stellar mass  formed during the thick disc phase, in the first 4-5 Gyr. This phase  was followed by a significant  dip in the star formation activity (at 8-9 Gyr) and a period of roughly constant lower level star formation for the remaining 8 Gyr.  The thick disc phase has produced as many metals in 4 Gyr as the thin disc in the remaining 8 Gyr.     Our results suggest that a closed box model is able to fit all the available constraints   in the inner disc.  A closed box system is qualitatively equivalent to a regime where the accretion rate, at high redshift, maintains a high gas fraction in the inner disc. In such conditions,  the SFR is mainly governed by the high turbulence of the ISM. By z$\sim$1 it is possible that most of the accretion takes place in the outer disc, while the star formation  activity in the inner disc is mostly sustained by the gas not consumed during the thick disc phase, and the continuous ejecta from earlier generations of stars.
The outer disc follows a star formation history very similar to that of the inner disc,   although initiated at z$\sim$2, about 2 Gyr before the onset of the thin disc formation in the inner disc.
}


\maketitle

\section{Introduction}

The basis of Galactic Archaeology  is that the signatures of the formation history  of  the Milky  Way  are  encoded  into the  distribution  and properties  of  populations  of  stars.  Thus,  the  history  of  star formation  should  be  impressed   into  the  chemical  properties  of stars.  Translating the  observed metallicity  of  stellar populations into a star  formation history (SFH) is, however,  non-trivial, and is an ongoing process  because of the need for  improved observations and better models.

The  overwhelming majority  of present  chemical evolution  models are based  on the  observation that  the disc  in the  solar neighbourhood contains far too few metal-poor stars - here, `metal-poor' means about 1/3 of the solar metallicity, \citep[e.g.][]{Binney1998,Tinsley1974,Schmidt1963,Pagel1997,Pagel1975,vandenBergh1962} - to have formed from  a  very gas rich, low metallicity, ISM at early times.  This is the  so-called `G-dwarf' problem.   To overcome this lack of intermediate metallicity stars most studies have  endorsed the idea that  the gas must  have been conveyed onto the disc on a  relatively long time scale, maintaining a small gas reservoir, a limited dilution and, consequently, a rapid rise of the metallicity in the ISM before many stars formed.  

This approach constitutes the basis of current Galactic  Chemical  Evolution (GCE)  codes  \citep[e.g.][]{Chiosi1980, Chiappini1997, Fenner2002},  which assume an  open box  model, where  gas falls  into the  galaxy over  time. The density  of the gas  is then  used to  calculate the  star  formation rate, according   to   a   Schmidt-Kennicutt   relation   \citep{Kennicutt1998}.  The rate of  infall is then  tuned to recover  the surface density of  the Milky  Way, the  current gas fraction, the current star formation rate (SFR), etc.  At this  point, the chemical  evolution  of the  system  is  examined,  usually against  a metallicity-[$\alpha$/Fe] distribution. The particular form of the gas inflow(s) (or outflow(s)) can  then be varied in order to improve  the fit to the data. Thus, we  have dual infall models \citep{Chiappini1997, Fenner2002}, three infall models  \citep{Micali2013}, two inflows and an  outflow \citep{Brusadin2013} etc.  These  models also  often require  tuning   in  terms  of   the  star  formation   (e.g. \citealt{Chiappini1997} double the value  of their star formation tuning parameter during the halo-thick disc phase compared to the thin disc phase). \\

However, in the last couple of years a number of new results are forcing us to revise our current picture of the Galaxy, and of the link between its stellar populations and their mass budget. 
There is now evidence for a direct chemical and kinematic continuity between the thick and thin discs, both on large \citep{Bovy2012c} and local  \citep{Haywood2013}  scales. It  would  be  extremely difficult to  imagine that such a large  structure, with similarities to both the halo and the bulge, could have been accreted \citep{Abadi2003}.  Moreover, recently measured structural parameters of the thick disc by \citet{Bensby2011} and \citet{Bovy2012c} indicate that this population is massive (see discussion hereafter, and \citet{Fuhrmann2012}).  
Hence, there is now increasing evidence that a significant amount of low metallicity stars (in the sense defined above, and which correspond to the thick disc metallicity) exist in the Galaxy, although most must be confined  at Galactic radii less than the solar circle (in the inner disc).  \\

The most  intense phase of star formation in the universe occurred at redshifts between 2 and 3 \citep{Hopkins2006,Madau1996,Lilly1996,Madau1998,Madau2014}. The thick disc represents  the only stellar population in the Galaxy which corresponds to  this star formation phase, since the classical spheroid, if present, makes only a minor contribution to the total mass \citep{Carney1990,Shen2010, Kunder2012, DiMatteo2013}. Yet, the thick disc is either absent from  most chemical evolution models, or its role is severely underestimated \citep[however, see][]{Gilmore1986}.  The reason for this is  that models are fitted to the solar neighbourhood, where the thick disc  represents only about 10\% of the local stellar density \citep{Juric2008}.  However, the population fractions measured in the solar vicinity are not representative of the entire Galaxy (see Section \ref{Discussion} and  Section \ref{data}). While the use of population fractions in the solar neighbourhood may, therefore, be misleading in efforts to recover the SFH of the early stages of the Galaxy,  and the corresponding stellar mass formed at those times, chemical abundances are not: indeed,  thick disc stars found at the solar vicinity were born at very different radii \citep[e.g.][]{Reddy2006}, making them representative of the whole thick disc population,  see Section \ref{data}. Their chemical enrichment patterns, as found at the solar neighbourhood, are the imprints left by a formation phase that involved the entire disc. \\

The present study is, therefore, an attempt to re-examine the question of the chemical evolution  of the Galaxy from the very beginning, and from its simplest
form, in light of the most recent results obtained in the field of  Galactic stellar populations. 
In an effort to reduce the number of assumptions 
to a minimum, we model the  inner disc system using a closed box model,  providing physical arguments to show that this is more than an academic exercise,  
and that it must be considered as a good, first order, representation of the evolution of the Milky Way disc.  The outer disc system is modelled as essentially the same sort of closed box system, but with a single infall event in order to reduce the metallicity.
In \citet{Snaith2014}, we described our first results on the derivation of the star formation history of the inner Galactic disc(s).
In the present study, we provide a full description of our model, an in-depth assessment of the various sources of errors in the derivation
of the SFH, and a derivation of the SFH of the outer disc. 
By means of the chemical data, and their evolution with time, recently discussed in \citet{Haywood2013}, we will show that the ``thick disc'' formation phase accounts for about half the stellar material present in the Milky Way. This points to a revision of the mass budget in the Galactic thin and thick discs; of the amount of star formation required in the early phase of the Galactic evolution; and of the gas reservoir that must have been available at high redshifts in the Galaxy.

We first present the  model (Section \ref{Model}). We then outline the key points of the data, and our understanding of recent observational results 
of importance to our study in Section \ref{scene}.
We explore, in  a general way, how changing the  SFH affects the chemical evolution history of the ISM, 
with regards to the silicon and iron abundances  (Section \ref{Patterns}).  We determine the SFH which best  
fits the observational silicon data in  Section \ref{SFR}, examine how the results of the model are affected 
by our choice of parameters and dataset (and chosen elements) in Section \ref{Robust}. Finally we will discuss our results in Section \ref{Discussion}.

\section{Model}\label{Model}
\subsection{Modelling philosophy}

Our basic approach was to develop a simple model, using the fewest possible  assumptions. We discuss below why these assumptions are not unrealistic.

A GCE model is a very simple model of the chemical history of a galaxy, and uses a set of simple recipes to follow galaxy evolution. A strength of this approach is that the effect of different scenarios can be very quickly studied, while more elaborate models and simulations can make it harder to disentangle the effects of different processes.   

We therefore developed a model where: (1) the ISM is considered to be always  well mixed; (2) the IMF is time independent; (3) the initial metallicity of the ISM is negligible; (4) the system is closed  in the inner disc; (5) the outer disc experiences a single accretion event at a look back time of 10 Gyr. This accretion event is a simple dilution of the insitu gas by  gas with primordial abundance, and is required to match observations, e.g. \citet{Haywood2013}. 

We do not assume instantaneous recycling, and take into account the lifetime of stars. The impact of this depends on the time step of the simulation. If the time step is large then all the gas and metals from SNII are returned at once anyway. \\ 
\\
Are these assumptions realistic?\\

\begin{enumerate}

\item We  know,  from  several  indicators \citep{Cartledge2006},  that the ISM is homogeneous on scales of a few hundreds of parsecs. There are indications that it has remained so in the past, with evidence from presolar grains formed during the thin disc formation   \citep[4.6 Gyr ago,][]{Nittler2005}. Similar homogeneity can be inferred for the thick disc  from the tight age-metallicity and age-alpha abundance ratios that have been measured by \citet{Haywood2013}. As explained in \citet{Haywood2013}, the apparent dispersion in the  age-metallicity relation of thin disc stars in the solar vicinity (age$<$8 Gyr) can be explained if the Sun lies at the interface between the inner and outer discs. It is, therefore, polluted by stars from both  discs through the epicyclic radial oscillation of stars. 
 It remains, however, that the assumption of an homogeneous ISM is a reasonable one.\\

\item Even the most  sophisticated chemical evolution models  consider the IMF to be  time-independent, and  it is only for the  very first  generation of stars  that any convincing evidence has been proposed that   the  IMF  could  be different (top-heavy) \citep{Yamada2013}.\\

\item The initial metallicity of the thick disc is at least two orders of magnitudes below solar  ([Fe/H]$<$-2 dex, \citep{Morrison1990, Reddy2006, Kordopatis2013}), which, in practice, means that essentially all  Galactic chemical enrichment - and therefore most  of the metallicity  distribution - is the result of the (thin+thick) disc formation phase. Our assumption that the model starts with zero initial metallicity can be considered a good, first order approximation.  \\

\item Can the  inner disc system be considered closed? In this form the question  looks largely  academic: we know  the  system was most probably ``open'' at  an  epoch  when interactions  with the  environment must  have been  intense.   This issue can, however, be  reformulated as follows: how much  gas was  available for rapid consumption in the first few Gyrs?  The time scale of the gas accretion is unknown. What is observed, however, is  that discs at redshift $\sim$ 2 are rich in gas \citep{Tacconi2010}. In the present study, we assume that the closed box model either approximates a situation where most of the accretion in the inner disc (R$<$10~kpc) has occurred early in the evolution of the Galaxy, or a situation where the gas accretion maintains a gas fraction sufficiently high that the disc evolution is not directly dependent on the accretion history.  That copious amount of gas must have been available in the first Gyrs of the Galaxy is therefore in accordance with the observation of discs at high redshift, and with the recent estimate of the mass of the thick disc in the Milky Way \citep[see comment in][]{Haywood2013, Haywood2014,Snaith2014}. This is at  variance with standard chemical evolution modelling,  where gas accretion is parsimonious.

\end{enumerate}

By making use of these assumptions we explore the characteristics of the chemical tracks of stars at the solar neighbourhood, and their evolution with time \citep[see][and Sect.~\ref{data}]{Adibekyan2012, Haywood2013}. This is done in order  to recover the SFH that best fits the data. We identify the best SFH by constraining it with the chemical characteristics of stars in the solar vicinity, and are able  to quantify the chemical enrichment track that a given SFH produces. To this end,  we have developed a chemical evolution code and describe it in Sects.~\ref{partA} and \ref{partB}. The code can be broken up into two sections. The first (part A) reads the chemical yields and converts them into a chemical evolution for a simple stellar population (SSP) of given initial metallicity, Z.  The second (part B) uses these tracks and traces the chemical enrichment of the ISM due to a given SFH. 

In a departure from standard chemical evolution codes, we do not assume any specific form of the Kennicutt-Schmidt  \citep{Schmidt1959, Kennicutt1998} law to link the gas and SFR densities. The two quantities are decoupled. This means that at any given time, part of the gas in the system is used to make stars, while the remaining part can be seen as the gas reservoir for subsequent star formation. Because the SFR is the quantity we wish to constrain, this approach is justified, and will allow us avoid developing any hypothesis on the precise form of the specific Schmidt law \citep{Schmidt1959}.

 When we call inner disc `closed' we mean that the gas present at the beginning is both eligible to form stars, and acts as a reservoir of primordial gas into which the metals are injected. We do not specify whether this gas is cold gas in the disc of the galaxy (the cold ISM), warm gas in the Galactic corona, or hot gas in the halo. This means that the closed box system consists of gas which can cool to form stars, and can act to dilute the metals ejected by stars. 

In the following (Sects.~\ref{partA} and \ref{partB}), we describe the key features of our chemical evolution code. 

\subsection{Making chemical evolution tracks (Part A)}\label{partA}

This first part of the code generates the chemical evolution tracks. The tracks provide the normalised mass (if the initial stellar population has a mass of 1) of metals or gas released from a stellar population of given metallicity after a given amount of time. We use a grid of nine metallicity values ranging from 0 to Z$_\odot$, and calculate the cumulative amount of metals released over 14 Gyr. We assume that metals are produced from SNII, SNIa and AGB stars. The time step is exponentially increasing with the age of the population. This is because most metals are released within the first few Gyrs, and so we must trace the early release with greater resolution. 

The steps involved in generating these tracks are:
\begin{enumerate}
    \item Choose an IMF \citep[i.e.][]{Kroupa2001}.
    \item Choose the yields for SNII, AGB and SNIa.
    \item Get stellar life times.
    \item For each Z in the grid calculate the mass fraction of stars which are SNII and AGB at a given time $t$.
    \item Multiply step 4 by the yields.
    \item Define the functional form of the SNIa time delay, multiply by the SNIa yields.
    \item Cumulatively sum the yields with increasing time.   
\end{enumerate}

This part of the code allows us to calculate the amount of gas returned, the mass of the various species of metals, the number of SNII or SNIa at a given time etc.  for a stellar population for a given initial metallicity.

 We use \citet{Nomoto2006} yields for SNII \footnote{In section \ref{diffel} we also use \citet{WW1995}. These yields are modified according to \citet{Timmes1995} where the iron yield is reduced by a factor of two.}. This is required if the [Si/Fe] is to have a high enough maximum value at the beginning of the model's run. At the lower end of the SNII mass distribution, we assume that the minimum mass of SNII stars is 8 solar masses \citep[e.g.][etc.]{Few2012,Kawata2003}, unfortunately, the SNII yields only go down to approximately 12 solar masses. We, therefore, extrapolate down to 8 by scaling proportionally to the mass. In other words, we take the yield of stars at 12 solar masses and multiply them by 2/3 to get the expected yield at 8 solar masses \citep{Few2012}. We also extrapolate the \citet{Karakas2010} AGB yields down to 0.1 solar masses in the same way.  By scaling the lower end of the iron yields by the stellar mass we introduce an uncertainty of $\sim$15\%, which is a relatively small difference.

For the SNII yields, the minimum metallicity in the theoretical yields tables is Z=0, meaning a pristine star, while the maximum is Z=1$Z_\odot$ or solar metallicity. Where the metallicity of the star exceeds 1$Z_\odot$ we assume that the stellar yield is the same as if it was still at 1$Z_\odot$. Similarly, where the AGB metallicities drop above or below the metallicities given in the tables, we assume that any chemical yields are the same as they are at the extremity of the table. 

None of the theoretical yields for SNII we used provide values for stellar masses greater than 40 solar masses.   Alpha elements scale strongly with mass, between 8 and 40 solar masses, while the iron value does not.  If we assume that stars with masses greater than 40 solar masses all have yields the same as stars at 40 \citep[as in ][]{Few2012,Kawata2003}, we cannot arrive at a solution for silicon. We therefore only include stars with masses up to 40 solar masses in our SNII yield tables.  The uncertainties in the various yield tables \citep[e.g.][]{Nomoto2006,Limongi2002,WW1995} means that this is no worse an assumption than alternative approximations.  Ideally, we await yields which extend to 100 solar masses, and greater than solar metallicity, in order to reduce this avenue of uncertainty.

The tables of yields are given according to the mass of the star ejecting those metals. We use a simple analytical fit of \citet{Raiteri1996} to calculate the stellar lifetimes :

\begin{equation}
\log(t_*)= a_0(Z) +a_1(Z)\log{M}+a_2(Z)\log{M}^2
\label{egn:timemass}
\end{equation} 

\noindent
where, \\
$a_0(Z) = 10.13 + 0.07547\log{Z}-0.008084\log{Z}^2 $ \\
$a_1(Z) = -4.424 + 0.7939\log{Z}-0.1187\log{Z}^2 $ \\
$a_2(Z) = 1.2662 + 0.3385\log{Z}-0.05417\log{Z}^2 $ \\

\noindent
where Z is the metallicity and M is the mass of the star. Equation \ref{egn:timemass} is valid within the metallicity range of $7\times10^{-5}$ to $3\times10^{-2}$, so for metallicities greater (less) than this we set them to $3\times10^{-2}$ ($7\times10^{-5}$) for use in this equation. We rearrange Eqn. \ref{egn:timemass} to solve for $M$, in order to determine the mass of stars which end their lives in a given time step. 

We also convolve the yield tables with an IMF. We use the \citet{Kroupa2001} IMF in sections \ref{Patterns}, \ref{SFR} and most of section \ref{Robust} (normalised between 0.1 and 100 solar masses) because it is a simple, basic and commonly used IMF. The particular form of the IMF is,

\begin{equation}
\Phi(m) = 
           \begin{cases}
                   0.332 m^{-0.8} & 0.1 < m \le 0.5\\
                   0.178 m^{-1.7} & 0.5 < m \le 1\\
                   0.178 m^{-1.3} & 1 < m \le 100
           \end{cases} 
\end{equation}
\noindent

A discussion of the particular choices of IMF and yields used in the model is discussed in the Section \ref{imfvar}.

In order to put the yields, stellar lifetimes and IMF together, we use:

\begin{equation}
 \Gamma(t=now) = \int_{t=0}^{t=now}{\frac{\Phi(M(t))}{M(t)}\gamma(M(t))dt}
 \label{eqn:cumey}
\end{equation} 
\noindent
where $ \Gamma(t)$ is the integrated yield with time, $\Phi(M(t))$ is the IMF of the stars dying at $t$ and $\gamma(M(t))$ is the stellar yield of the stars dying at t. 

Finally, we include the chemical yields of the SNIa from \citet{Iwamoto1999}. We utilise the supernova rate given in \citet{Kawata2003}, based on the paper by \citet{Kobayashi2000}. This approach divides the potential SNIa host systems between those which contain a white dwarf and a red giant, and those which contain a white dwarf and a main sequence star. In effect, this delays the addition of elements from SNIa by 1 to 2 Gyrs. Here, we have to assume that a SNIa returns 1.38 solar masses to the ISM, completely destroying the white dwarf. We also assume that the amount of hydrogen fed back into the ISM is 0.01 solar masses/supernova. The actual contribution of hydrogen from SNIa is negligible compared to the SNII and AGB contributions, so this value is essentially everything that is not metals. In general, however, this value is unimportant for deriving the [Fe/H] value from the model.  Several authors have used a `minimum metallicity' below which SNIa are assumed not to occur \citep{Kobayashi2000}. In general we neglect this, but will discuss the effect it has on our model in section \ref{Robust}. 

\begin{figure}
\centering
\includegraphics[width=3.in, clip]{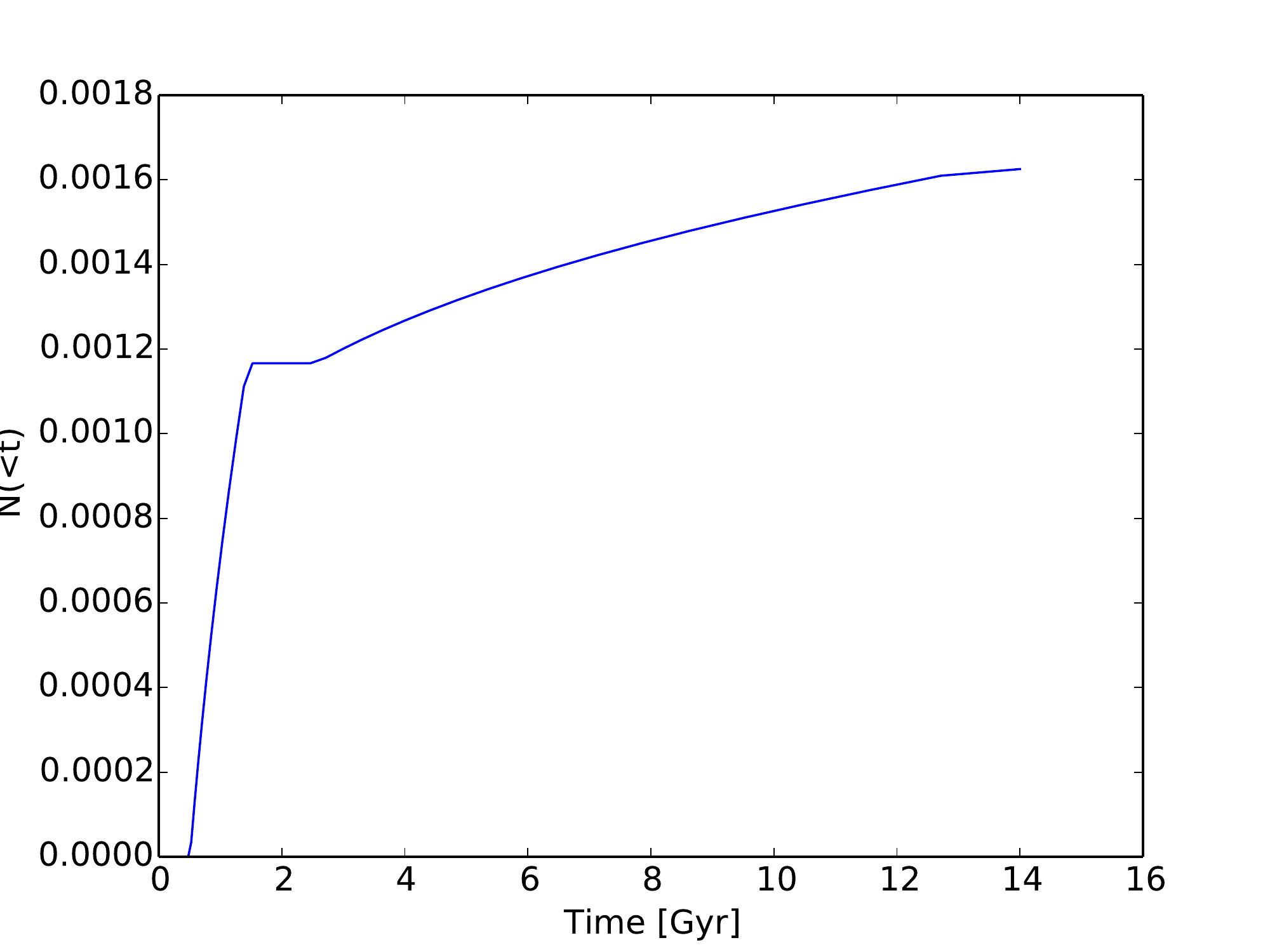}\\
\caption{The time delay distribution of SNIa for solar metallicity stars. The blue line shows the cumulative number of SNIa expected from an SSP.}
\label{Fig:Tdelay}
\end{figure}

The form of the SNIa time delay distribution is still debated, with numerous functional forms available in the literature (e.g. \citealt{Kobayashi1998, Greggio2005, Matteucci2009}). We choose the following form, given in \citet{Kawata2003}:

\begin{align}
N_{SNIa}(t) &=  f_p(m) \times(g_{h,MS}(m)+g_{h,RG}(m)) \\
f_p(m)&=  \int^8_3{\frac{\Phi(M)}{M}}dm \\%\left(\int^{100}_{0.1}{\Phi(M)}dm \right)^{-1}\\
g_{h,MS}(m)&=  b_{MS} \frac{\int^{2.6}_{max(1.8,m(t))}{\Phi(m) dm}}{\int^{2.9}_{1.8}{\Phi(m)}dm}\\
g_{h,RG}(m)&=  b_{RG} \frac{\int^{1.5}_{max(0.9,m(t))}{\Phi(m) dm}}{(\int^{1.5}_{0.9}{\Phi(m)}dm}
\end{align}

\noindent 
where $N_{SNIa}(t)$ is the number of supernovae with time, $\Phi(m)$ is the IMF, 8 and 3 are the masses of the stars which evolve into white dwarfs, (2.6, 1.8) and (1.5, 0.9) are the mass range of stars which fuel the white dwarf, and are either main sequence stars or red giants, respectively.  In this single degenerate model, we neglect the contribution of binary white dwarf pairs. These pairs may provide another path to SNIas. Such a `double degenerate' model could provide a drawn out tail of SNIa with very long time delay \citep{Greggio2005}. SNIa produce 0.75 solar masses of iron, 0.16 solar masses of silicon. The cumulative number of SNIa ($N_{SNIa}(<t)$) is shown in Fig. \ref{Fig:Tdelay}.

At every time, $t$,  since the formation of the stellar population we can calculate the cumulative yield of various chemical elements (O, Mg, Si, Fe),  the total metallicity (Z), and total gas released since $t=0$. Having defined the yields at a given mass, and the relation between mass and stellar lifetime, we can now produce the cumulative metal release for a given stellar population, with a given IMF for a given set of stellar yields. 

We sum the amount of metals produced in descending order of mass and plot this against stellar lifetime. This is then used to generate a table of how a given stellar population ejects metals with time. 

\subsection{Calculating the GCE (Part B)}\label{partB}
Part B then utilises the tables created by the above approach, and calculates the chemical evolution of the system a given SFH produces. 

The GCE code models the galaxy, and follows the mass of metals and gas present in the ISM with time. Of particular interest to this paper is the mass of $\alpha$ elements (e.g. Si, Mg, O), Fe, H and total gas present. From these data we can follow the evolution of the metallicity, gas fraction, abundance ratios etc at each time step.

At $t=0$ the system contains pristine gas of mass $1$, composed of 0.75 hydrogen and 0.25 helium. We normalise the SFH such that the integral  of the SFR over 14 Gyr is equal to 1. Because mass is released back into the gas phase due to supernovae and stellar winds, the total stellar mass at the end of the evolution is never itself equal to 1, but rather,

\begin{equation}
   M_*(t=0)= \int^{14 Gyr}_{0}{\Psi(t)} dt -M_{recy}(t)
\end{equation} 
\noindent
where $M_*(t=0)$ is the stellar mass , $\Psi(t)$ is the SFR such that,

\begin{equation}
   \int^{14 Gyr}_{0}{\Psi(t)} dt = 1
\end{equation} 
\noindent
and $M_{recy}(t)$ is the cumulative gas release.

For each time step the code looks at the provided star formation history and calculates the amount of mass which is converted from gas into stars, and subtracts it from the ISM. The mass of O, Mg, Fe, Si, H, total gas released corresponding to the abundance of the ISM at that time is also removed from the ISM, and used to calculate the chemistry of the stellar population:

\begin{equation}
\Delta M_{(O, Mg, Si, Fe)}=\frac{\Psi(t)\Delta t}{M_{ISM}}\times M_{(O, Mg, Si, Fe)} 
\end{equation}
\noindent
where $\Delta M_{(O, Mg, Si, Fe)}$ is the mass of O, Mg, Si or Fe removed from the ISM, $\Psi(t)$ is the SFR, $\Delta t$ is the time step, $M_{ISM}$ is the mass of the ISM and $M_{(O, Mg, Si, Fe)}$ is the mass of O, Mg, Si, Fe present in the ISM.

The code then looks at the stars created in every previous time step, and calculates the amount of gas and metals which is returned to the ISM. This is  added to the ISM, and the fraction of recycled gas is removed from the stars and returned to the gas component. This is done by interpolating the cumulative yields produced by part A for a stellar population with t= stellar age, and a metallicity equal to the formation metallicity. We assume the instantaneous mixing approximation.

We also keep account of the mass of a stellar population at its formation. It is this quantity which is used at each time step to calculate the mass a population returns to the ISM. The mass of the stellar population taking account of the gas release is simply an output. 

Our model, therefore, differs from standard GCE codes \citep[e.g.][]{Chiappini1997} in that it is a closed box model, while other GCE codes assume gradual gas infall over the whole of cosmic time. We have decoupled the SFR from the gas surface density, which other GCE codes implement using a Schmit law \citep{Kennicutt1983}.  In our case the SFH is derived purely from fitting the age-[Si/Fe] data with no accretion,  while we address the question of fitting the MDF in a separate paper (Haywood et al., in prep).

\section{Setting the scene}\label{scene}

\subsection{A general scheme}

As important as the assumptions in the model is the question of exactly what we want to model. Data can be diversely interpreted, and one must decide on a general  scheme prior to any modelling. 

The scheme we adopt has been sketched out in \citet{Haywood2013}  and can be summarized as follows. 

We describe the disc of our Galaxy as being made of  two systems, the inner and outer discs, which roughly divide at R$_{GC}$=7-10 kpc, the Sun being  in the transition zone between the two. This suggestion is different from modelling the Galaxy as one system with properties that vary gradually with galactocentric radius, as it has been described up to now \citep{Chiosi1980, Chiappini1997}. The  data in the solar vicinity reflect the existence of these two systems,  and we advocate that two chemical evolutionary paths are necessary to describe fully the solar  neighbourhood content.

\begin{figure}
\centering
\begin{tabular}{c} 
\includegraphics[width=2.75in]{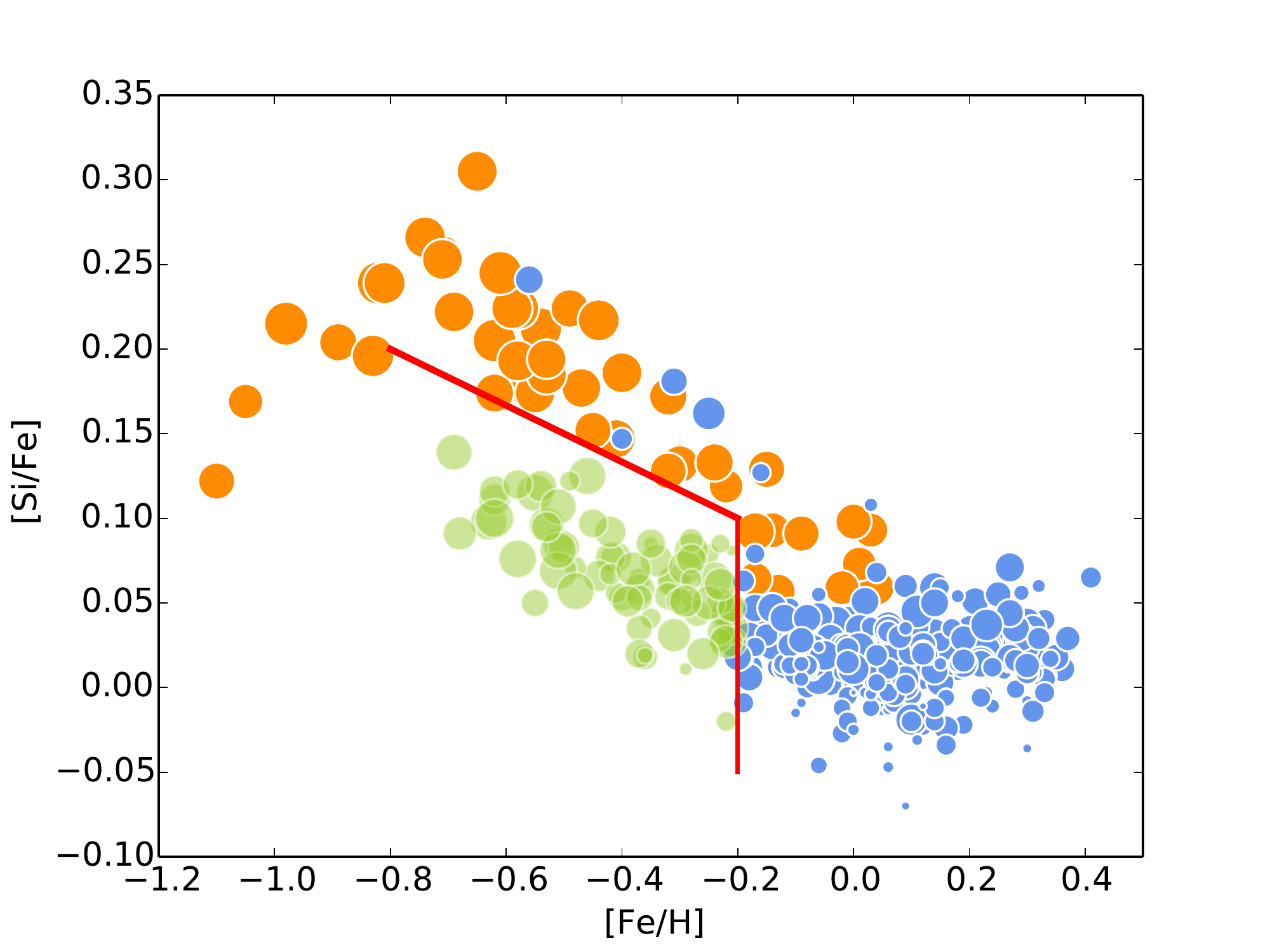}\\
(a) \\

\includegraphics[width=2.75in]{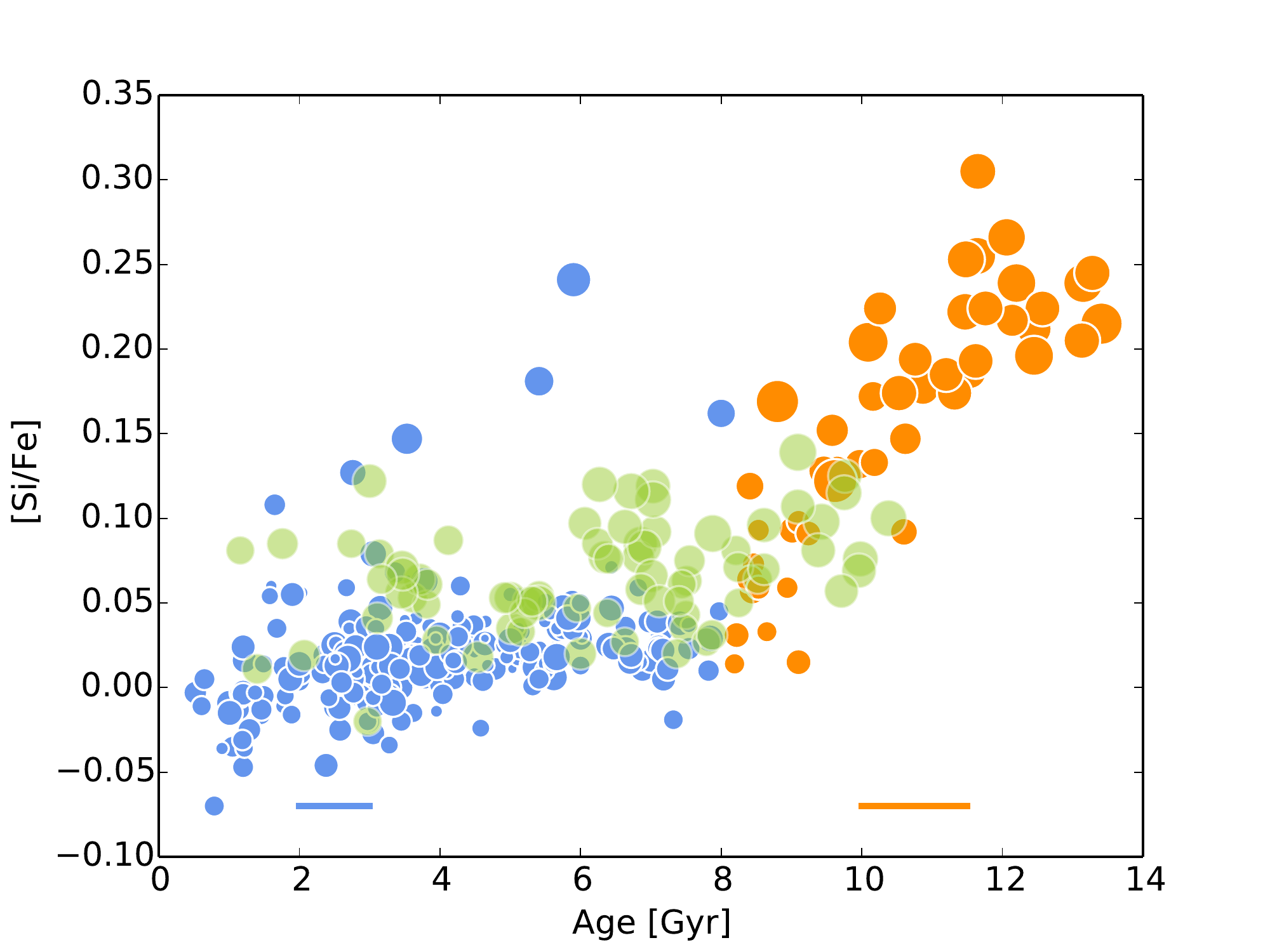}\\
(b) \\
\includegraphics[width=2.75in]{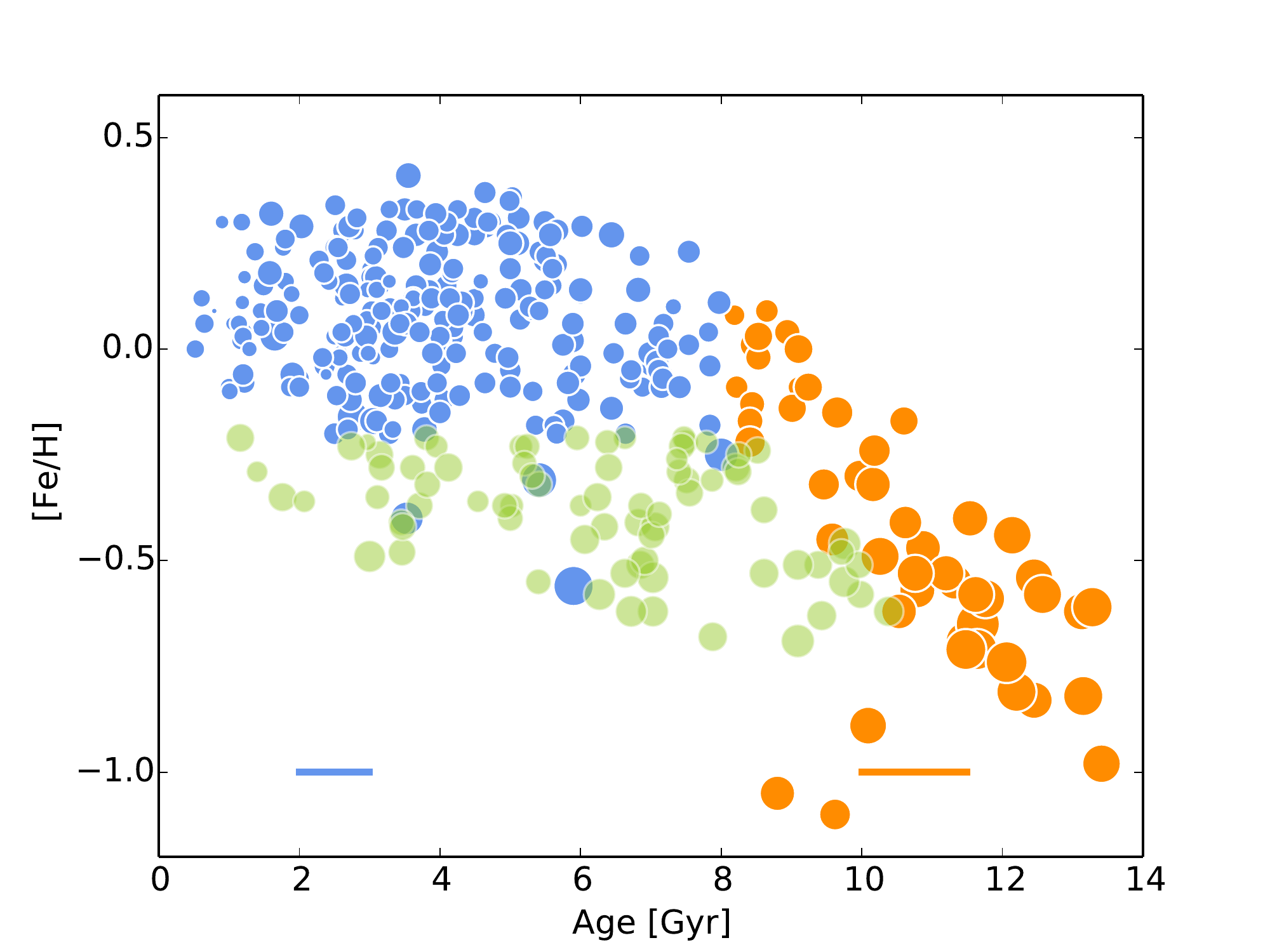} \\
(c) \\
\end{tabular}
\caption{[Si/Fe] versus [Fe/H] (panel (a)), [Si/Fe] versus age (panel (b)) and [Fe/H] versus age (panel (c)) for the \citet{Adibekyan2012}/\citet{Haywood2013} data adopted in this work. In each panel, the orange points represent thick disc stars, green points ``outer'' thin disc stars and blue points  inner thin disc stars. The red line in panel (a) shows the selection criterion of outer thin disc stars, as defined in Eqn. \ref{Eqn:Select}. The size of the points is age for panel (a), [Fe/H] for panel (b) and [Si/Fe] for panel (c).  The horizontal lines give the estimated error on the ages for young (blue, 1 Gyr) and old (orange, 1.5 Gyr) stars.}

\label{Fig:Dataobshaywoodetal}
\end{figure}

\emph{First path -- } The inner disc (R$_{GC}<$7-10 kpc) is composed of the thick and thin discs \footnote{Here we refer to the thick and thin discs as the `chemical thick and thin discs' outlined in \citet{Haywood2013}.}, which are essentially in  continuity, although a marked transition between the two is encoded in the evolution of $\alpha$ elements  and metallicity with age (see Fig. \ref{Fig:Dataobshaywoodetal}), due to a change in the regime of star formation at that epoch (as discussed in Section \ref{SFR}).  A tight age-metallicity relation has been found in the thick disc \citep{Haywood2013}, testifying  that the ISM at this epoch was well mixed. The data show that the chemical evolution continued steadily during a period lasting 4 to 5 Gyr, up to metallicities and alpha abundances well in accordance with those of the thin disc 8 Gyr ago. The spread in metallicity visible in the thin disc sequence, which contrasts with the tight relation obtained for the thick disc, can be explained by the position of the Sun at the division  between the inner and outer disc. This was argued in \citet{Haywood2013},  who further elaborate that the simple effects of blurring (radial  excursions of stars on their orbits due to epicyclic oscillations) are sufficient to contaminate the  solar vicinity with a few percent of stars coming from either the inner or outer regions. This then explains the low and high metallicity tails observed in the local distribution.  In other words, it is the position of the Sun, at the division of two systems -- the inner and the outer disc -- with markedly different chemical history -- that can explain the origin of the spread in metallicity observed in the thin disc at the solar vicinity. 
Hence, as explained in \citet{Haywood2013}, we think that no radial migration is necessary to explain 
the chemical patterns seen in the solar vicinity, see below for further justifications (end of section \ref{data}). 
\\

\emph{Second path -- }The outer disc (R$_{GC}>$9-10 kpc) started to form stars 10 Gyr ago, i.e when the thick disc formation was still on-going in the inner Galaxy. The similarity in alpha-element abundance between the thick disc and the outer  disc at identical ages ($\sim$10 Gyr) suggests that the gaseous material from which the outer disc started to form may have been polluted by gas expelled from the forming thick disc. The substantially lower metallicity of the outer disc stars at those epochs (reaching [Fe/H]$\sim$ -0.8 dex), compared to the already high metallicity of the thick disc at the same time ([Fe/H] $\sim$-0.4 dex), suggests that gas in the outer disc resulted from a mixture of pristine gas present in the outskirts, and gas processed in the inner parts \citep{Haywood2013}.

To summarize, we thus assume that the  inner (thick +thin) discs and the outer disc are two systems that  can be mostly considered  to be the result of two  distinct chemical evolution paths, and will be modeled as such in this paper.  In the next Section, we present the data with which we compared our model, and the  main characteristics that our best fit SFH -- and associated chemical evolution -- must be able to reproduce.

\begin{figure}
\centering
\begin{tabular}{c} 
\includegraphics[width=9.5cm]{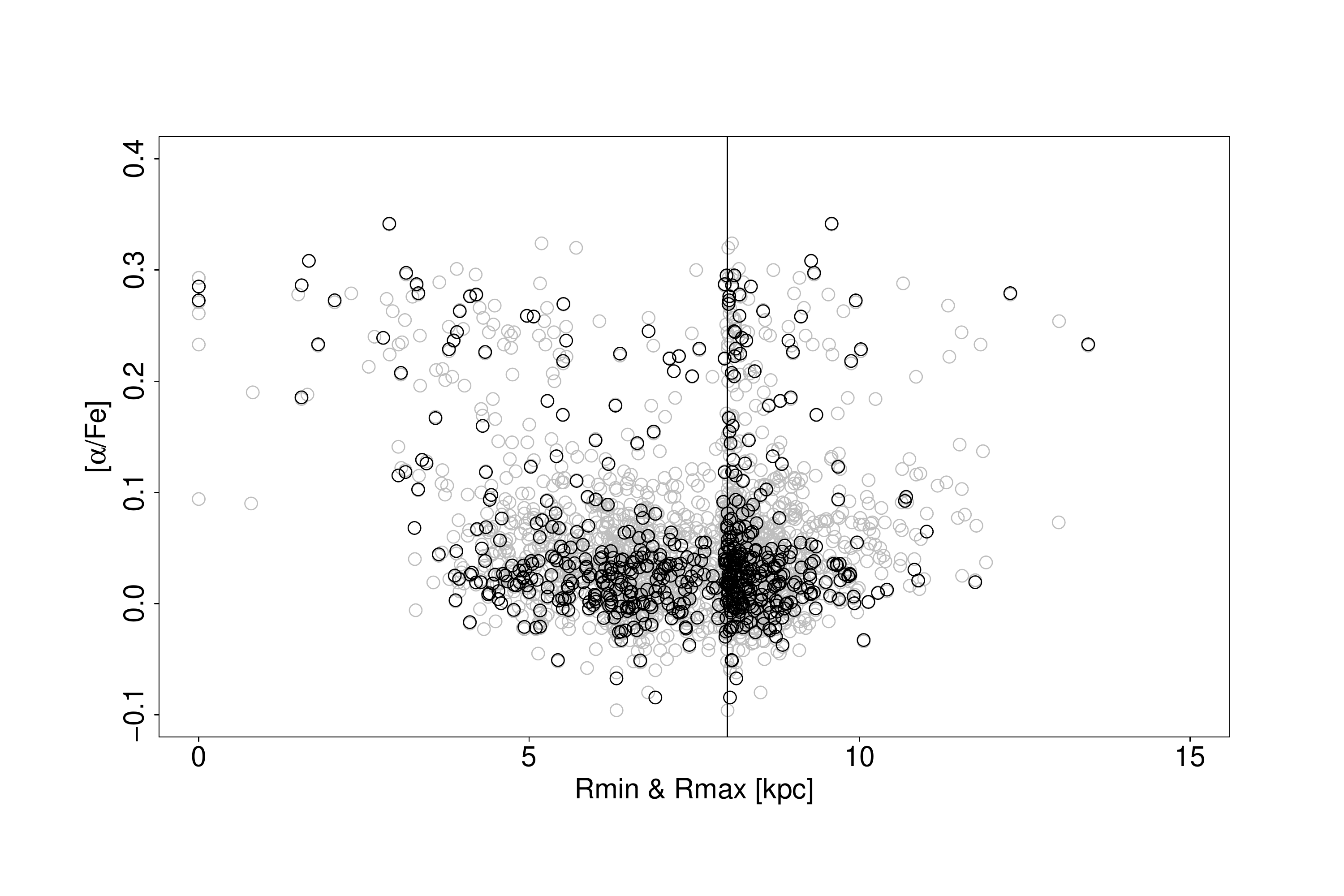}\\
\end{tabular}
\caption{Pericenters (at R$<$8kpc) and apocenters (at R$>$8 kpc) $vs$ [$\alpha$/Fe]
for stars in the Adibekyan et al. sample. The stars with age determinations used in the present paper
are shown in black. The data illustrate that the thick disk 
stars at [$\alpha$/Fe]$>$0.2 dex span the whole range of radii from the galactic center 
to $\sim$10kpc.
}
\label{orbits}
\end{figure}

\subsection{Data}\label{data}

The original observations used in this paper are from \citet{Adibekyan2012}. This was a spectroscopic survey aimed at providing atmospheric parameters and chemical compositions for nearby stars for the purposes of extrasolar planet research. This produced a sample of 1111 stars providing values for temperature, elemental abundances, stellar velocities, and  associated errors. 

Since our results are derived from fitting the age-alpha element distribution, {\it not stellar densities}, no bias due to the volume 
definition of the sample is expected. Three other biases could however be important: the first is due to the measured abundances. 
More than any other sample, we expect this one to have excellent abundance measurements because of the very high signal to noise ratio
advocated in \citet{Adibekyan2012}. Moreover, comparison of abundances with other state-of-the-art spectroscopic samples show
good agreement.
The second type of biases could come from the age estimates. The uncertainties for the ages -- both systematic and random -- were estimated
in \citet{Haywood2013}, and we refer the reader to \citet{Haywood2013} for the full details of how the ages were derived for the sample. We will, however, provide a brief summary here. We derived individual ages using the bayesian method of \citet{Jorgensen2005}, adopting the Yonsei-Yale (Y2) set of isochrones \citep[version 2,][]{Demarque2004}. 
\citet{Haywood2013} estimated that the uncertainties in the atmospheric parameters translate into uncertainties of about 0.8 Gyr 
for the younger stars (age $<$ 5 Gyr) and 1.5 Gyr for the oldest (age $>$ 9 Gyr). Because systematic errors on the effective temperature scale
induce large errors on the derived ages \citep[see][]{Haywood2006}, the effective temperatures from \citet{Adibekyan2012} were also checked 
with the V-K-T$_{eff}$ scale from \citet{Casagrande2010}. An attempt to estimate systematics from both uncertainties in stellar models (by adopting
another set of isochrones) and method (bayesian estimate compared to $\chi^2$ fitting) was also carried out, and yielded absolute (systematic) uncertainties of 1 
to 1.5 Gyr, and random uncertainties of  0.8 to 1.5 Gyr.
The final source of potential bias we considered could be induced by incomplete coverage of the alpha elements over the
whole age interval. However, within the age interval considered here,  the age-alpha relation is extremely well defined. A consequence of this, however, was a severe pruning of the sample to keep only the best ages.

The \citet{Adibekyan2012} sample was used by \citet{Haywood2013} to derive ages for a cleaned sample of 363 stars with their associated error estimates. 

In their paper, \citet{Haywood2013} developed a scenario where the Galaxy is composed of several distinct components, an inner and outer thin disc, and a thick disc. Their interpretation has been discussed extensively in the previous subsection, and is modeled and tested in this paper. 

In Fig. \ref{Fig:Dataobshaywoodetal}  we show the distribution of stars, comparing  [$\alpha$/Fe] vs [Fe/H], [$\alpha$/Fe] vs age and [Fe/H] vs age. We
differentiate the inner (thick and thin) disc and outer disc.

We use silicon in order to follow the $\alpha$ element evolution. The reason we chose this element and not the $\alpha$ as defined by the \citet{Haywood2013} paper is that the abundance ratios of the various $\alpha$ elements are much larger for the theoretical yields than in the observations. This means that using $\alpha$ as described in \citet{Haywood2013} is not possible using \citet{Anders1989} solar normalisations. Thus, we cannot use the definition of $\alpha$ used in \citet{Haywood2013}, and we cannot expect much agreement between the chemical tracks of the different $\alpha$ elements.  Silicon is, however, different from other alpha elements, such as magnesium, in that it is produced in significant quantities by both SNII and SNIa. This is an important difference, because more of the $\alpha$ total value is produced in a few Myr for magnesium, while the stars release silicon over longer time periods.

The [Si/Mg] value is significantly larger using the theoretical yields than in the data (assuming \citet{Anders1989} solar values). We chose silicon because it is the element where our model can fit both the [$\alpha$/Fe]-age distribution and the [$\alpha$/Fe]-[Fe/H] distribution using the same SFR. This inability to match magnesium evolution to the data may be due to either shortcomings in the model itself or the theoretical yields. However, although several GCE papers \citep[e.g.][]{Francois2004} have studied several elements and compared them to data they tend to only fit the [$\alpha$/Fe]-[Fe/H] distribution and not to the level of precision we require\footnote{A stringent test of the chemical yields is beyond the scope of this paper.}. Silicon allows us to achieve a good fit to the data, while the other elements tend to fail to produce a fit in [$\alpha$]-age-[Fe/H]. 

We will discuss in greater detail, in section \ref{diffel}, the reason for choosing to follow silicon rather than an alternative $\alpha$ element.

\subsubsection*{[Si/Fe] versus [Fe/H] relation}

In the [Si/Fe] vs [Fe/H] plane (Fig. \ref{Fig:Dataobshaywoodetal}, panel (a)), the inner thick and thin discs still show two distinct patterns, with the thick disc forming a sequence of decreasing [Si/Fe] for increasing metallicities. At the end of this sequence, for [Si/Fe]$\sim$ 0.05 dex and [Fe/H]$\sim$ -0.1 dex, the thin disc appears as a nearly flat track, with nearly constant [Si/Fe] values (with a mean around solar) and [Fe/H] between -0.2 dex and 0.4 dex. The low metallicity tail of the thin disc sequence ([Fe/H]$<$-0.2-- -0.3 dex) is the locus of outer disc stars. 
Interestingly those stars have [Si/Fe] values larger than those characterizing inner thin disc stars, reaching as high as 0.15 dex. This places the outer disc as an intermediate sequence (in $\alpha-$ abundances) between the thick and thin discs. 

As anticipated, we separate the outer disc and treat it as a separate component from the point of view of chemical evolution. Formally, outer disc stars are defined as those which lie inside the region  defined by the relation :
\begin{align}
&\rm{[Si/Fe]} < -\rm{[Fe/H]}/6.+1/15.~dex \\
& \rm{[Fe/H]} < -0.2~dex
\label{Eqn:Select}
\end{align}
\noindent
shown in Fig. \ref{Fig:Dataobshaywoodetal}, panel (b).

These limits are somewhat arbitrary. An a posteriori justification (which came after the present work was submitted) for this choice comes from the APOGEE survey \citep{Nidever2014}
which has confirmed the scheme presented by \citet{Haywood2013} that the outer and inner disk are the result of two different paths of chemical
evolution, and as is modeled here. The APOGEE data clearly shows that the inner disk stars of solar alpha abundance have essentially
metallicity higher than -0.2 dex (see. Fig. 11 of \citealt{Nidever2014} or Fig. 14 of \citealt{Anders2014}). We give further details below
on the origin of the stars in our sample.

\subsubsection*{[Si/Fe] versus age relation}
The thick disc is made of stars with ages greater than 8 Gyr, and shows a rapid evolution of [Si/Fe] with time, almost 0.3 dex in 6 Gyr (see Fig. \ref{Fig:Dataobshaywoodetal}, Panel (b)).  The idea that the ISM is well mixed is indicated by the low scatter in [Si/Fe] with age for old stars in the thick disc (Fig. \ref{Fig:Dataobshaywoodetal}, \citet{Haywood2013, Adibekyan2012}).

The (inner) thin disc is made of stars younger than 8 Gyr and its [Si/Fe] evolution vs time is significantly flatter than that of the thick disc, only about 0.05 dex over 8 Gyr. A key characteristic of the transition between the thick and thin discs in the [Si/Fe] vs age plane is that it is apparently very sharp. It forms a knee where the gradient of [Si/Fe]-age evolution drops very suddenly. 

Finally, the oldest stars in the outer  disc form at approximately 10 Gyr and evolve parallel to the inner thin disc, but with a higher [Si/Fe] ratio. It is as if the outer  disc turns off the thick disc sequence earlier than the (inner) thin disc.

\subsubsection*{[Fe/H] versus age relation}

Once outer disc stars are separated from inner disc stars, the thick disc sequence appears in the [Fe/H] vs age plane (Fig~\ref{Fig:Dataobshaywoodetal}, panel (c)) as a well defined track. This track is characterized by a small scatter around the mean value for all ages between 8 and 13 Gyr.  As extensively discussed in \citet{Haywood2013}, the metallicity homogeneity that characterizes the  thick disc phase can be interpreted as the signature of an intense episode of star formation at those times. As a result, the feedback from SNe explosions efficiently mixed metals, and assured a homogeneous ISM during the whole phase of thick disc formation. This efficient mixing must also have limited the formation of any radial metallicity gradient: the small scatter around the mean [Fe/H] vs age relation of the thick disc is evidence that any gradient, if present, must have been weak, otherwise the spread in the relation would appear much higher \citep{Haywood2013} .

The signature of the existence of a steep metallicity gradient in the transition zone (7-9 kpc) between the inner
and outer thin disk  
is imprinted in the significant scatter observed in the [Fe/H] vs age relation at ages $<$ 8 Gyr. 
The APOGEE survey \citep[see][] {Anders2014, Nidever2014} has 
shown that stars with [Fe/H]$>$+0.2 dex and [Fe/H]$<-0.2$ dex are  present at just R$<$7 kpc and R$>$9kpc respectively,
which explains why simple blurring can bring these stars to the solar orbit.
 
The outer disc, as already noted in the $\alpha/Fe$ vs age plot, starts to appear about 10 Gyr ago at metallicities $\sim$ -0.7 dex.  Its subsequent [Fe/H] pattern is always below that characterizing the inner thin disc. The metal enrichment proceeds in the outer disc at a rate similar to that of inner disc stars, reaching values of about -0.3 dex, for ages $<$ 2 Gyr. 

\subsubsection*{Origin of the stars}

Fig. \ref{orbits} shows the alpha abundance of the stars in the sample as a function of their pericentre (at R$<$8 kpc)
and apocentre (at R$>$8 kpc). 
The orbits of the stars were computed from U,V,W velocities provided in the Adibekyan et al. catalogue,
and an axisymetric potential from \citet{Allen1991}. The mean orbital parameters were derived from orbits integrated for 5 Gyrs. 
The figure illustrates that thick disk stars are present at all radii, from the Galactic centre
to about 10 kpc, confirming that stars in the solar vicinity probably originated at a large range of radii. 
Note that, contrary to other studies, we do not assume that (thick or thin) disk stars have migrated.
In addition to the detailed justifications given in \citet{Haywood2013}, 
Fig. \ref{orbits} shows clearly that thin disk stars have pericentres that bring them to distances of less than 5 kpc from the centre (towards the Galactic centre) to distances of more than 9 kpc (towards
the anticentre).

The metal-poor stars are characterised by having a tendency to travel along their orbit faster than the local standard of rest (LSR), as shown in \citet{Haywood2008} -- whereas the thick disk, at the same metallicities, lags the LSR. 

These stars have been characterised on larger orbits, see \citet{Bovy2012c}, figure 7, which shows that these objects have 
average guiding centers larger than the solar radius ($R_{mean} > 9$~kpc).

\subsection{The Ram\'irez et al. (2013) Sample}

\begin{figure*}
\centering
\includegraphics[width=7.in, clip]{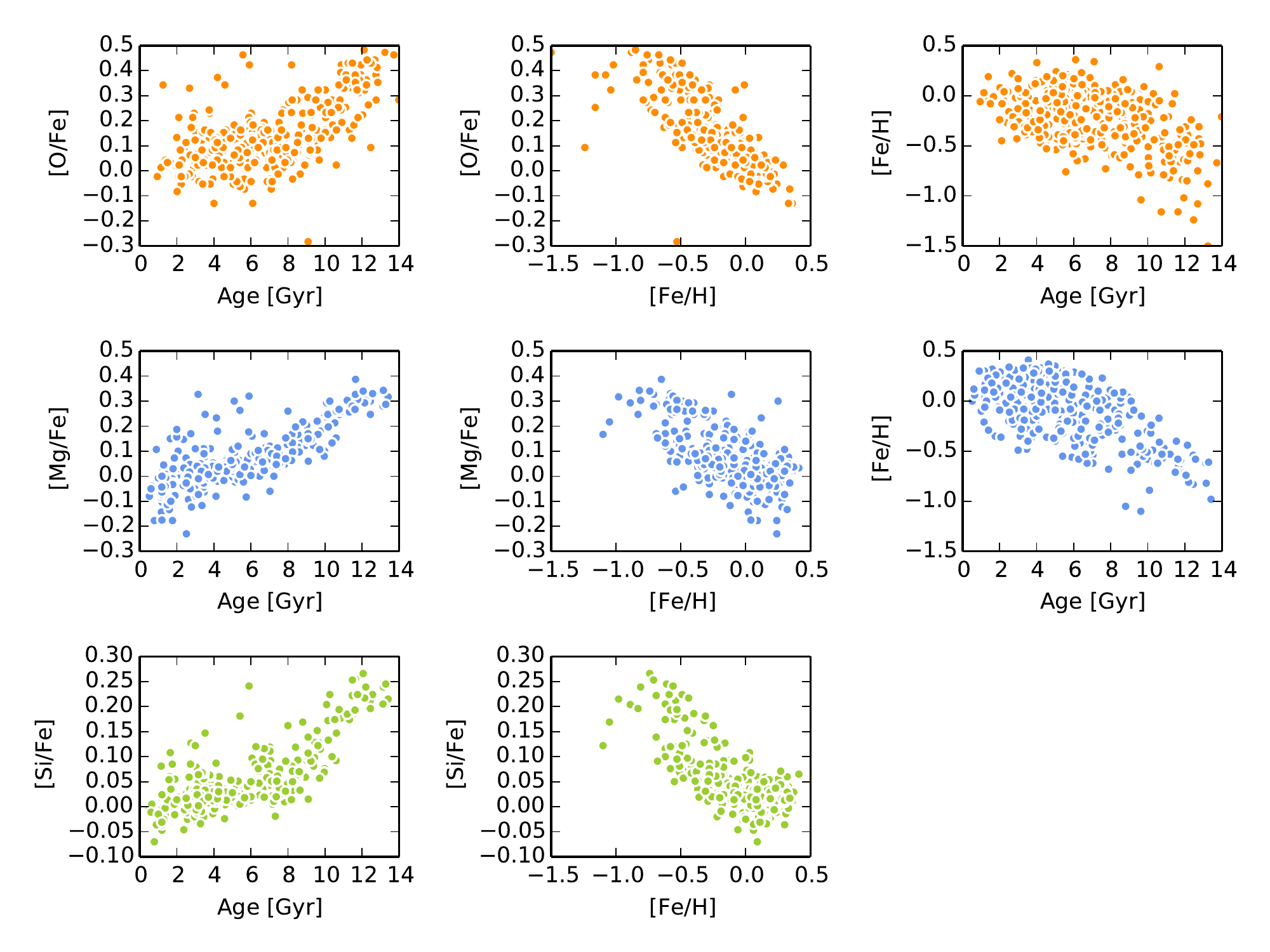}\\
\caption{Comparison of the different $\alpha$ species and datasets. From top to bottom we show the \citet{Ramirez2013} oxygen data (orange) and the \citet{Adibekyan2012} and 
\citet{Haywood2013} magnesium (blue) and silicon (green) data. The columns are for age-[$\alpha$/Fe], [Fe/H]-[$\alpha$/Fe] and age-[Fe/H] distributions. The [Fe/H]-age distributions are the same for both silicon and magnesium so are only shown once.}
\label{Fig:Combes}
\end{figure*}

We have mentioned above that we have chosen silicon as our marker for the $\alpha$ abundance for the majority of this paper.  Figure \ref{Fig:Combes} compares the silicon and magnesium and age 
from \citet{Adibekyan2012} and \citet{Haywood2013} with the oxygen abundances and ages presented in \citet{Ramirez2013}  for a sample of 391 FGK dwarfs and subgiants from the solar vicinity. 
The mean error in age is of the order of 2-3 Gyr, and about 0.05-0.06 dex on [O/Fe], as quoted by the authors. The remarks concerning possible biases in our
sample also apply to \citet{Ramirez2013}.
The oxygen data have a considerable scatter compared to silicon and magnesium, due to the larger uncertainties. The data with the tightest correlations are clearly the silicon. 
However, the key features of the [$\alpha$/Fe] evolution is visible in each plot. We see a steep early time slope with very little scatter followed by a sharp transition to a shallower slope. 
Unfortunately, the split between inner and outer discs visible in the [$\alpha$/Fe]-[Fe/H] distribution of the \citet{Haywood2013} dataset cannot be seen for the \citet{Ramirez2013},
but the change of slope due to the transition from thick to thin disk in oxygen abundance is clearly seen at about 8 Gyr, although both the spectroscopic data, element, and
ages are different.

~\\
\section{Impact of star formation histories on chemical abundance patterns}\label{Patterns}

Over the next few years a wealth of high quality spectroscopic data, together with highly accurate measurements of stellar ages, are expected as a result of on-going spectroscopic surveys, and the {\it Gaia} mission. 

These data should allow us to measure the evolution of most chemical species in the Milky Way in detail, and so reconstruct the  SFH of the Galaxy. However, we still have very little knowledge of which processes generate the chemical abundance  patterns observable in a galaxy. The features in the abundance data for the Galaxy are usually only discussed qualitatively, or over a restricted range of parameters which are tuned to fit solar neighborhood data.

In this section we present how the chemical evolution of the model is affected by the SFH we provide.  We will explore how the various parameters and ingredients generate, and affect, the chemical  patterns which result.

\subsection{An exponential star formation rate}\label{exp}

\begin{figure*}
\centering
\begin{tabular}{cc} 
\includegraphics[width=3.5in, trim={0cm 0cm 0cm 0cm}, clip]{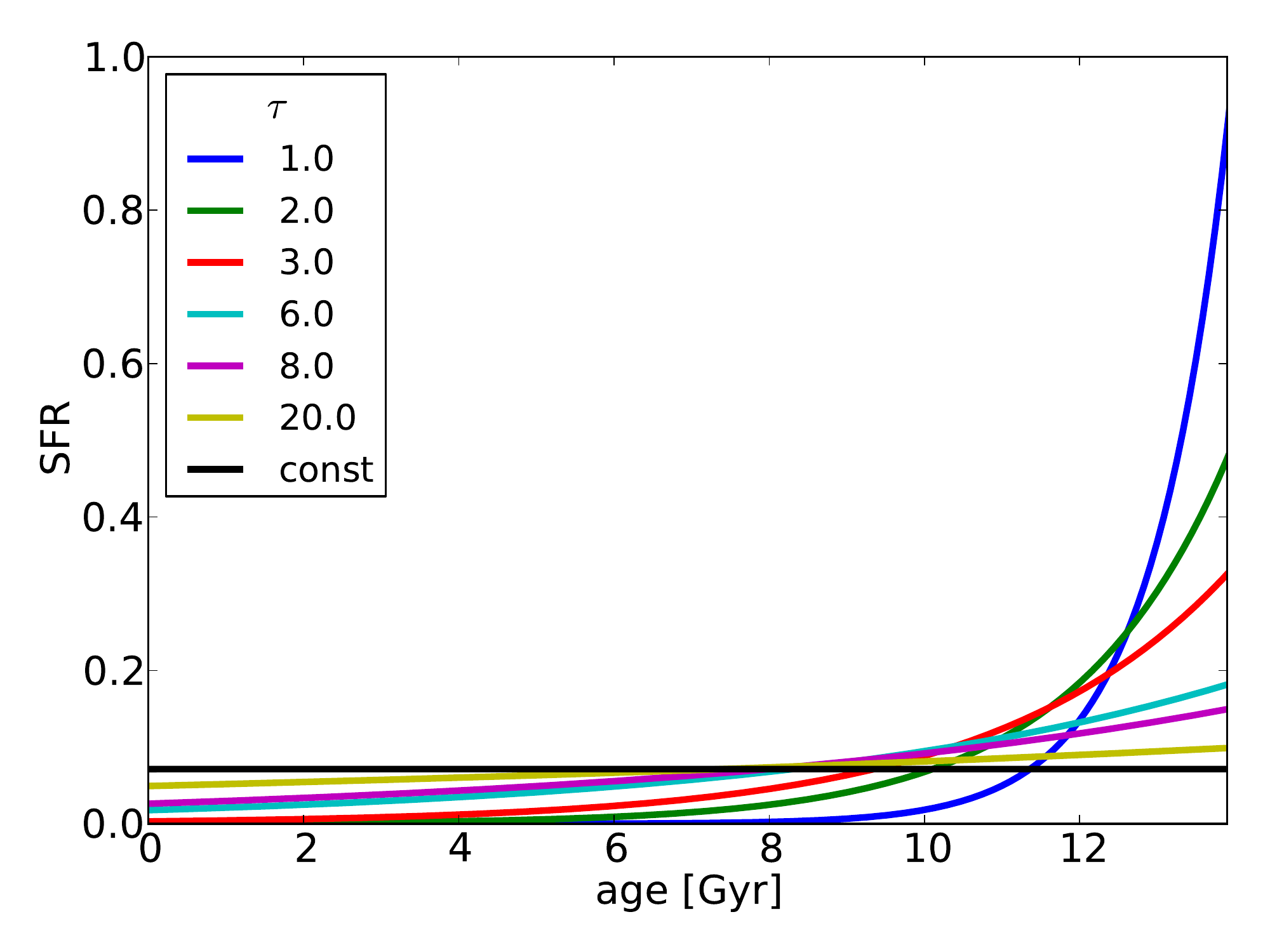} &
\includegraphics[width=3.5in, trim={0cm 0cm 0cm 0cm}, clip]{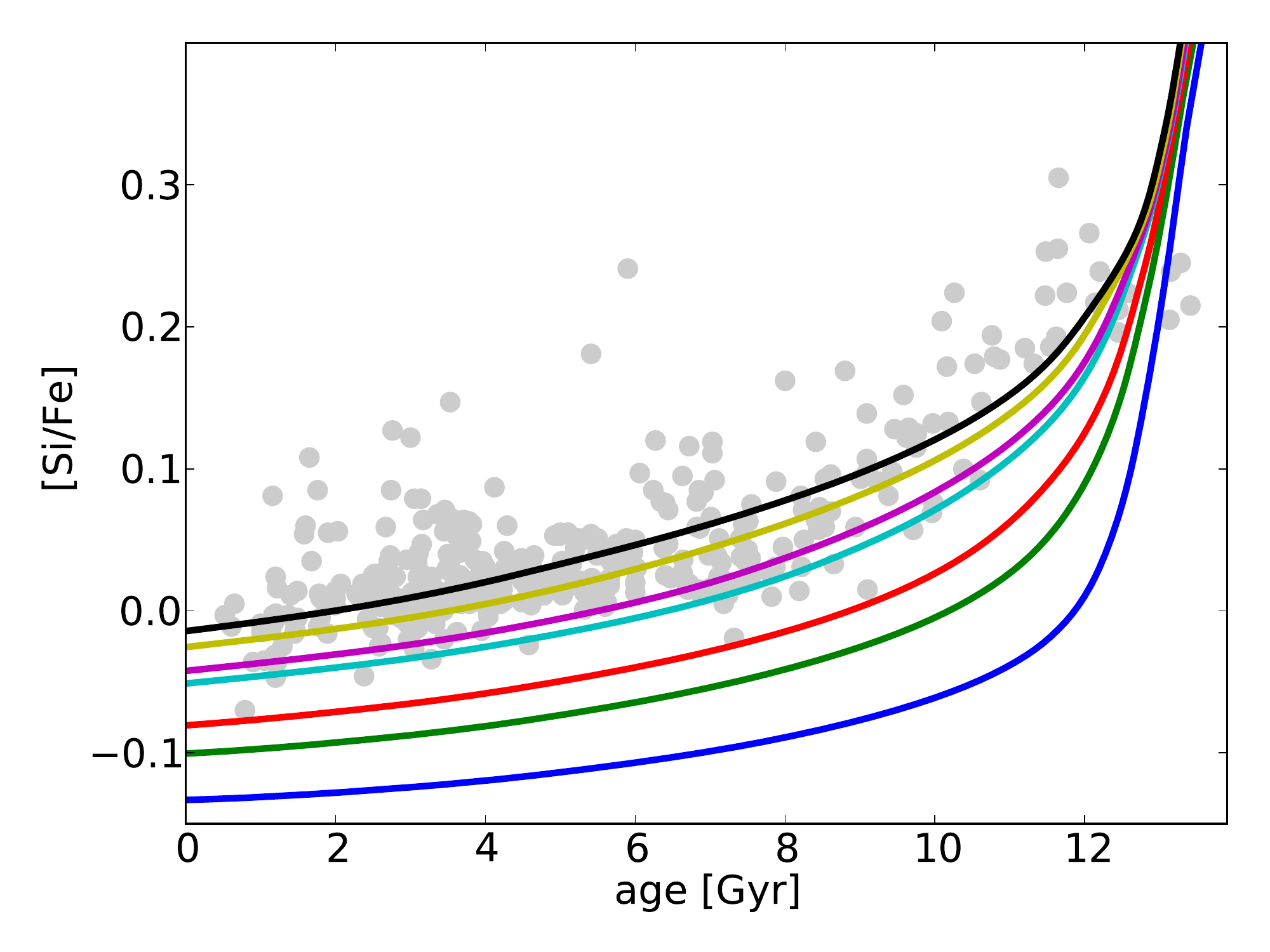}\\
(a) & (b)  \\
\includegraphics[width=3.5in, trim={0cm 0cm 0cm 0cm}, clip]{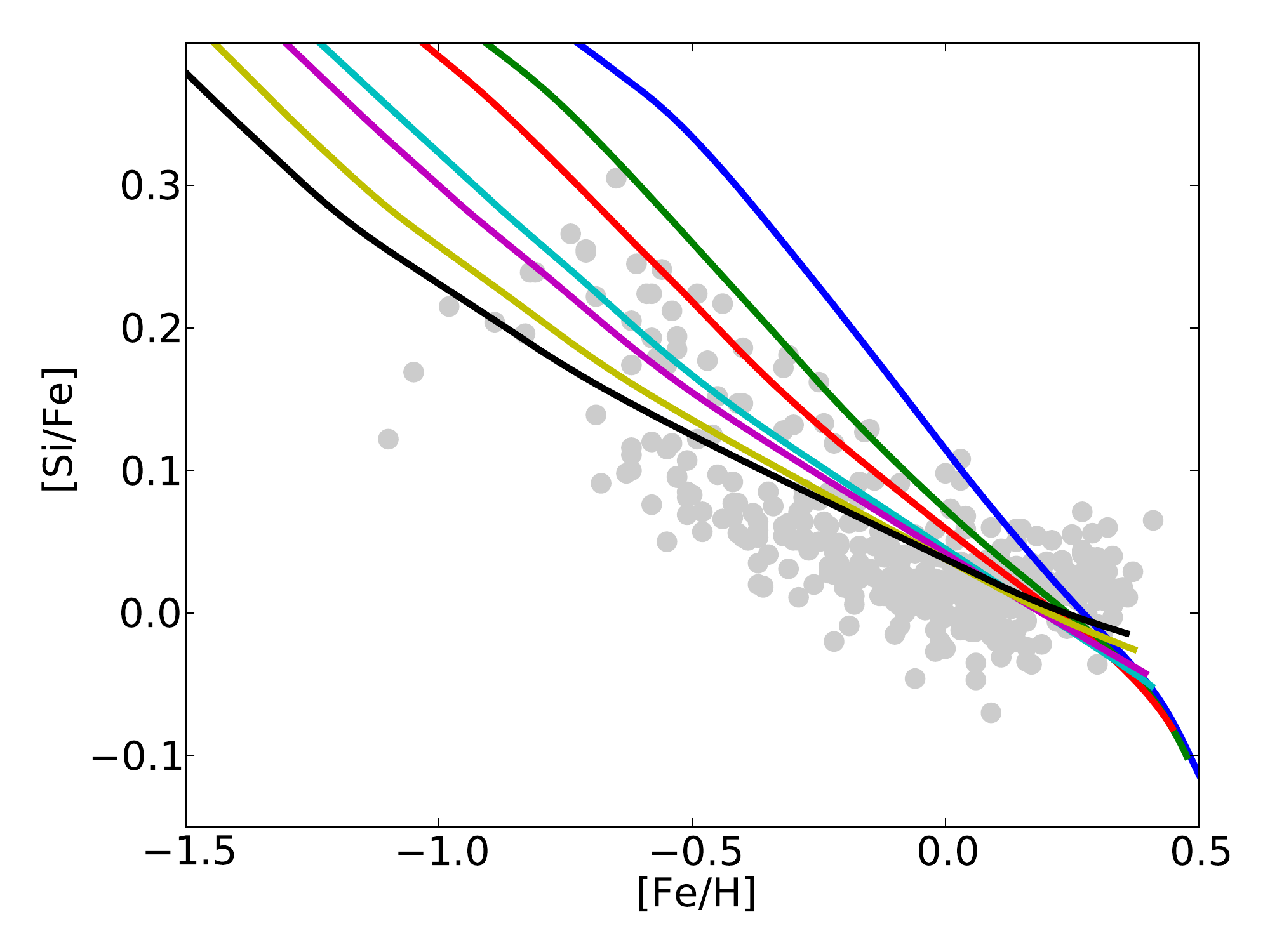} &
\includegraphics[width=3.5in, trim={0cm 0cm 0cm 0cm}, clip]{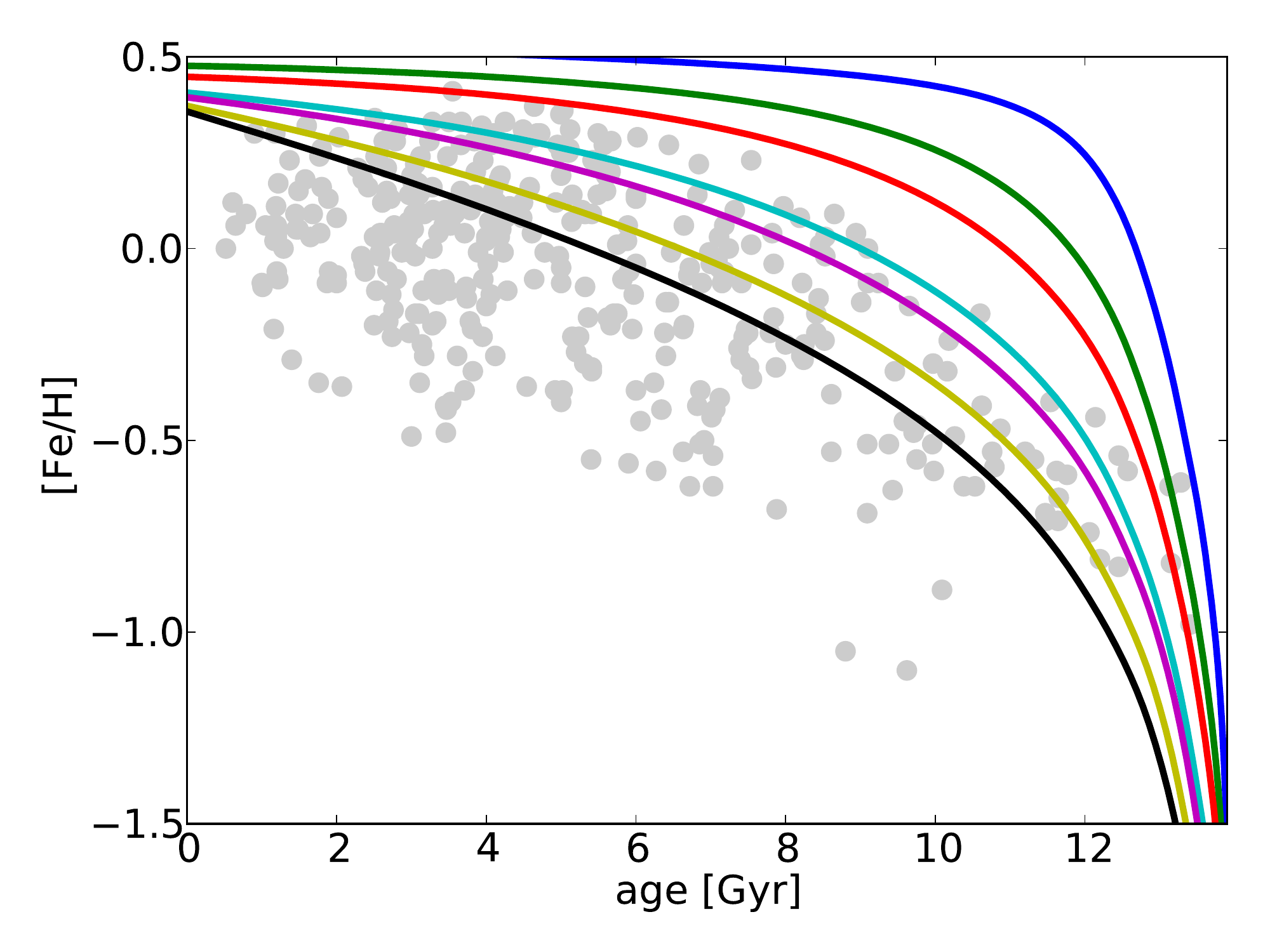}\\
(c) & (d)  \\
\end{tabular}
\caption{Chemical evolution tracks in the [Si/Fe] vs age (panel (b)), [Si/Fe] vs [Fe/H] (panel (c)) and [Fe/H] vs age (panel (d)) planes, produced by exponential SFRs, with different $\tau$ timescales, (panel (a)).   The grey points are the solar neighborhood data  from \citet{Haywood2013} and include both inner and outer disc stars. }

\label{Fig:SFRexponential}
\end{figure*}

We start by using a SFR  with an exponential distribution of the form,

\begin{equation}
SFR(t) = A \exp(-t/\tau),
\label{Eqn:NormSFR}
\end{equation}

\noindent
where $t$ is the time in Gyrs since the beginning of the `universe', $\tau$ is the characteristic time scale, and A is a constant. We vary A so that the total mass of stars formed over the whole time interval  is equal to 1, meaning that in the absence of recycled gas from stellar evolution, all the ISM is consumed. We vary the value of $\tau$ between 1 Gyr and 10000 Gyrs so that the SFR ranges from a very high initial value followed by a rapid decline to an essentially constant function (see  Fig.~\ref{Fig:SFRexponential}, panel (a)).  

We chose an exponential form for the SFR at this stage because it is  a simple function and closely mimics the most commonly occurring forms in the literature e.g. \citet{Matteucci1989}, \citet{Chiosi1980}, \citet{Chiappini1997}, and \citet{Fenner2002} for a SFR based on the surface density of infalling gas and the Schmidt-Kennicutt relation \citep{Kennicutt1998}.

From Fig. ~\ref{Fig:SFRexponential} we can deduce some important features of the evolution of the SFR and the chemistry:
\begin{enumerate}
\item A high initial SFR followed by a rapid decline (e.g. $\tau=1$) induces rapid fall in the [$\alpha$/Fe] ratio, while higher values of $\tau$ produce a more gentle decrease of the [$\alpha$/Fe] ratio with time.
\item A  low $\tau$ SFH also generates a too rapid enrichment in [Fe/H], such that the metallicity reaches solar values after only about 1 Gyr. Such a rapid increase in the metal content for low $\tau$ values is also the cause of the rapid decrease in [$\alpha$/Fe] discussed at the previous point.
\item The higher the characteristic time, $\tau$, the closer the corresponding tracks are to the [Si/Fe] and [Fe/H] versus age data (see panels (b) and (d)). The constant SFR, in particular, seems a good solution that reproduces reasonably well the chemical patterns as a function of age. However, it is able to reproduce neither the thick disc nor the thin disc sequence in panel (c). 
\item In the age-[Si/Fe] plot (panel (b)) it is clear that the `knee' feature, which can be seen in the data (Fig. \ref{Fig:Dataobshaywoodetal}), is most evident in the model tracks where the   initial SFH is large and the sharpness of the contrast between the early and late SFR is greater. This suggests that a sharp transition is required in the SFH of the Galaxy in order to mimic the chemical abundances of the \citet{Adibekyan2012} stars, and that the contrast between early and late time Galactic SFR must be considerable. 
\item In most GCE models the [$\alpha$/Fe]-[Fe/H] distribution (as in our panel c) is most commonly used in order to test the conformity of models to observations. There is considerable variation in the tracks followed by the model in [$\alpha$/Fe]-[Fe/H] space as a result of changing $\tau$. The figure clearly shows that a wide range of $\tau$ values are a fairly good fit to the data. More specifically, panel (c) shows that SFHs with characteristic times $\tau$ between 3 and 6 Gyrs pass through the thick and thin disc sequences, with $\tau=3$ perhaps the best solution. However, none of these values can adequately represent  the sequences in the [Si/Fe]-age and [Fe/H]-age plane. This indicates that [$\alpha$/Fe]-[Fe/H] patterns are not sufficient to disentangle the SFH of the Galaxy, and that the knowledge of ages is essential for this purpose. 
\end{enumerate}
 
We note that the $\tau$=6 Gyr SFR is  the steepest exponential  allowed by the data. If more stars are formed at earlier times, then the line no longer passes through the [Si/Fe]-age distribution or the upper sequence of the [Si/Fe]-[Fe/H] panel. Choosing $\tau=6$ results in an evolution of [Fe/H] with age which covers the top envelope of the data. 

However, the exponential functional form does not  fit the old `thick disc' stars in panel (b). The [Si/Fe]-age evolution of the Milky Way data can be fit by two straight lines plus a knee. This is not what we see in the chemical tracks generated by exponential SFRs,  because the change in slope is much more gradual. This discrepancy suggests that another functional form is more appropriate. \\

\begin{figure*}
\begin{tabular}{cc} 
 \includegraphics[width=3.5in, trim={0cm 0cm 0cm 0cm}, clip]{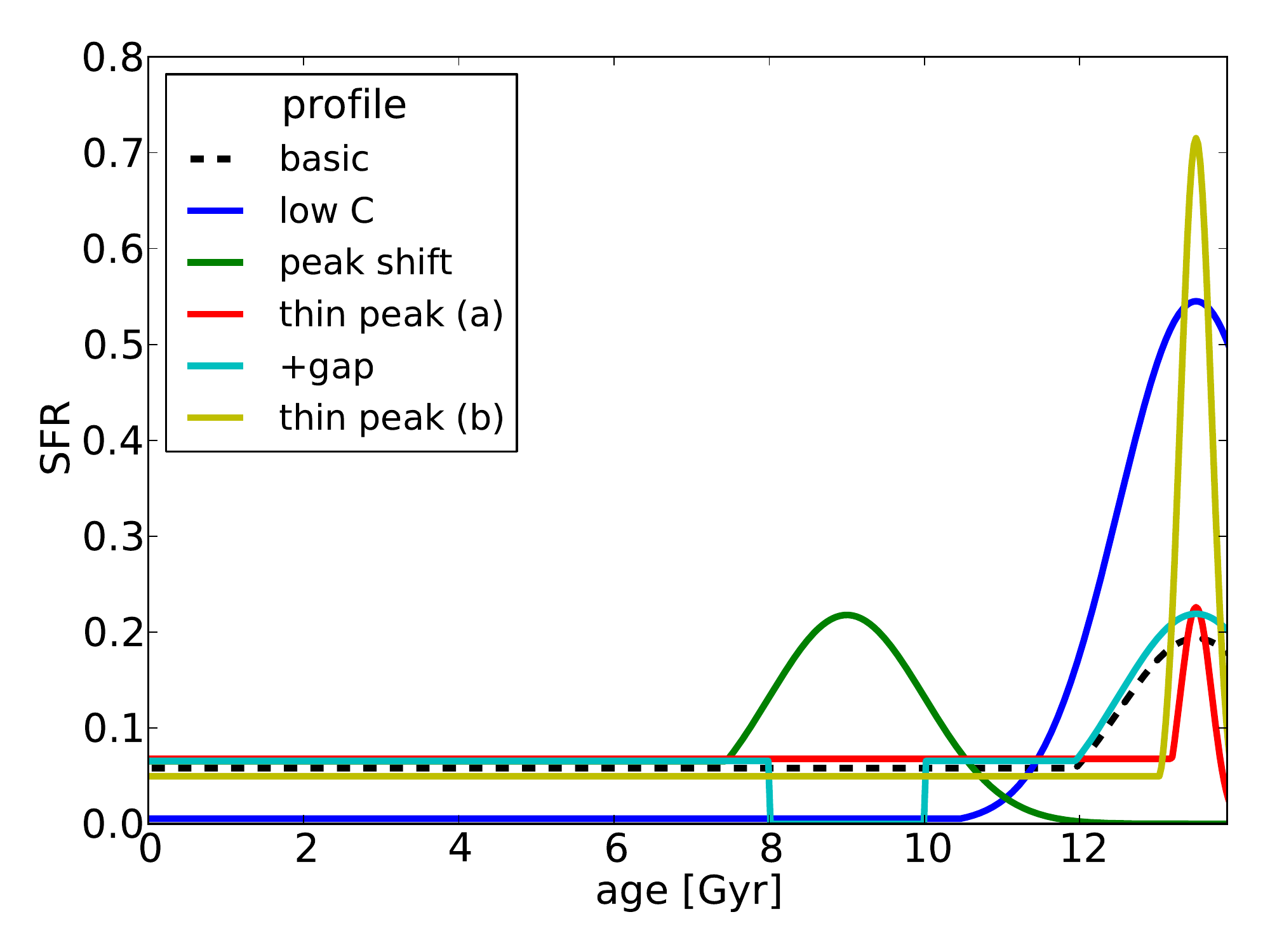} &  
 \includegraphics[width=3.5in, trim={0cm 0cm 0cm 0cm}, clip]{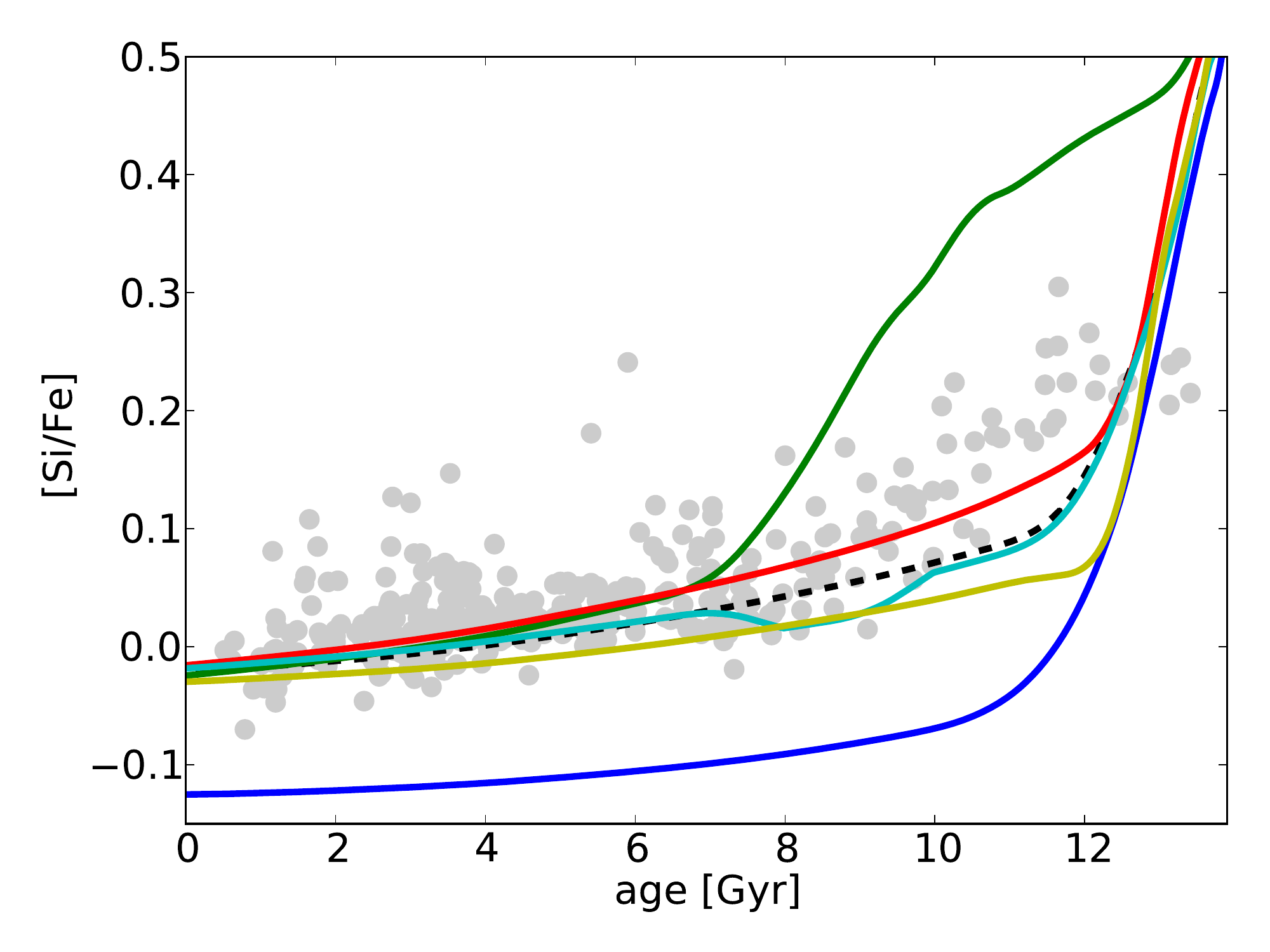}\\
(a) & (b)  \\
\includegraphics[width=3.5in, trim={0cm 0cm 0cm 0cm}, clip]{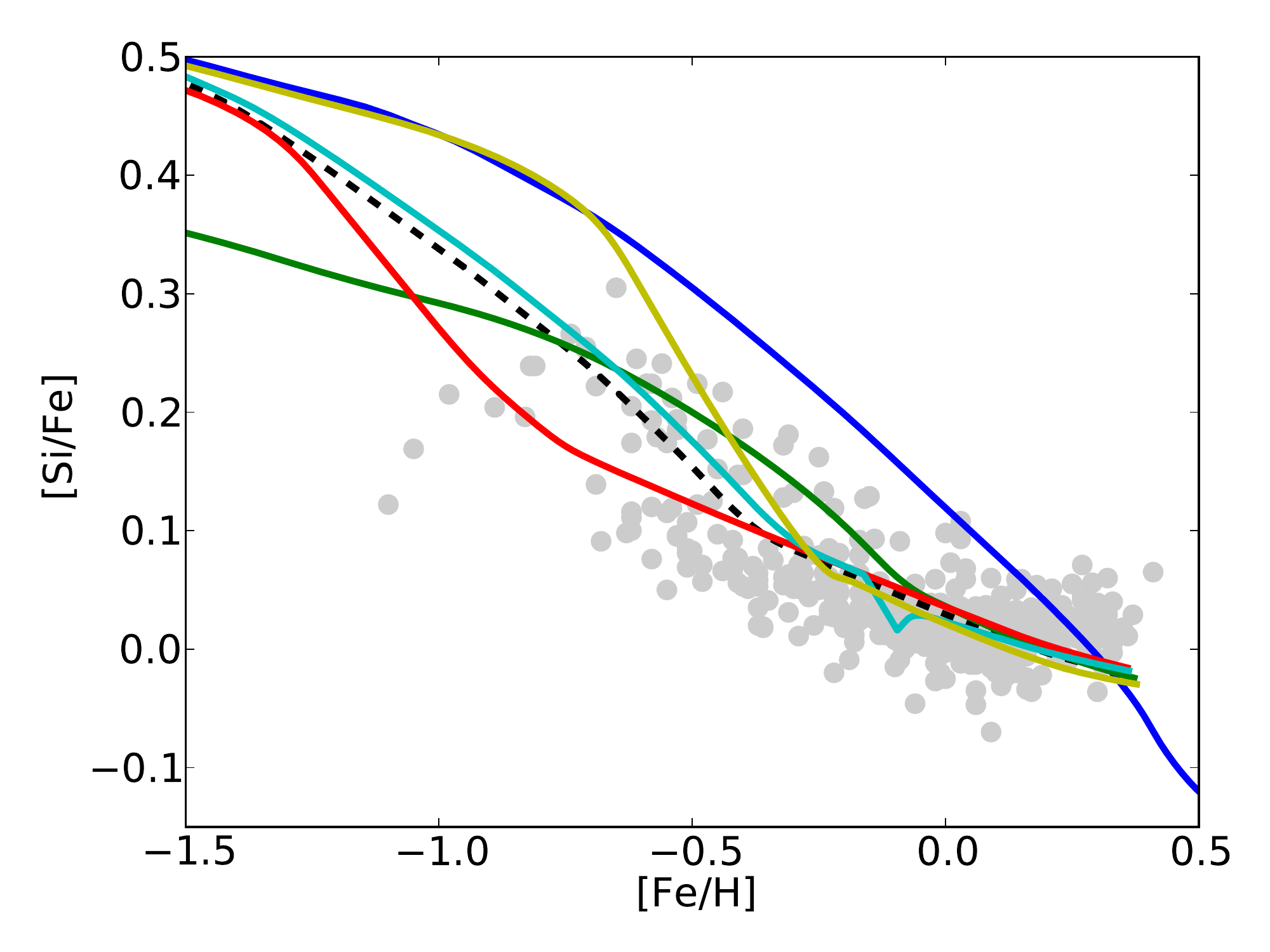} & 
\includegraphics[width=3.5in, trim={0cm 0cm 0cm 0cm}, clip]{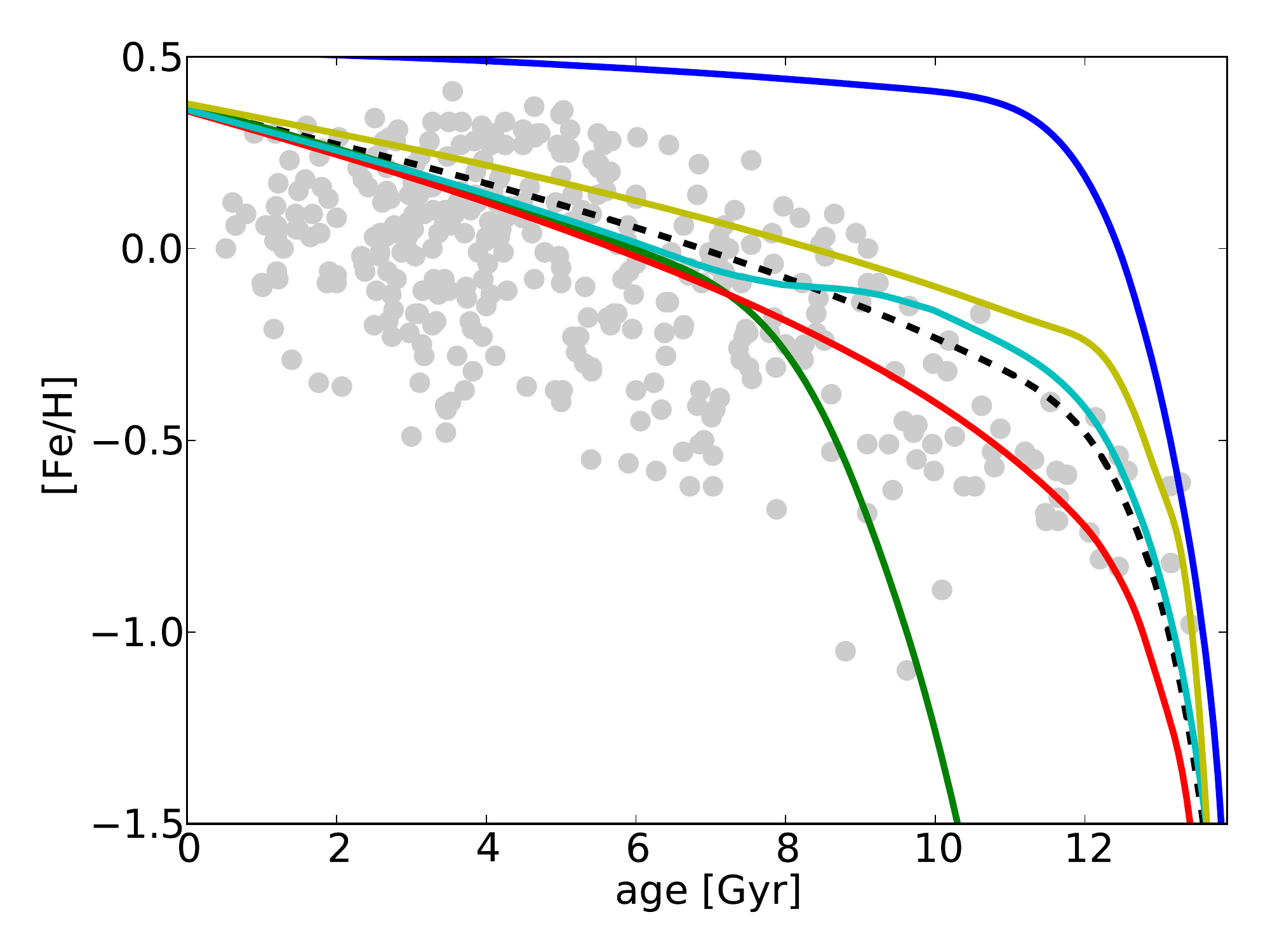}\\                         
(c) & (d)  \\
\end{tabular}
\caption{Chemical evolution of the model depending upon a selection of different SFHs (panel a, Table \ref{Tab:Favourites}). Panels  (b), (c) \& (d)  show the evolution of [Si/Fe] with age, [Si/Fe] with metallicity and the metallicity with age. The grey points are the \citet{Haywood2013} data, including inner and outer disc stars. }

\label{Fig:SFRfavouries}
\end{figure*}

\begin{table}
\centering
\begin{tabular}{cccccc}
\hline
 &\multicolumn{2}{c}{Gaussian} & \multicolumn{3}{c}{Other} \\
 \hline
Name & $\mu$ & $\sigma$ & gap width & gap position & c \\
 \hline
basic                 & 0.5 & 1.0 & 0. & N/A & 0.30\\
low C                 & 0.5 & 1.0 & 0. & N/A & 0.01\\ 
peak shift            & 3.0 & 1.0 & 0. & N/A & 0.01\\       
thin peak (a)         & 0.5 & 0.2 & 0. & N/A & 0.01\\ 
+gap                  & 0.5 & 1.0 & 1. & 9   & 0.01\\
thin peak (b)         & 0.5 & 0.2 & 0. & N/A & 0.01\\
  \hline
\end{tabular}
\caption{The parameters for the different SFH's shown in Fig. \ref{Fig:SFRfavouries}. The SFH consists of a Gaussian peak, with $\mu$ and $\sigma$ corresponding to the mean and width of the peak (see Eqn. \ref{eqn:compSFH}). Low C indicates that the constant part of the SFH has a low value. The thin peak SFRs (a \& b) change the width of the peak. Thin peak (a) shows a SFH where the peak is narrow and the height of the constant part is the same as for the  basic case. Thin peak (b) is where constant is adjusted so that the mass fraction formed in the peak is the same as for the basic case.}
\label{Tab:Favourites}
\end{table}

\subsection{A more complex star formation rate}

In Fig. \ref{Fig:SFRfavouries} we show the various forms of a more complicated SFH. We choose a form consisting of an initial burst of star formation, modeled with a Gaussian, followed by a quiet period, modeled using a constant star formation rate. This SFH is of the form,

\begin{equation}
 SFR= \left\{
  \begin{array}{ll}
    \exp(-(t-\mu)^2/2\sigma^2) & \quad \text{if SFR$>$C and t$<\mu$ }\\
    C & \quad \text{otherwise,}
  \end{array} \right.
 \label{eqn:compSFH}
\end{equation}
\noindent
where $t$ is the stellar age, $\mu$ is the position of the peak in Gyr, $\sigma$ is the width of the Gaussian and C is the height of the constant part of the SFH. We also allow for the existence of a gap in the SFR such as the one introduced by \citet{Chiappini1997}. This functional form is more similar to the SFRs recovered by some $\Lambda$CDM cosmological simulations \citep[e.g.][]{Brook2013, Stinson2013, Aumer2013}. Table \ref{Tab:Favourites} shows the various parameters used for the different SFHs adopted in Fig. \ref{Fig:SFRfavouries}. These parameters were chosen to provide, as much as possible, an overview of the different chemical features produced by SFHs of this type.

While none of these SFRs is a good representation of the data, we can see that:
\begin{enumerate}

\item Many of them  reproduce a `knee'  feature in the [Si/Fe] vs age plot (panel b) and the timing of the knee appearance is related to the  end of the burst phase (described by the Gaussian). Moreover, the higher the ratio between the peak of the SFR in the burst phase, and the value in the constant SFR regime, the sharper the knee feature. There is no sign of such a feature in the  lower $\tau$ SFRs in Fig. \ref{Fig:SFRexponential}. 

\item  The `knee' arises when the SFR, and thus the SNII rate, falls rapidly or becomes zero (for example the cyan line in  Fig. \ref{Fig:SFRfavouries}). When this occurs the contribution of silicon from the SNII is reduced within a few Myrs (essentially in step with the SFR) but, because of the longer time delay distribution for SNIa, the iron continues to be produced at almost the same rate for over a Gyr. Thus, the amount of iron continues to build up, due to SNIa from stars which have previously formed. It is only when SFR subsequently increases do the alpha elements start to enrich the gas again. This causes a drop in the [Si/Fe] and thus makes the `knee' sharp.  

\item  A delayed star formation burst (peaking at $t\sim$ 9 Gyr) implies too high [Si/Fe] values for the whole thick disc phase and also a too slow metal enrichment. The delayed star forming burst does, however, fits the metallicity-[Si/Fe] distribution (panel c). 

\item A 1-2 Gyr halt in the star formation causes a decrease in the [Si/Fe] values at about the same time, as well as a local minimum in the [Fe/H] vs age plot. This is for the reasons outlined in point 2.

\item  The data are essentially bracketed by SFRs that have an initial peak of star formation at $t=$9 Gyr and $t=$13.5 Gyr, respectively: both fit the [$\alpha$/Fe]-[Fe/H] (panel c) sequence, but provide too low or too high alpha values and metallicities in the thick disc phase. However,  the thin disc phase is fitted equally well for both observables.
\end{enumerate}

From these experiments, it is clear that changing the SFH has a considerable effect on the chemical evolution of a galaxy. Moreover, as already noticed in Sect.~\ref{exp}, we see that using a [$\alpha$/Fe]-[Fe/H] distribution alone is not sufficient to properly distinguish between different SFHs. For a good comparison between models and data age information is required.

\begin{table}
\centering
\begin{tabular}{ll}
  \hline
  \multicolumn{2}{c}{Yields} \\
  \hline 
	 SNII & \citet{Nomoto2006}\\
	 AGB & \citet{Karakas2010}\\
	 SNIa & \citet{Iwamoto1999}\\
	\hline
    IMF & \citet{Kroupa2001}\\
    \hline
     \multicolumn{2}{c}{Solar abundances}\\
     \hline
	Si & 7.55 \citep{Anders1989} \\
	Fe & 7.51 \citep{Anders1989} \\
\hline	
\end{tabular}
\caption{Summary of the parameters used in Section 5.}
\label{Tab:summarymodel}
\end{table}

\section{Reconstructing the SFR of the Milky Way}\label{SFR}
Based on our exploration of the effect of a given SFR on the chemical abundance history, we now attempt to use the chemical evolution of the Galaxy, as traced by \citet{Haywood2013}, to recover the SFH. After describing our fitting procedure, we first derive the star formation history of the inner (thick+thin) disc, then that of the outer disc. We assume the inner and outer discs to be two systems which evolve independently, as described in Section \ref{scene}. The properties and ingredients of our  model are given in table \ref{Tab:summarymodel}. The solar iron normalisation is the `best' value for the iron normalisation and is not the one used  to normalise the \citet{Adibekyan2012} data  (the value used is  7.47 based on \citet{Gonzalez2000}). It matches \citet{Anders1989} and allows a very good fit to both the age-[Si/Fe] and metallicity-[Si/Fe] distributions

\subsection{Fitting procedure}

To recover the star formation  history of our Galaxy in the last 13 Gyr, we have chosen to fit the evolution of [$\alpha$/Fe] with age. This choice is motivated by the fact that this relationship is much more dependent on the SFH than the [$\alpha$/Fe]-[Fe/H] distribution (Fig. \ref{Fig:SFRfavouries}). 

As shown in the previous section (see Fig. \ref{Fig:SFRexponential}), two SFHs are likely to bracket all other possible tracks: the constant SFR, and  a sharp initial burst containing the vast majority of the star formation, which results in both a very rapid increase in the metallicity, and a sharp decrease in the [$\alpha$/Fe] distribution. We found that of the three elements we have chosen to follow (silicon, magnesium and oxygen) only the silicon and oxygen chemical tracks generated by these two limiting 
SFHs bracket the observed abundances, when adopting  the same (photospheric) solar abundances used in the data. For this reason, and because oxygen data is considerably noisier than silicon, we deduce the SFH from fitting our model to the evolution of [Si/Fe] with age.

Our fitting procedure is based on a simple $\chi^2$ approach. The algorithm requires an initial guess SFH, and a user supplied convergence criterion. For our initial condition we use a SFH defined by the function,
\begin{equation} 
     f(t)= A/(0.01t+0.1) 
\end{equation}
\noindent
where A is the normalisation and t is the time. We also tested other SFH forms, such as a sharp initial burst, constant star formation or a  high initial SFR followed by a linearly declining SFR.  We find that the SFHs all converge to the same result, except for the constant SFR. The constant SFR converges to {\it almost} the same age-[Si/Fe] distribution and final SFH but it lacks the dip and therefore the sharp break in the age-[Si/Fe]  track. However, using an alternative fitting algorithm (Powells algorithm, see Fig.  \ref{Fig:withpowell}) this difference vanishes. This algorithm seems more robust when fitting, but takes considerably longer to converge. 

We divide the age range (0-14.0 Gyr) into 28 nodes separated by 0.5 Gyr and use a $\chi^2$ fitting procedure to calculate the SFR. Once we have provided an initial SFH we allow minimisation to take place using this initial guess. At each iteration an entire SFH is created, and the chemical evolution code is used to calculate the corresponding chemical track. We then calculate the $\chi^2$ value using the difference between the model value of [Si/Fe] at the age of each star in the data, and the observed [Si/Fe],  for all stars with ages lower than 13 Gyr. We set this maximum age limit because there are very few stars with ages greater than this, and so this region is difficult to constrain reliably.

The algorithm uses an N dimensional simplex and follows a series of steps, attempting to find a minimum. Convergence is considered to be where a step of the algorithm is smaller than a given minimum, or that the decrease in $\chi^2$ is less than a given value. 

Once  a best fit SFH is found on the whole data sample, we use a bootstrap procedure to quantify the sensitivity of the SFH to the data.
We then average the various solutions obtained in order to smooth the star formation history.

As discussed previously, we normalise the SFH such that initially the ISM has a mass of 1 and, over the 14 Gyrs of evolution the integral of the SFH is equal to one. This means, that without gas release, the total mass of stars would be unity, and the mass of the ISM would be zero. However, because gas is released  from stellar populations over time, the final gas mass is not zero, and the final stellar mass is not one. After 14 Gyr, the total mass present in the ISM is equal to the mass of gas released from the stars. That is not to say that all the gas present has been recycled through stars, because at each stage the gas and metals released are mixed with the ISM. However, at all times, the total baryonic mass (gas and stars) in the system is one. Once the total amount of star formation is fixed, the only freedom in the system is the actual form of the SFH, not the total amount of stars formed.

\subsection{The inner discs: Thick \& Thin}\label{sec:inner}

\begin{figure*}
\centering
\begin{tabular}{cc} 
\includegraphics[width=3.0in, trim={0cm 0cm 0.0cm 0cm}, clip]{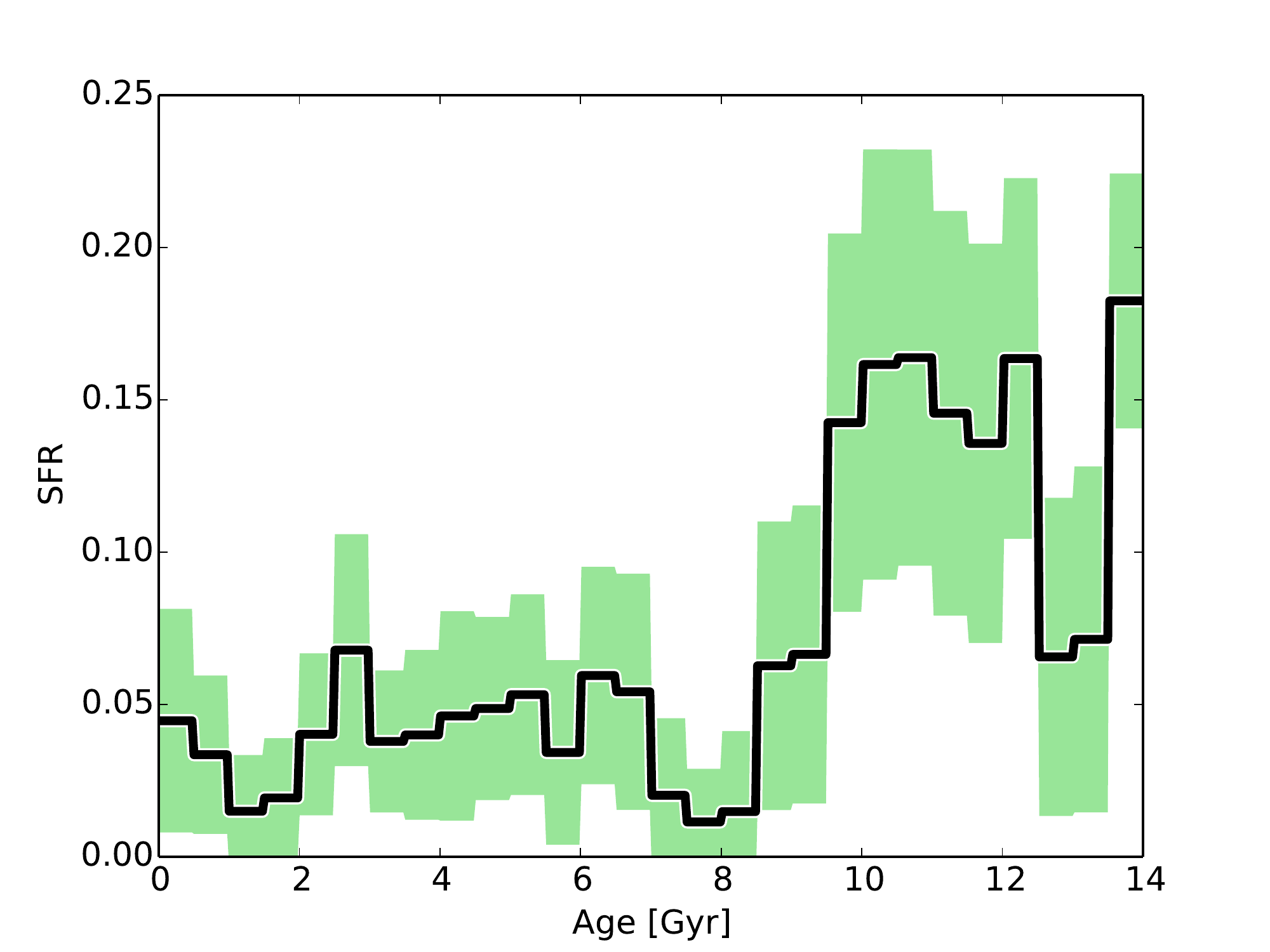}&
\includegraphics[width=3.0in, trim={0cm 0cm 0.0cm 0cm}, clip]{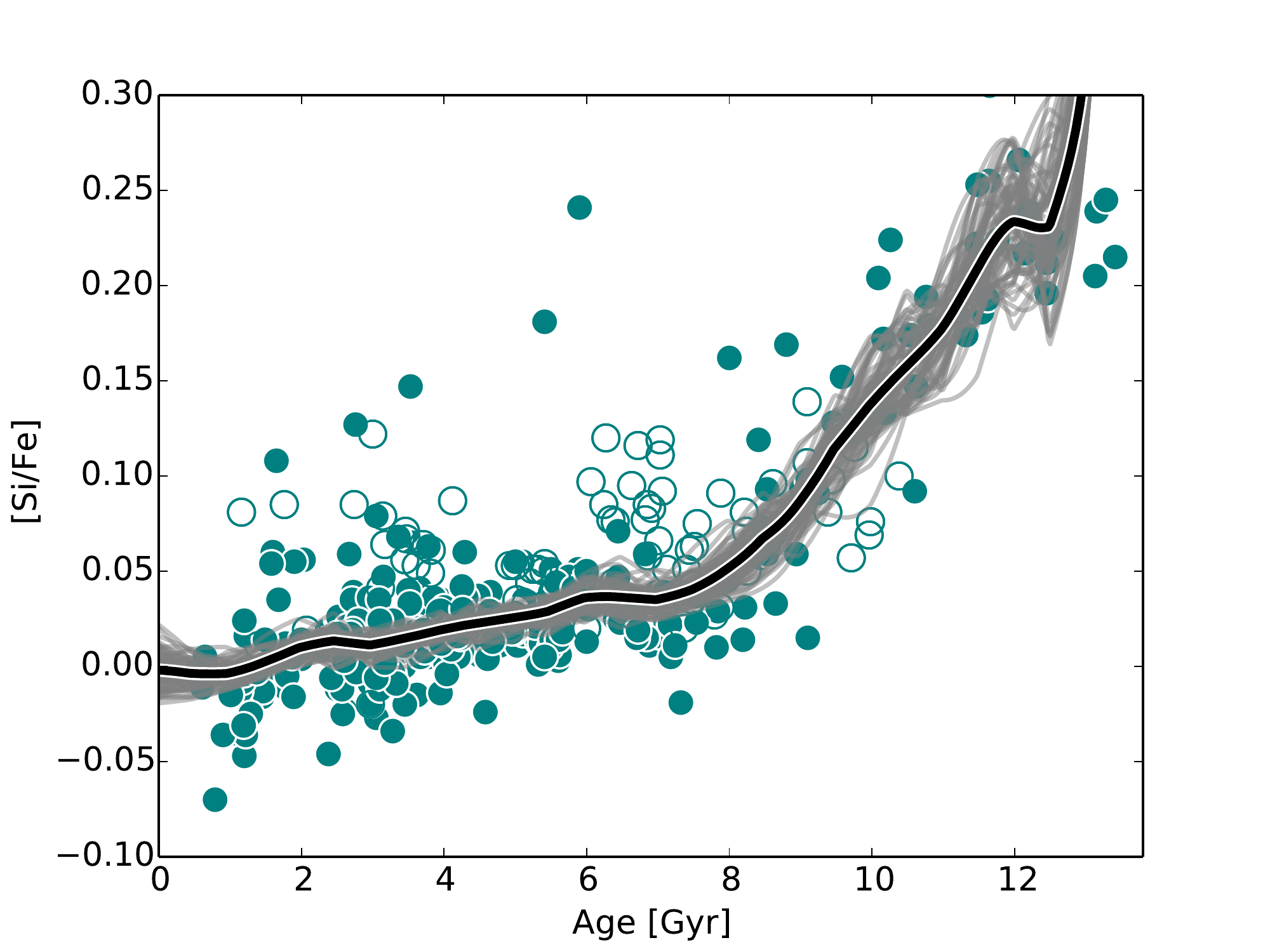} \\
(a) & (b) \\
\includegraphics[width=3.0in, trim={0cm 0cm 0.0cm 0cm}, clip]{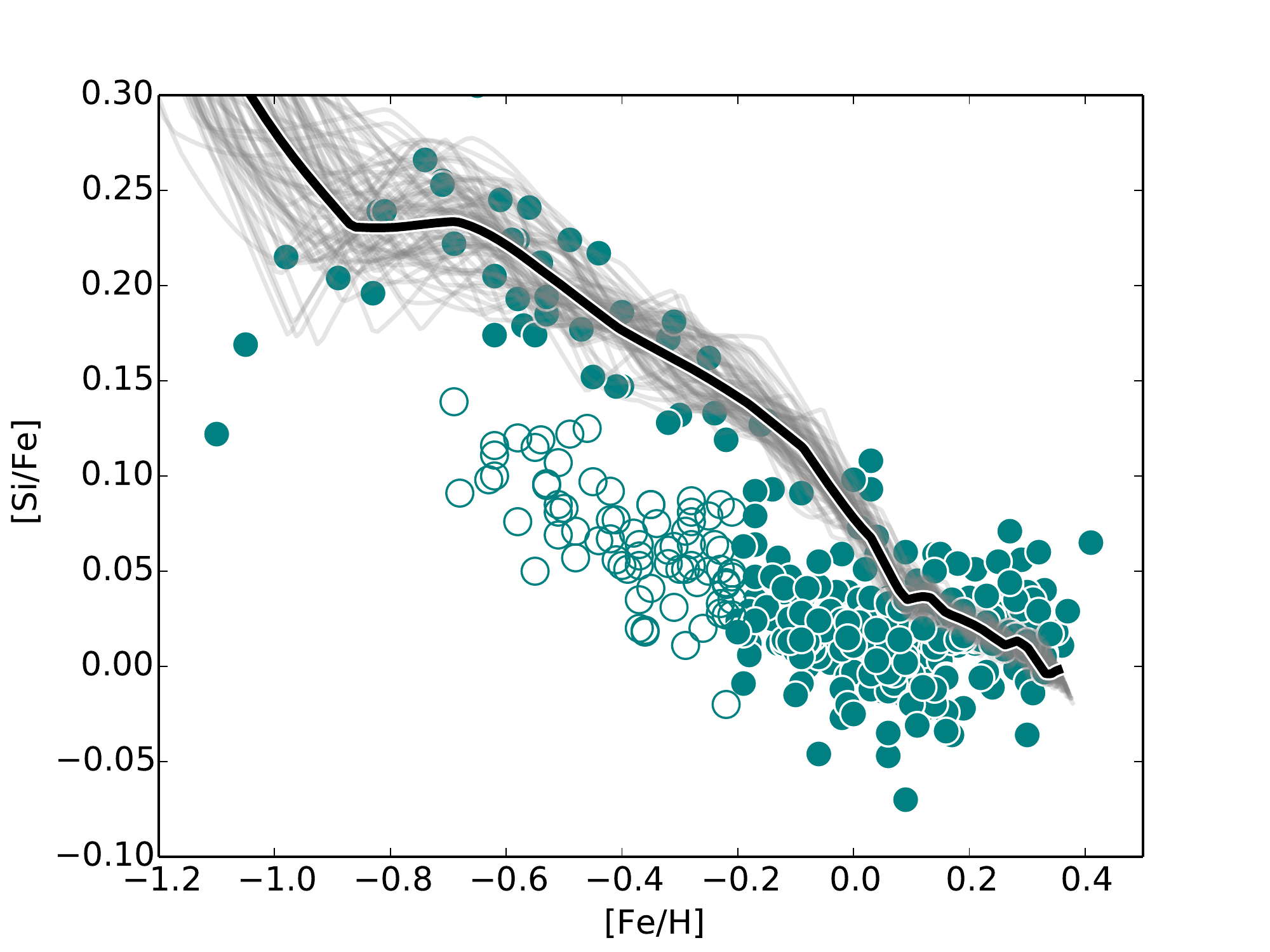} &
\includegraphics[width=3.0in, trim={0cm 0cm 0.0cm 0cm}, clip]{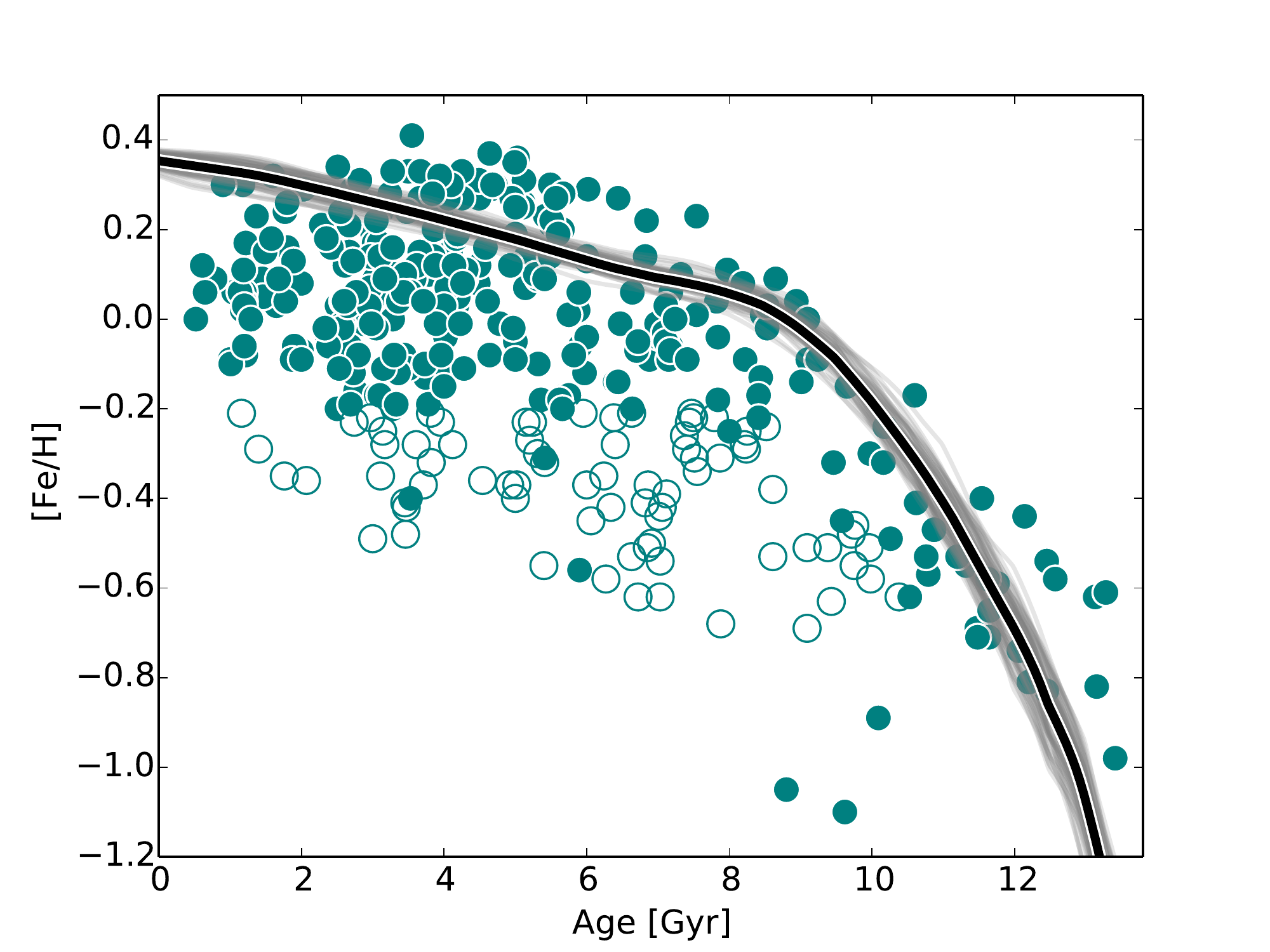} \\
(c)  & (d) \\
\end{tabular}
\caption{The chemical evolution of the best fit SFH for silicon of the inner disc using \citet{Anders1989} solar abundances. The grey lines are the results of the various bootstrapped best fit SFHs, and the black line is the mean value. Filled circles are the inner disc stars, empty circles are the outer disc from \citet{Haywood2013}. In panel (a) the green region is the standard deviation on the values in each bin. These values are not independent and the variance in the total SFH is considerably smaller than in each individual bin. Panels (a), (b), (c) and (d) show the best fit SFH, the evolution of [Si/Fe] with age, the evolution of [Si/Fe] with metallicity and the evolution of the metallicity with age. }

\label{Fig:bestfitsiandersmet}
\end{figure*}

The SFH resulting from fitting  the [Si/Fe]-age sequence of inner disc stars  is given in Fig. \ref{Fig:bestfitsiandersmet} (panel (a)). We show the results of each bootstrap, the mean recovered SFR and an estimate of the variation in each bin in panels (b, c  \& d). It is important to note that while there is considerable variation in each bin (green shaded area), the shape of the overall SFH is more constrained. The  spread of SFRs in each of the 28 bins shown in panel (a) is not independent, and is more constrained globally than in each individual step.\\

\begin{figure}
\centering
\includegraphics[width=3.0in, trim={0cm 6cm 1.5cm 8cm}, clip]{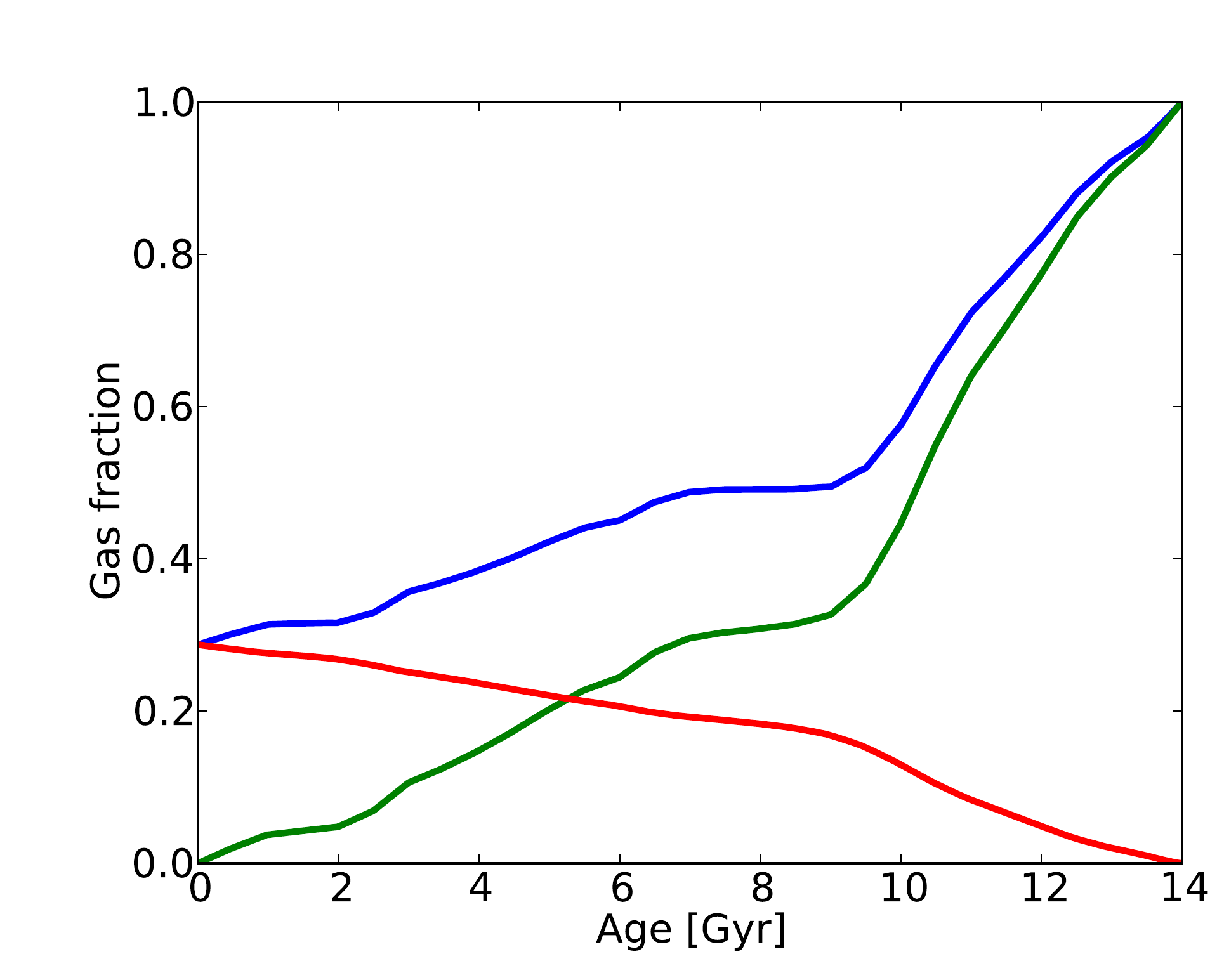}
\caption{The evolution of the gas fraction with time resulting from the mean SFH given in Fig. \ref{Fig:bestfitsiandersmet}. The blue line is the net fraction of gas in the system with time, the red line shows the fraction of recycled gas from stars, while the green line shows the evolution of gas fraction in the absence of gas recycling.  }

\label{Fig:bestfitsiandersmetgasrec}
\end{figure}

We can immediately see various features of the SFH of interest. 
\begin{itemize}
\item  As we are sampling stars from across the entire Galactic disc, and fitting their evolution, we can conclude that the thick disc phase  SFR is three times more intense than the thin disc phase. We therefore expect that the thick disc (formed from stars with ages between 8 and 13 Gyr) contains about 56\% of the total stellar mass of the Milky Ways. We emphasize that this is a {\it ratio} of thick to thin disc star formation rates across the discs rather than an absolute value. 

\item We can also see a dip in the SFR for about 1 Gyr at around 8 Gyr, which is required to form the sharpness of the `knee' between the two slopes.    We note that although is it a common feature of GCE codes -- in which it tends to occur earlier \citep[e.g.][]{Chiappini1997} -- it is the first  time  that it is measured from chemical data.  This dip indicates that some process is required to re-initiate star formation at later times.  Our model cannot suggest what event results in this, although suggestions include external gas accretion or  the cooling of hot gas to restore a depleted cold gas reservoir.

\item The mean SFH in the thin disc phase (age$<$7 Gyr) is compatible with a constant SFR at the level of  4.7 solar masses per year if our total SFH is normalized to the total Milky Way stellar mass from  \citet{McMillan2011}. This is compatible with the SFR measured in the young disc \citep{Diehl2006, Robitaille2010}  of between 4 M$_\odot$/yr and 1.4 M$_\odot$/yr. The last Gyr of star formation is the most poorly constrained, because of lack of stars, and because any given feature in the SFH only affects stars formed after it. This means that old stars have a stronger effect on the chemical evolution than young stars.

\item The error bars reflect the fluctuations in the fitting, due to the varying sample sizes along the [Si/Fe]-Age distribution. The [Si/Fe]-age distribution is less populated in the thick disc phase where the error bars are larger. However, our metallicity is less sensitive to the SFR than in a GCE model which includes infall, because there is a more substantial ISM in which to dilute the metals in our model.

\item Panel (c) shows the model [Si/Fe] vs [Fe/H] track compared to the data. In the thick disc regime (for [Si/Fe]$>$0.05 dex), the model describes the  [Si/Fe] vs [Fe/H] sequence perfectly. Note that although the variation of [Si/Fe] vs [Fe/H] in the thick disc is {\it de facto} a temporal sequence,  this is much less the case for thin disc stars, where the variation of $\alpha$ with metallicity reflects the contamination of the solar vicinity by stars of the inner and outer disc of all ages, not an evolution.  {\it Therefore, the model is not expected to represent the variation seen in the data in the thin disc regime.}
\item The recovered SFH also fits the age-metallicity relation nicely, except for very young ages ($<$ 2 Gyr). We can see that the  relation obtained is above the data. This may be due to the fact that there are only few stars with ages smaller than 2 Gyr, to constrain the late chemical evolution and that once the stellar Z is greater than 1Z$_\odot$ the yields are extrapolations. 

\item The model recovers similar features for each bootstrap, except for some noise. The chemical evolution tracks are much narrower than the SFR, with only slight variations, thus illustrating that the chemical evolution is not as sensitive to short variations in the SFR but to the overall shape of the SFH.    
\end{itemize}

We can see from  Fig.~\ref{Fig:bestfitsiandersmetgasrec} that, at the end of the evolution,  the amount of recycled gas is 30\%. This corresponds to the idea that, of total amount of stars formed during the whole evolution of the system (1, in normalised units), only 70$\%$ is still locked in stars (or stellar remnants such as white dwarfs, black holes etc.) at the final time.  The remaining  30\% having been recycled as gas in the ISM.  This recycled gas fraction depends closely on the chosen IMF, and the quoted value is for the \citet{Kroupa2001} IMF. Other IMFs can give noticeably different values, for example, a  \citet{Scalo1998} IMF produces a recycled gas fraction of around 40\%. This will be discussed more extensively in section \ref{Robust}.

\subsection{The Outer Thin Disc}

\begin{figure*}
\centering
\begin{tabular}{cc} 
\includegraphics[width=3.0in, trim={0cm 0cm 0.0cm 0cm}, clip]{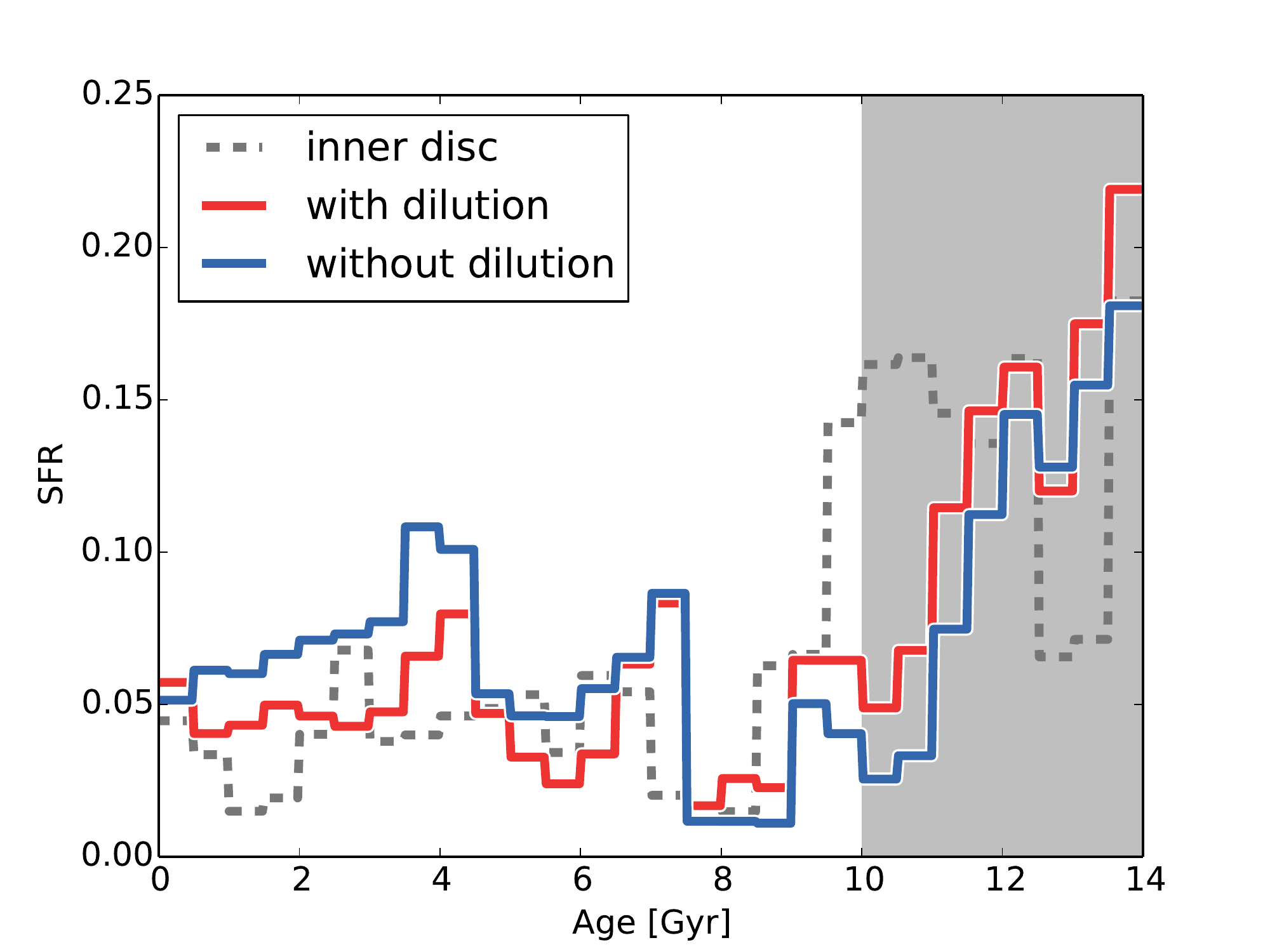}&
\includegraphics[width=3.0in, trim={0cm 0cm 0.0cm 0cm}, clip]{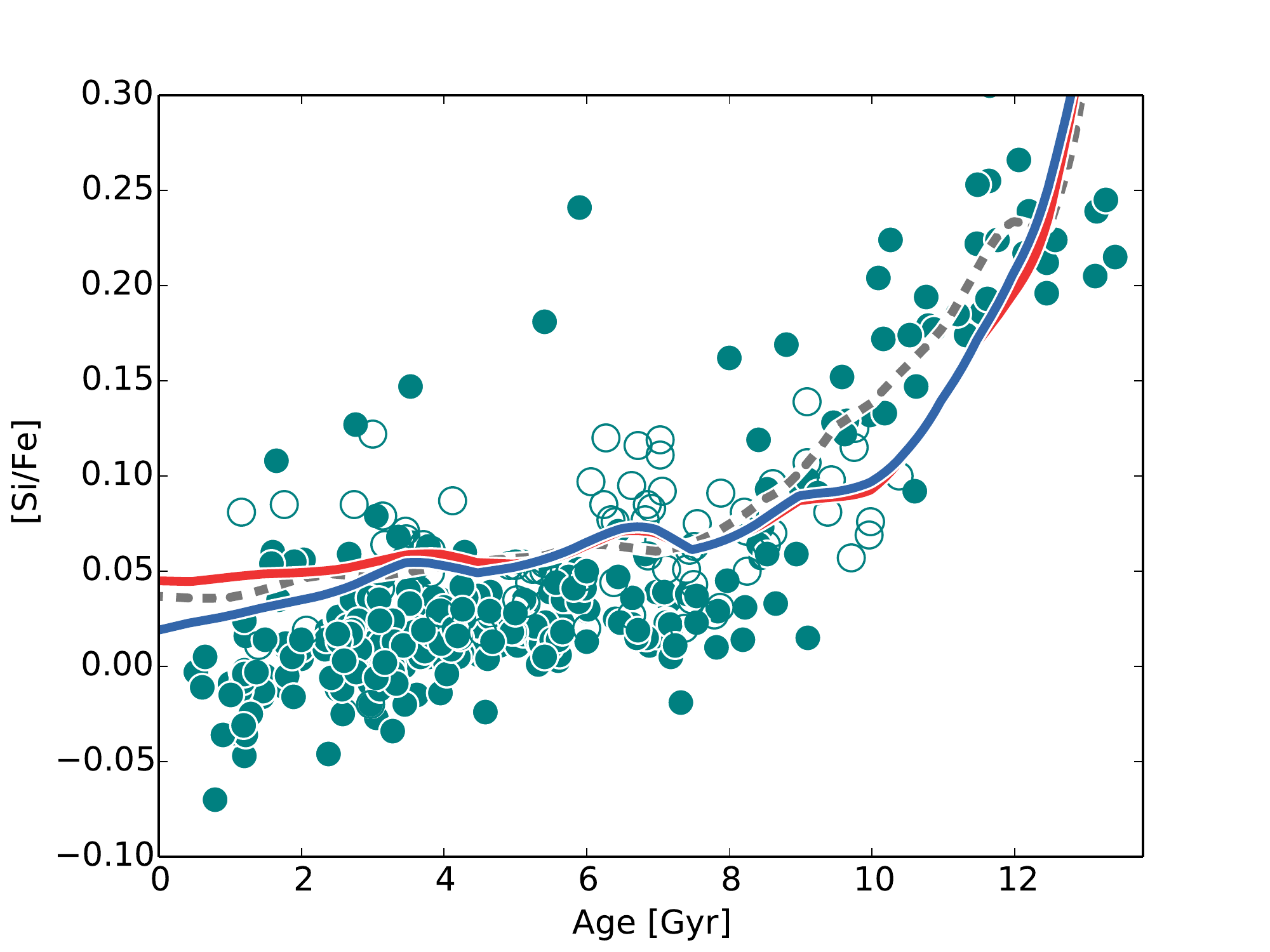} \\
(a) & (b)  \\
\includegraphics[width=3.0in, trim={0cm 0cm 0.0cm 0cm}, clip]{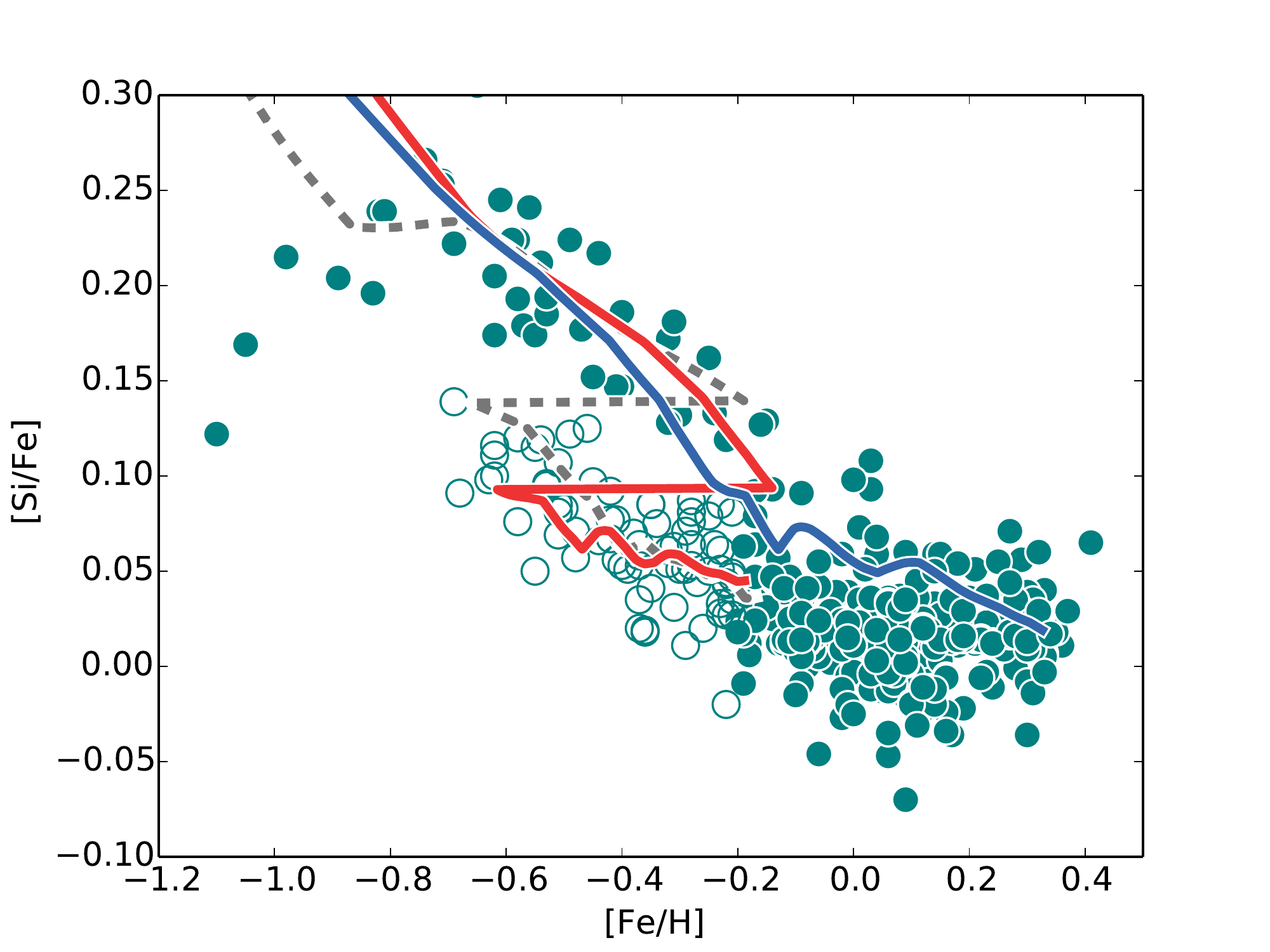}&
\includegraphics[width=3.0in, trim={0cm 0cm 0.0cm 0cm}, clip]{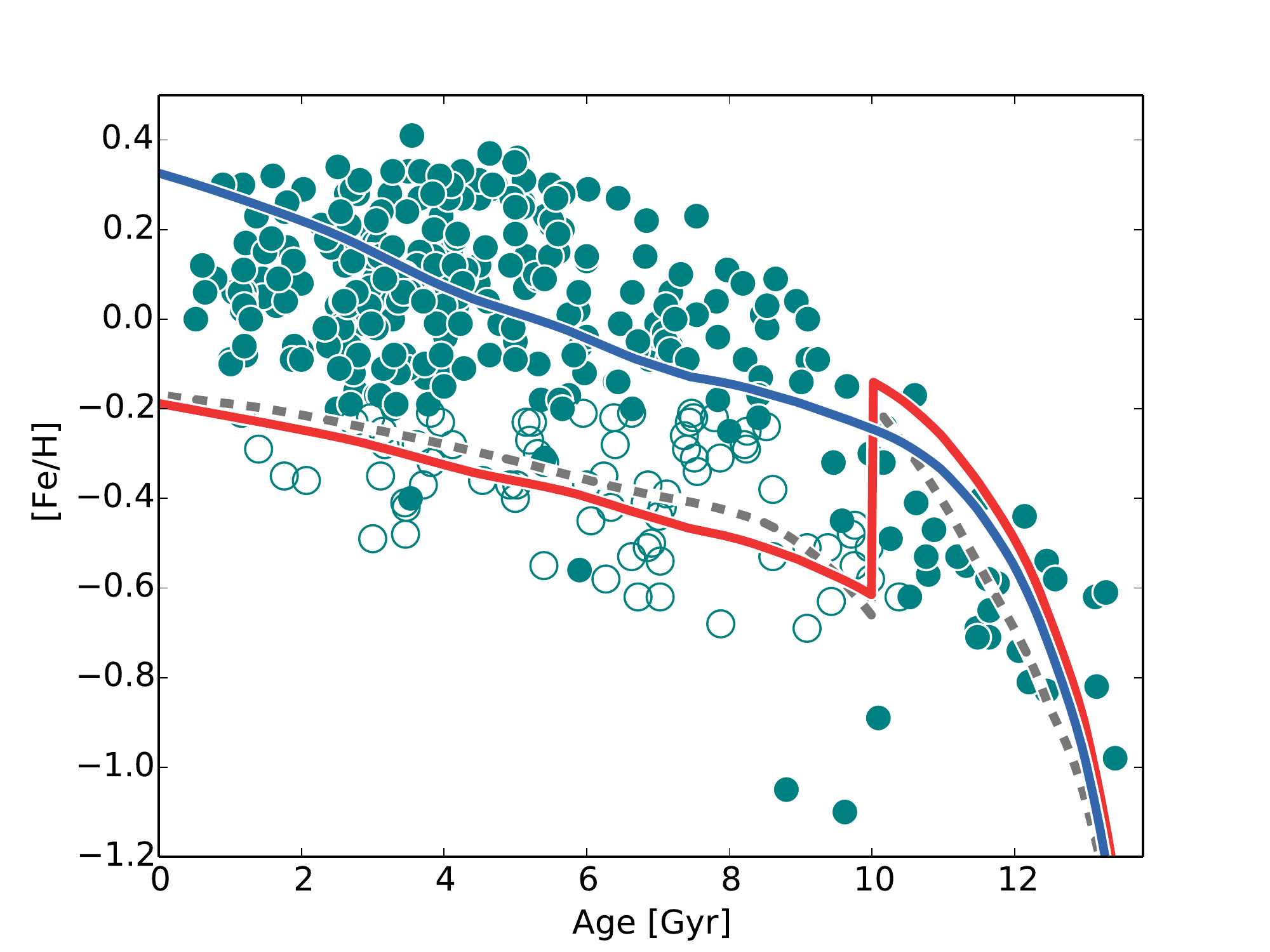} \\
(c) & (d)  \\
\end{tabular}
\caption{The chemical evolution of the best fit SFH to the outer disc, using silicon and the \citet{Anders1989} meteoric solar abundances. The blue line is the best fit to the outer disc stars using the standard model, while the red line is the chemical track  where the amount of hydrogen in the ISM is increased by 3 (the dilution model). The grey line is for the best fit line from the inner stars (Fig. \ref{Fig:bestfitsiandersmet}) using the `dilution' model. The grey region in panel (a) depicts the ages before the {\it outer disc} begins to form and is, therefore, essentially unconstrained. All three curves are the results of the mean of the bootstrapping method used in the previous subsection. Panels (b), (c) \& (d) are the [Si/Fe]-age, [Si/Fe]-metallicity and metallicity-age tracks of the model, and the corresponding data. The outer disc data is given as empty circles, while the inner disc is shown as filled circles.}

\label{Fig:bestfitmgandersmetouter}
\end{figure*}

The oldest metal-poor thin disc stars seen in the solar neighbourhood have ages of 9-10 Gyr and metallicities of [Fe/H]$\sim$-0.6 dex 
\citep[see][]{Haywood2013}.  They have similar levels of alpha element enrichment as thick disc stars of the same age, but a noticeably 
lower metallicity (at least 0.4 dex lower). While the origin of such differences needs to be fully understood, it was suggested  in \citet{Haywood2013} that
these lower metallicity stars result from outflows of enriched gas coming from the forming thick disc and pristine, 
accreted, gas falling in from the outer halo. In other words, metals coming from the inner disc may have been diluted by more pristine gas coming from
the halo, with the outer thin disc starting to form stars at a lookback time of 10 Gyr. So that contrary to what has been adopted for the inner disc, we assume that the initial metallicity of the outer disc
is not zero, but $\sim$-0.6 dex. 
In practice, we start computing the chemical track at 14 Gyr as before, but we introduce the dilution at 10 Gyr ``by hand'', 
see Fig.  \ref{Fig:bestfitmgandersmetouter}.  
The derivation of the SFH is made on the [Si/Fe]-Age distribution of outer disc stars, as defined in Sect.~\ref{data}.
The dilution has very little consequence on the derivation of the SFH for stars younger than 10 Gyr.
The only difference is the fact that we
use stellar yields at metallicities $\sim$0.4 dex lower than for the thick disc at the same age. This makes the system more silicon rich even for the same SFH, and the [Si/Fe]-age sequence is, therefore, naturally higher than for the inner disc. This upper [Si/Fe]-age sequence was pointed out in \citet{Haywood2013} and is shown to arise naturally as a result of the metallicity dependence  of stellar yields.

The results of the fitting procedure are shown in Fig. \ref{Fig:bestfitmgandersmetouter}.  The figure also shows the chemical tracks for the `dilution' model, using the SFH that was derived for the inner disc. This shows very similar evolution to the SFH fitted to the outer disc stars only. 
The recovered SFR of the outer disc is very similar to that of the inner disc in the last 10 Gyr, suggesting that, within the errors,  {\it the inner and outer discs had similar relative SFHs.}

We also fit the SFH calculated using outer disc stars without dilution and it is immediately clear that without dilution it is impossible to simultaneously fit age-[Fe/H], age-[Si/Fe] and metallicity-[Fe/H] at the same time.

Finally, a further word on the dip. 
The dip in the SFH of the inner disc is generated by the change of slope in the [Si/Fe]-Age plane, it should not appear 
in the SFH of outer disc stars, which do not show any change of slope. 

However, shortly after transition, several data points appear to have slightly higher than the local average [Si/Fe], 
which forces the fitting procedure to quickly reinitialize star formation. 
Whether this is real or due to the sampling of the data set awaits future studies.

\section{How robust are the results ?}\label{Robust}

In the following section we examine the robustness of the results presented in  the previous section, following a multi-step approach. 

We will firstly discuss how the various features in the derived SFH  are robust, by altering the recovered SFH, and then test how each of these adjustments changes the chemical evolution track. 
Secondly, because of the numerous uncertainties in the input physics of any GCE model \citep[see, for example,][]{Francois2004, Romano2005, Romano2010, Matteucci2009}, we will explore how changing the choices made when designing the basic model used in sections \ref{Patterns} and \ref{SFR} (summarized in Table \ref{Tab:summarymodel}) change the evolution of the chemical tracks. We will recalculate the SFH that best fits the data when these parameters are modified. We limit our analysis and discussion to the inner disc, the results being valid for the outer galactic regions as well. 

\subsection{Features of the derived SFH}\label{featuresSFR}

\begin{figure*}
\centering
\begin{tabular}{cc}
	\includegraphics[width=2.2in]{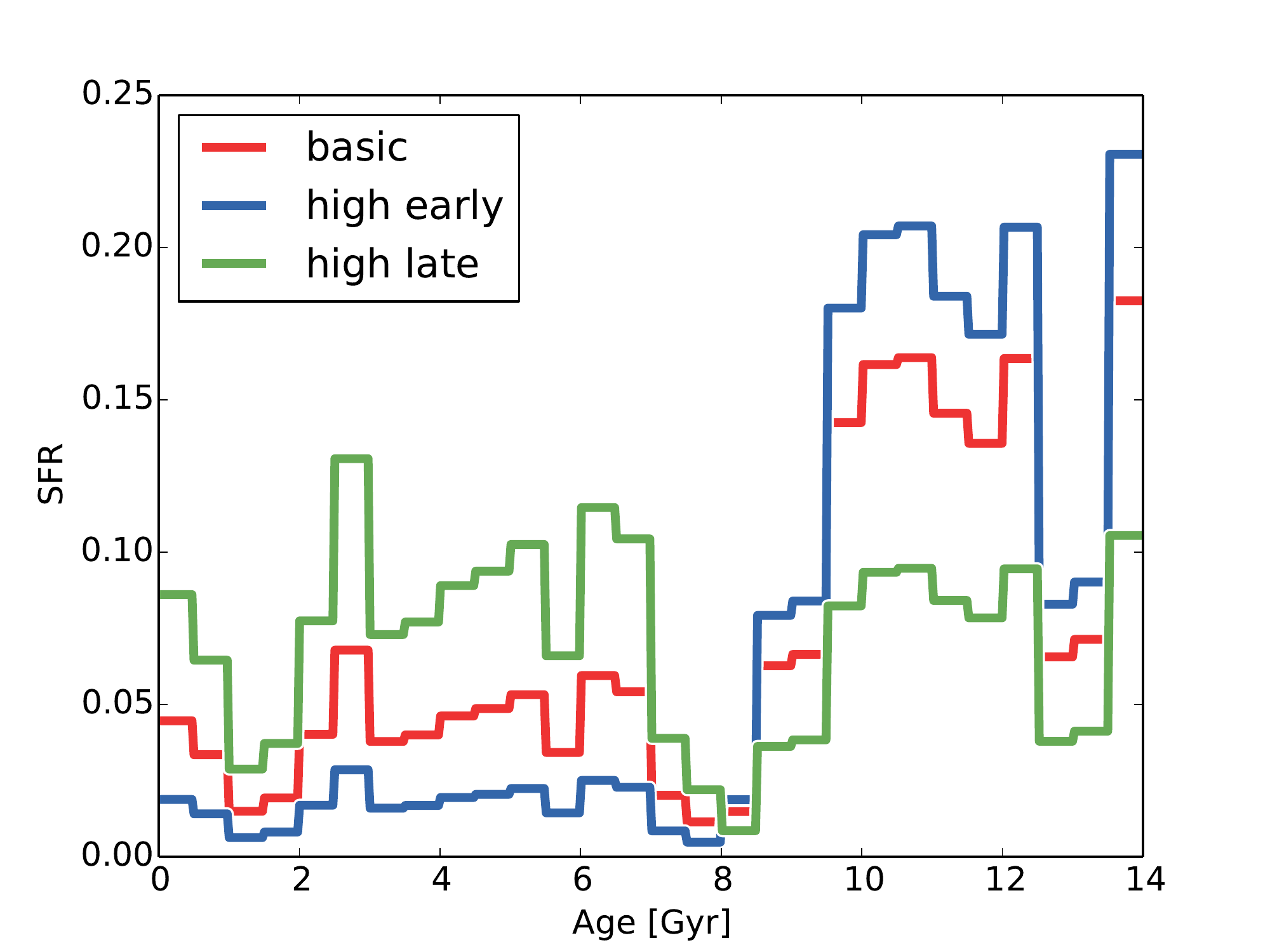} & 
        \includegraphics[width=2.2in]{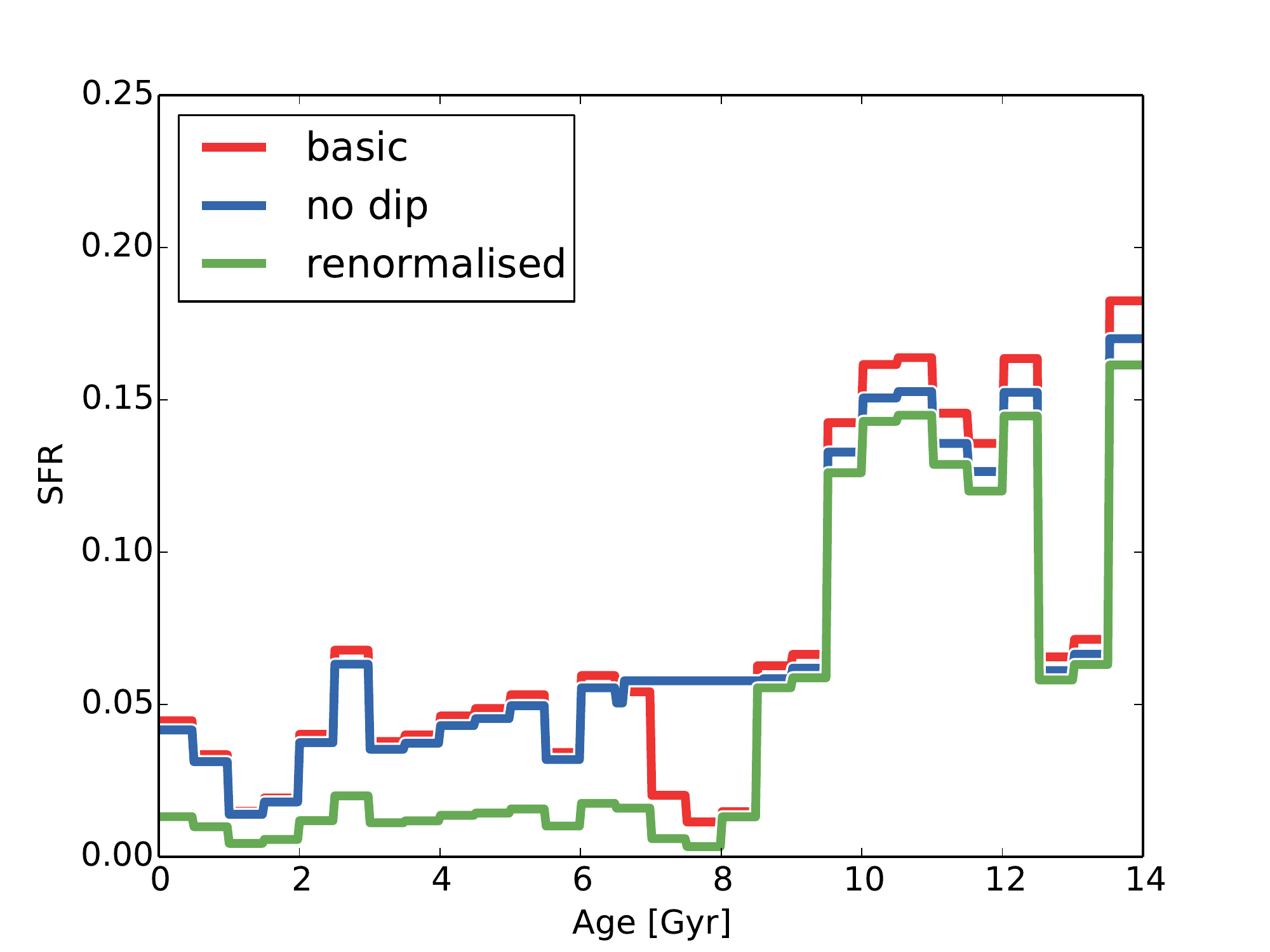}\\
        (a) & (b) \\
	\includegraphics[width=2.2in]{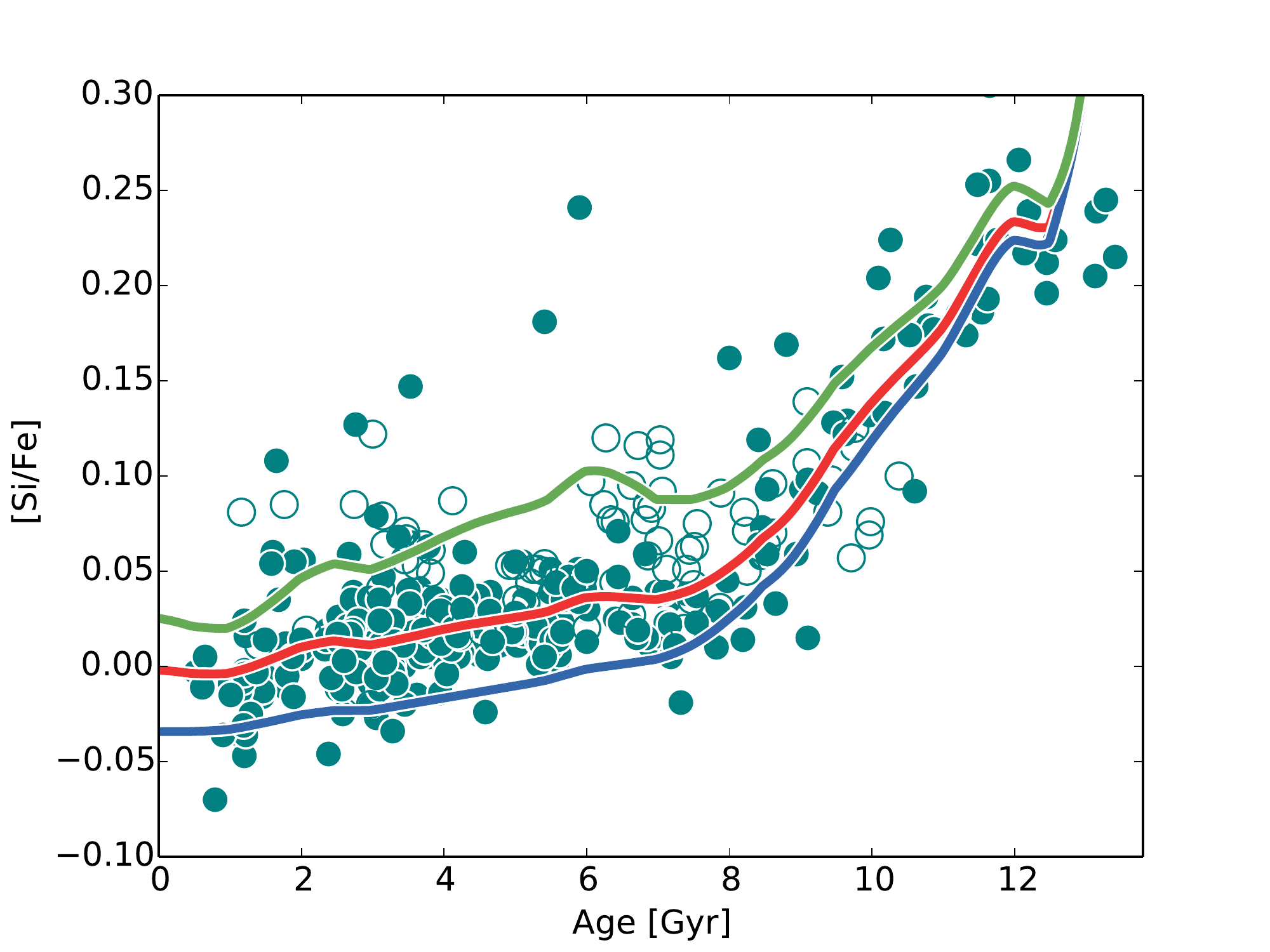} & 
    \includegraphics[width=2.2in]{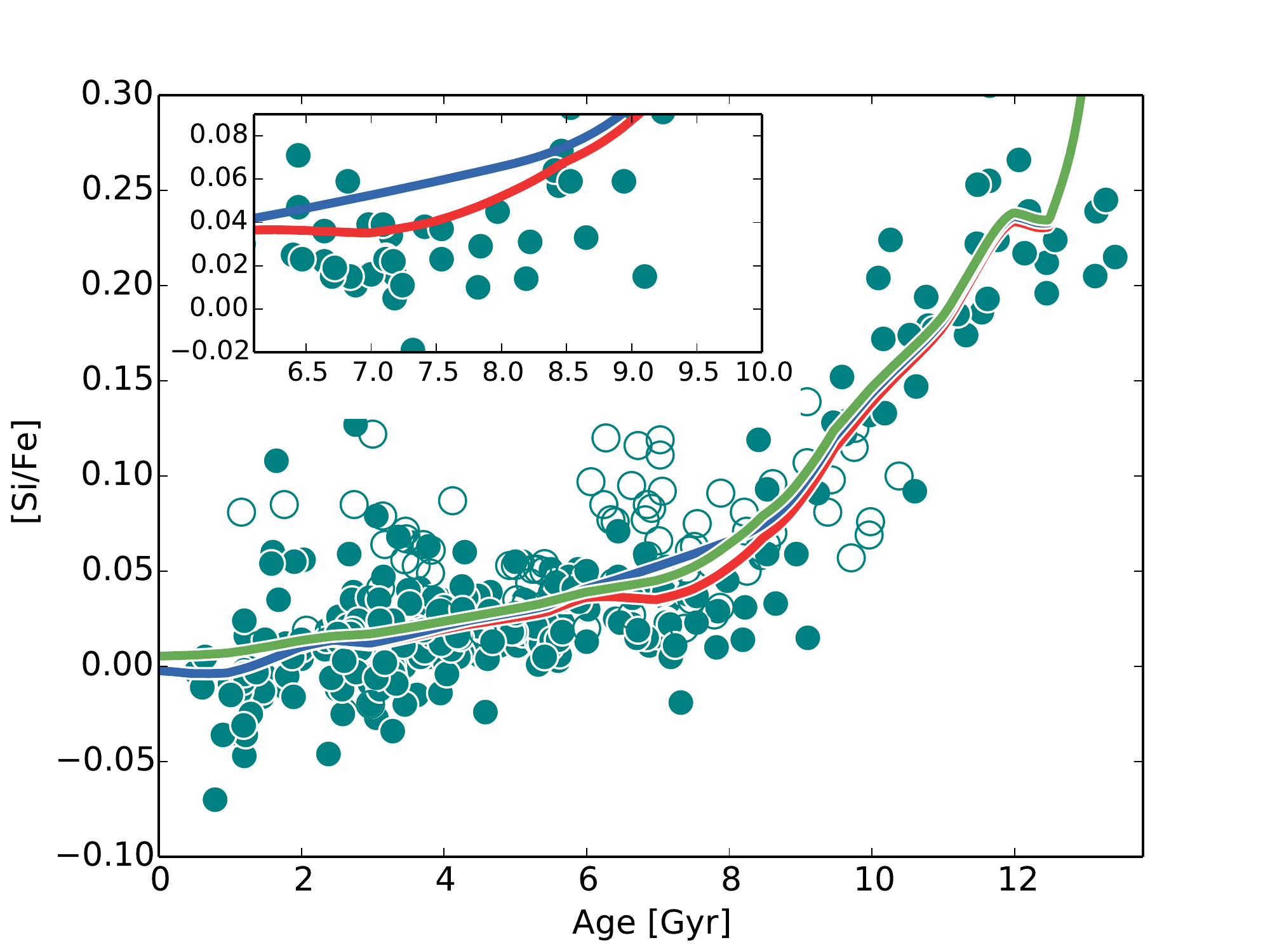}\\
        (c) & (d) \\
	\includegraphics[width=2.2in]{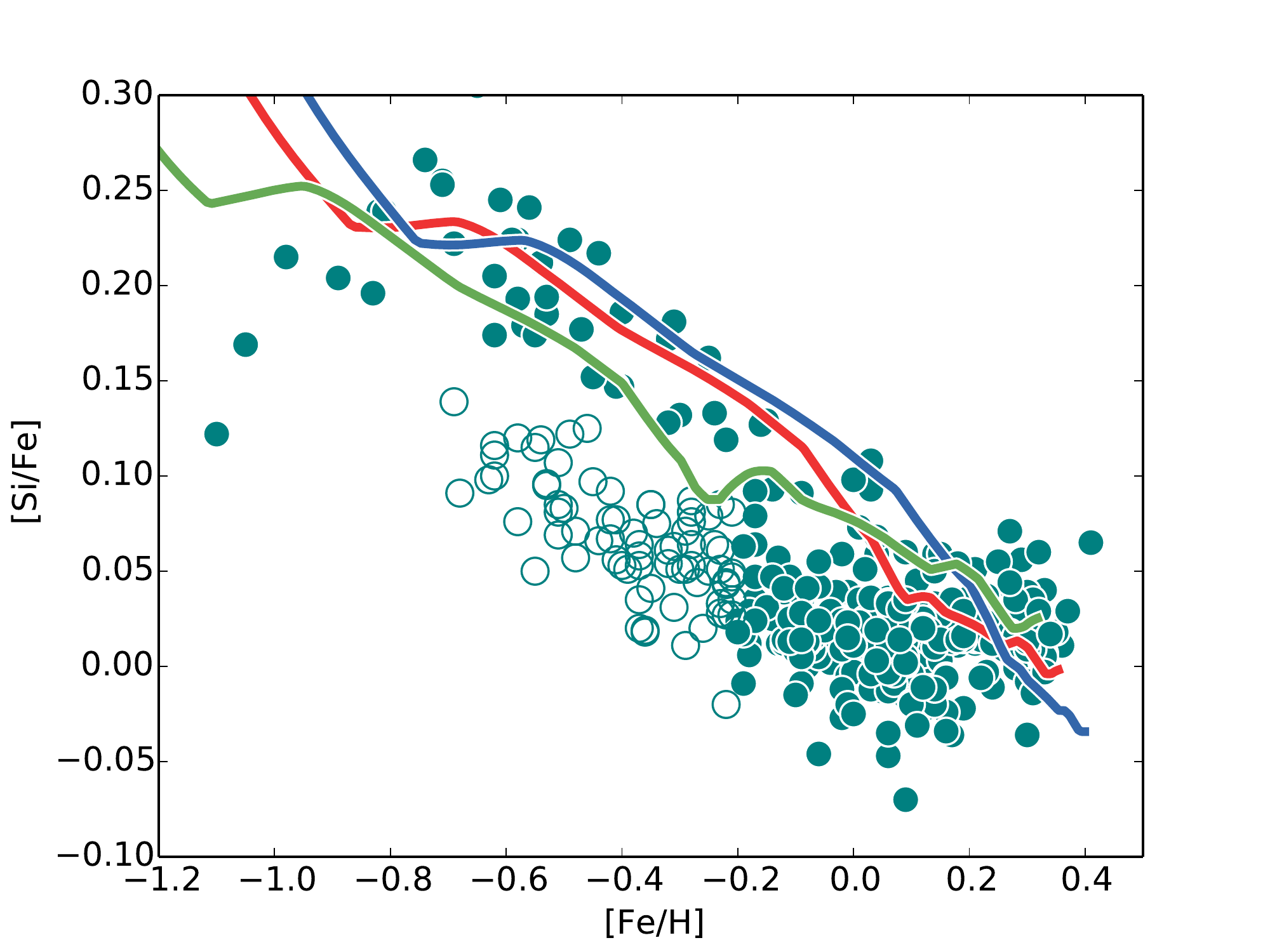} & 
        \includegraphics[width=2.2in]{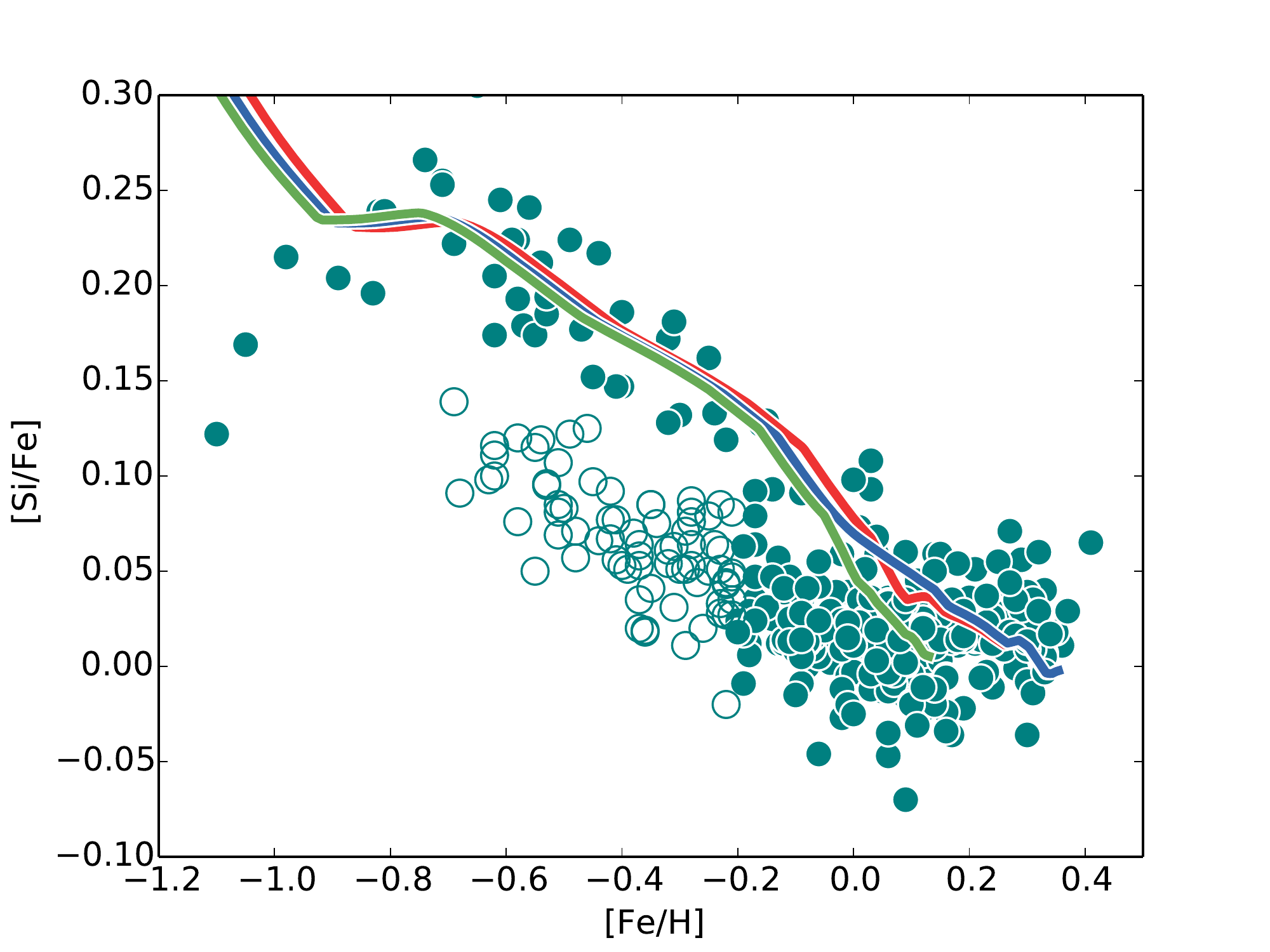}\\
        (e) & (f) \\
	  \includegraphics[width=2.2in]{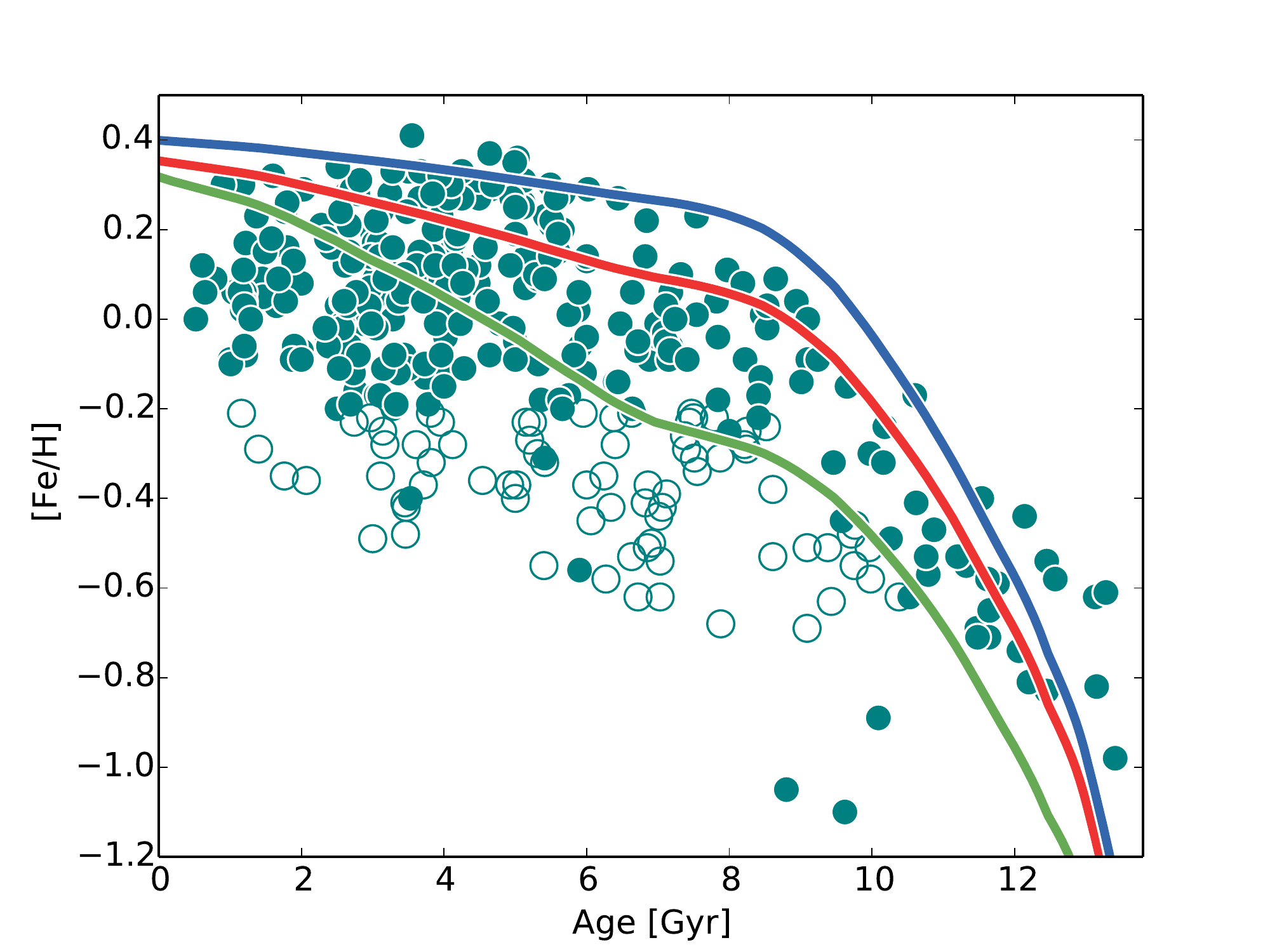} & 
        \includegraphics[width=2.2in]{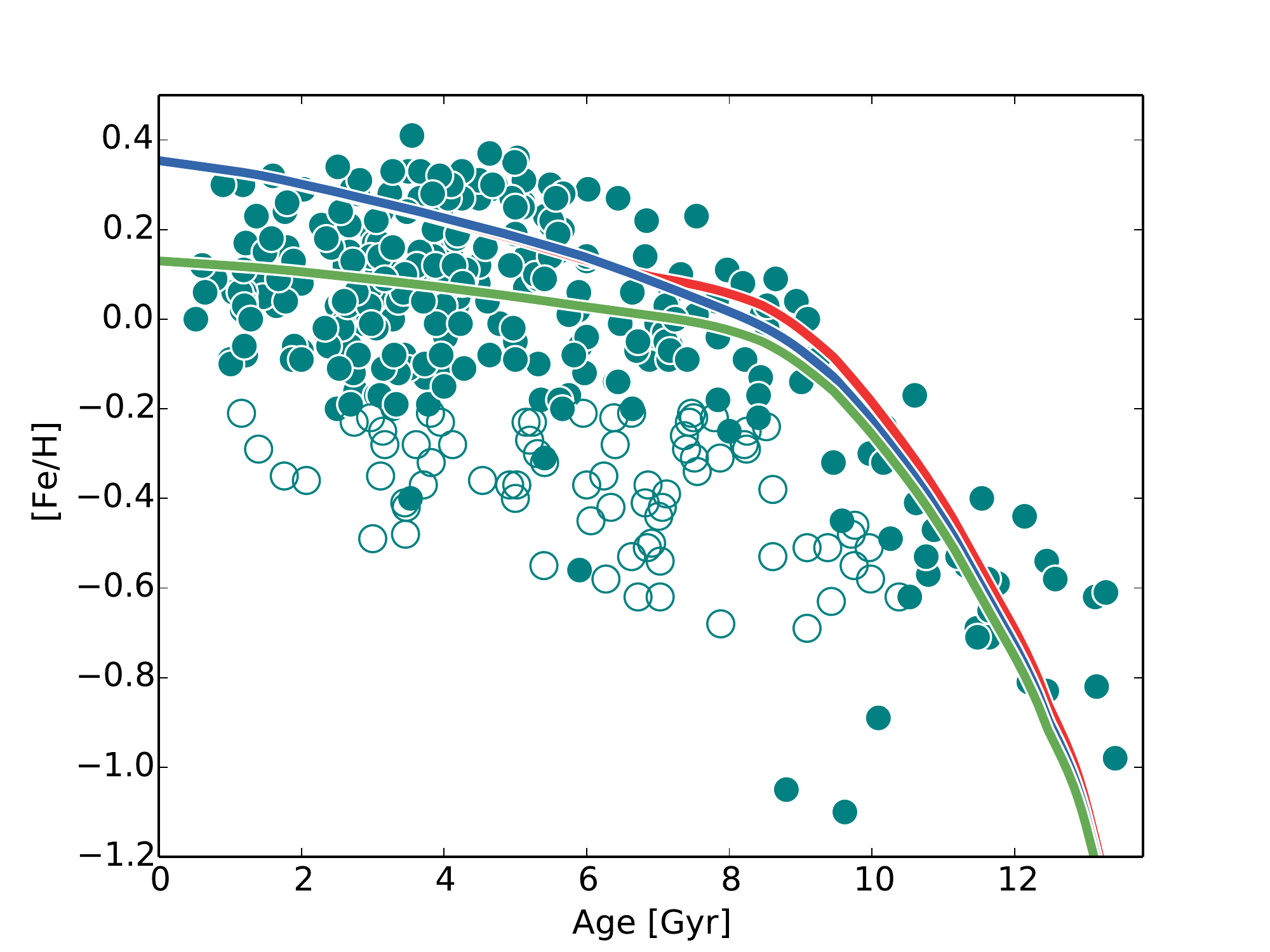}\\
        (g) & (h) \\
\end{tabular}
\caption{The effect of alterations to the best fit SFH on the chemical evolution of the model. Panels (a) and (b) illustrate the modifications to the SFR from Fig. \ref{Fig:bestfitsiandersmet}. The subsequent rows are the age-[Si/Fe], [Fe/H]-[Si/Fe] and age-[Fe/H] projections respectively.  The particular case of the green SFH in the right hand column is what happens if we reduce the SFH  normalisation (A in Eqn. \ref{Eqn:NormSFR}) to 0.7. The blue line is the SFH without the dip at 8 Gyr. In all plots the red line is the best fit SFH found for the inner disc stars (from Fig. \ref{Fig:bestfitsiandersmet}). The outer disc data is given as empty circles, while the inner disc is shown as filled circles. The inset in panel (d) emphasizes the effect of removing the dip (green line) from the SFH compared to the basic SFH (blue line) and only shows the inner disc data. }

\label{Fig:sfhfeatures}
\end{figure*}

In the previous section we derived the `best' star formation history for the data presented in \citet{Haywood2013}. In Fig. \ref{Fig:sfhfeatures} we show the result of six arbitrary alterations to the best fit SFH, for the inner thin and thick disc stars, and demonstrate how certain key features in the SFH affect the chemical history of the model (see panels (a) and (b)). This analysis allows us to better understand how the chemical evolution constrains the recovered SFH.

The left hand column of Fig. \ref{Fig:sfhfeatures} clearly shows that it is the [Si/Fe]-age slope which constrains the early time SFR. In panels (c), (e) and (g), we show the results of changing the relative contributions of the star formation on each side of the dip at 8 Gyr. Where the early SFR is too high, an {\it excess} of iron is quickly produced, while when the early SFR is too low {\it insufficient} iron enriches the ISM (panels (c) and (g)). Further, panel (g) illustrates that the thick disc data sets a lower limit on the early time SFR, as the [Fe/H]-age track is below the thick disc data for the low early SFR run, and above the data for the high early SFR. Panel (c) demonstrates that this variation in iron is stronger than the variation in silicon, as there is more silicon for a given amount of iron where the early SFR is high. Further, the high late time SFR produces a greater contrast on each side of the dip, effectively emphasizing this feature. Thus, we can see a `lurch' in the [Si/Fe]-age track. This allows us to gain an, at least qualitative, limit on the size of the dip. Finally, panel (e) demonstrates that despite considerable differences in [Fe/H] and [Si/Fe] with time the [Si/Fe]-[Fe/H] tracks  are not greatly dissimilar, and all overlap the thick disc sequence. The high early SFR produces more iron for a given value of [Si/Fe] suggesting that the evolution of iron is more sensitive to the relative contributions of the thick and thin discs than the silicon evolution.  

In the right hand column of  Fig. \ref{Fig:sfhfeatures} we present two further modifications of the SFH, which appear almost degenerate with the``basic'' form in the age-[Si/Fe] track in panel (d). 
The sharp transition from steep thick disc [Si/Fe]-age slope to shallow thin disc slope illustrates the need for the hiatus at around 8 Gyr. Where we remove the dip from the SFH the transition between slopes is more gradual, and occurs at higher [Si/Fe]. The SFH is constrained by the need for this sharp transition, the location of the transition and the size of the dip. The location of the dip is set by the transition from steep to shallow slopes, the depth of the dip is set by the need for a sharp transition. 

It could be argued that the `basic' chemical evolution track in panel (d) with the dip is still high (see the inset for a zoomed in view). The need for a potentially larger dip could be satisfied by a wider dip, as opposed to a deeper one, however, this is not recovered by our fitting procedure. 
If there was a smaller gas fraction, such as due to outflows, there would be a smaller dilution of iron ejected the SNIa. This would then emphasize that feature in the SFH, but would require us to `open the box' and allow gas inflows. This in turn would change the entire SFH, because if there is less gas present to enrich with metals then, to achieve the same level of enrichment, there must be less star formation. 

\begin{figure*}
\begin{tabular}{cc} 
\includegraphics[width=3.0in]{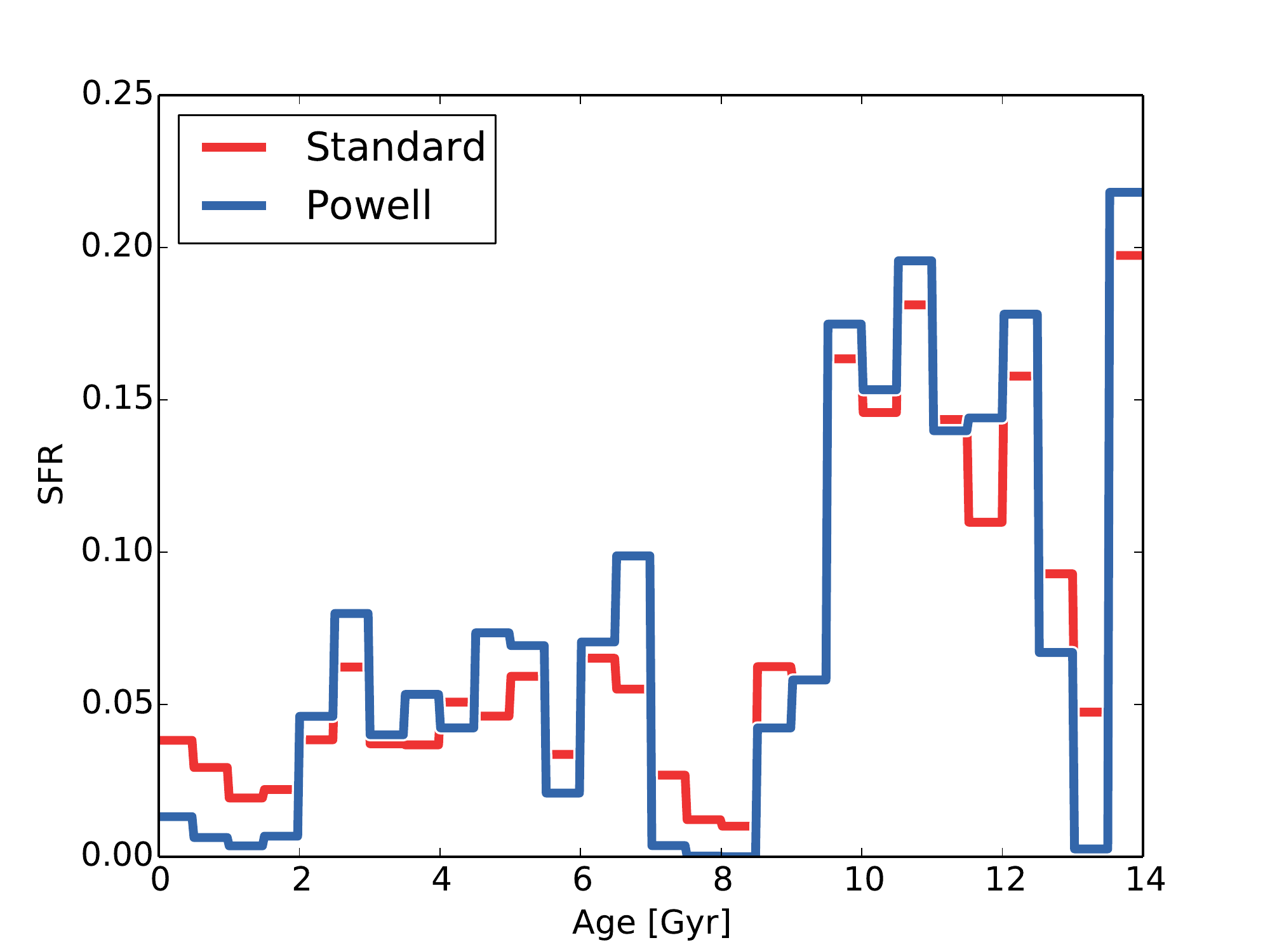} &
\includegraphics[width=3.0in]{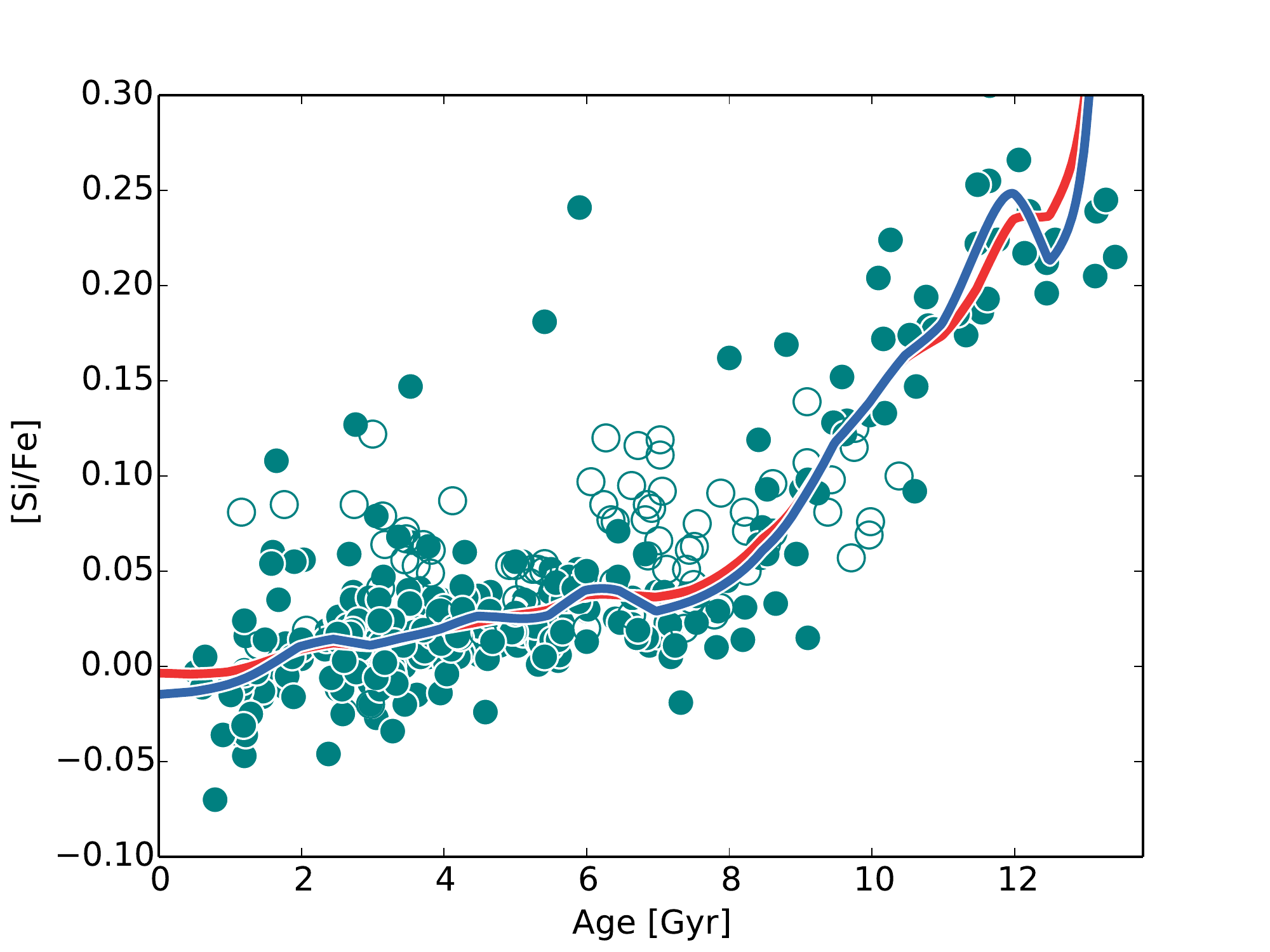} \\
(a) & (b)  \\
\includegraphics[width=3.0in]{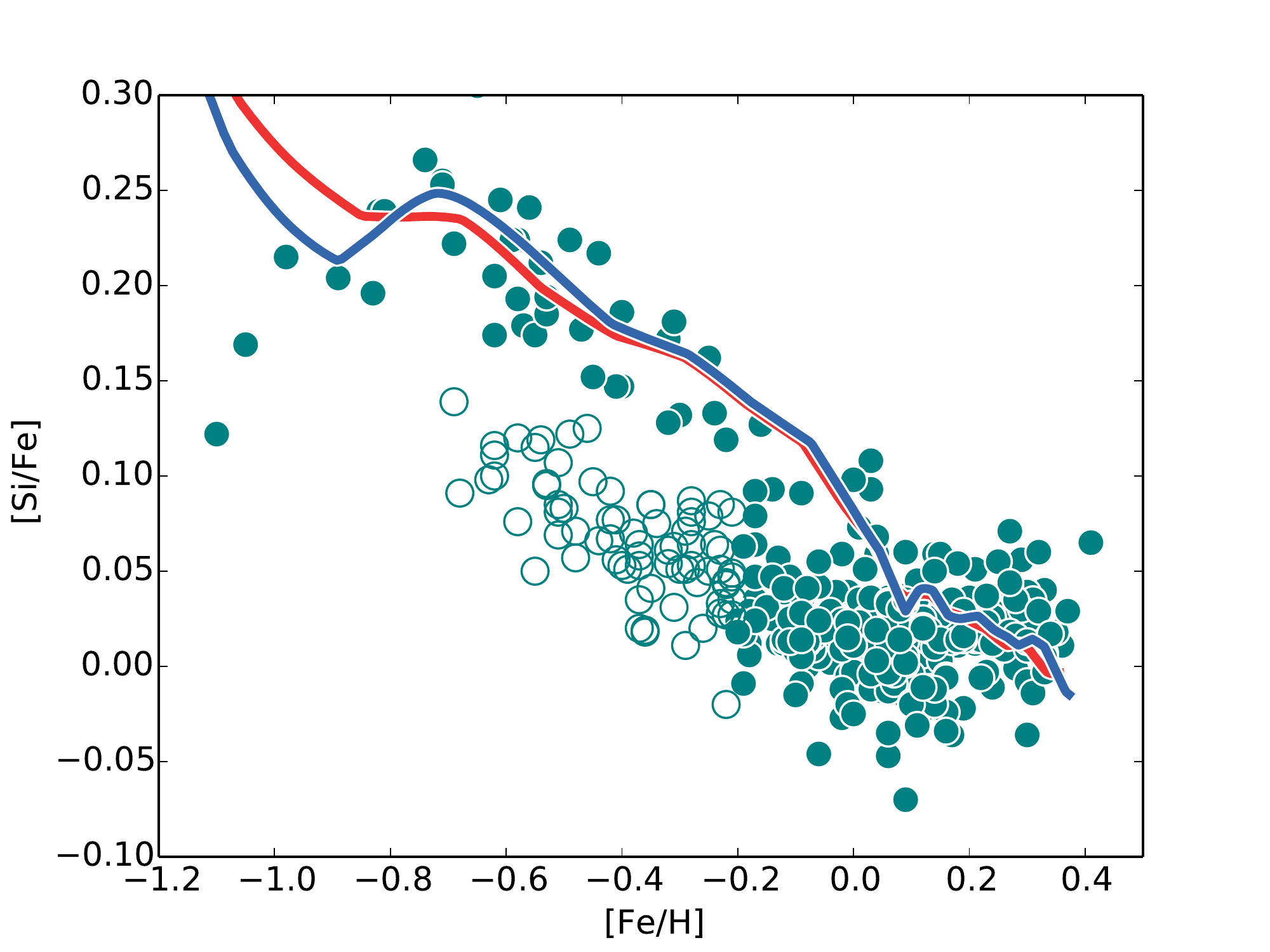} &
\includegraphics[width=3.0in]{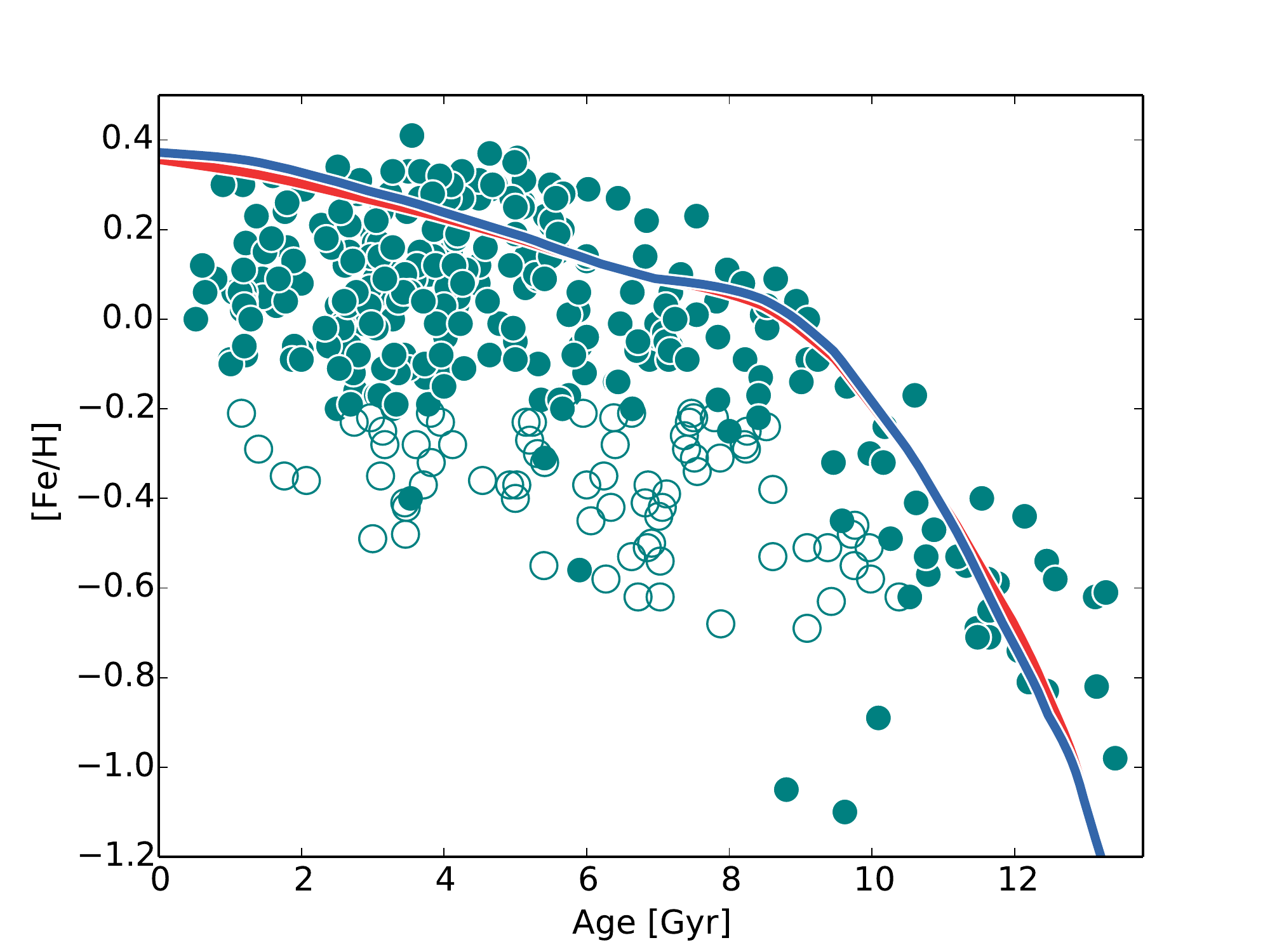} \\
(c) & (d)  \\
\end{tabular}
\caption{Comparison between the best fitting procedure for the standard fitting algorithm and the Powell algorithm for inner disc stars (solid circles). The outer disc stars are open circles and given for reference. }

\label{Fig:withpowell}
\end{figure*}

If we maintain our `closed box' model,  a complete hiatus in star formation might be required to match the data. Figure \ref{Fig:withpowell} shows a deeper dip and a sharper fall in [Si/Fe] at the time of the dip.  This uses an alternative fitting algorithm and essentially halts star formation for \~0.5 Gyr. The resulting drop is larger, and a slightly better fit to the data. Thus, from these observational data, the dip is required and robust.   

Figure \ref{Fig:withpowell} is the the  result of using our fitting procedure with a different fitting algorithm. It was derived by means of Powell's method \citep{Powell1964}, while the other fits in the paper use a Nelder-Mead simplex algorithm \citep{Nelder1965}. The figure shows that the choice of algorithm makes little apparent difference to the recovered SFH.

We have also repeated the ``high-early time" SFH, but reduced the total mass of stars formed over the whole time evolution to 0.7, instead of 1. The resulting chemical track fits the data points well, and leaves the total mass of stars formed relatively unconstrained by the data. The recovered SFH (panel b) does, however, give a very similar thick disc SFH to the canonical SFH, with the principle difference being in the thin disc. Here, the thick disc is very much the dominant galactic component. Panel (d) shows that this lower normalized curve does not produce a dip and fits the data less well at the transition between the thick and inner thin discs. This means that this `reduced' SFH is less favoured than the canonical SFH. The dip cannot be resolved in this SFH because of the very low late time SFR.  Another issue with this reduced SFH is that we do not produce the highest metallicity stars (panels (f) and (h)). While this SFH is, perhaps, a better fit to the centre of the [Fe/H]-age distribution it cannot recover stars with metallicity greater than 0.2 dex. Thus, the normalization value must be greater than 0.7 in order to produce these stars. 

This has implications for the origin of those stars. The model with the standard normalisation can fit those metal rich stars, but, if the normalisation is not equal to unity then the super solar metallicity stars must come from the inner regions of the galaxy. The effect of the normalisation will be discussed later in greater depth.

This figure makes it clear that the late time SFH is more poorly constrained than at early times. The late time chemical tracks can be made to follow the data for a range of changes to the SFH. The divergences from the data and the `basic' SFH tracks appear mainly due to changes to the SFR at early times. However, the requirement for a shape transition from the steep [Si/Fe] slope at early times to a shallower slope at later times means that we require the existence of the dip. In order for the dip to appear there needs to be sufficient contrast between the trough of the dip and the SFR on either side.

\subsection{Solar values}

The fitting procedure is extremely sensitive to the chosen solar abundance offsets. Our best fitting procedure fails match both the age-[Si/Fe] and the metallicity-[Si/Fe] distributions concurrently when the solar abundances are changed too greatly. The reason why the model is so sensitive to the solar values is best illustrated by  Fig. \ref{Fig:SFRexponential}, particularly panel (b). The narrow range (about 0.05 dex) of possible final values for the [Si/Fe] means that a small shift in the solar abundance normalisations can move the chemical tracks to regions of the parameter space where a good fit to the data is impossible. For example, the run where the solar iron is 7.45 lies off the thick disc sequence in panel (c).

We have used our best fitting procedure to recalculate the SFH using different solar normalisations  (Fig. \ref{Fig:bestfitsiandersmet}). We alter the solar iron abundance because this value differs between the standard \citet{Anders1989} which we prefer, and the value used for iron in the \citet{Adibekyan2012} data. 

\begin{figure*}
\centering
\begin{tabular}{cc} 
\includegraphics[width=3.0in]{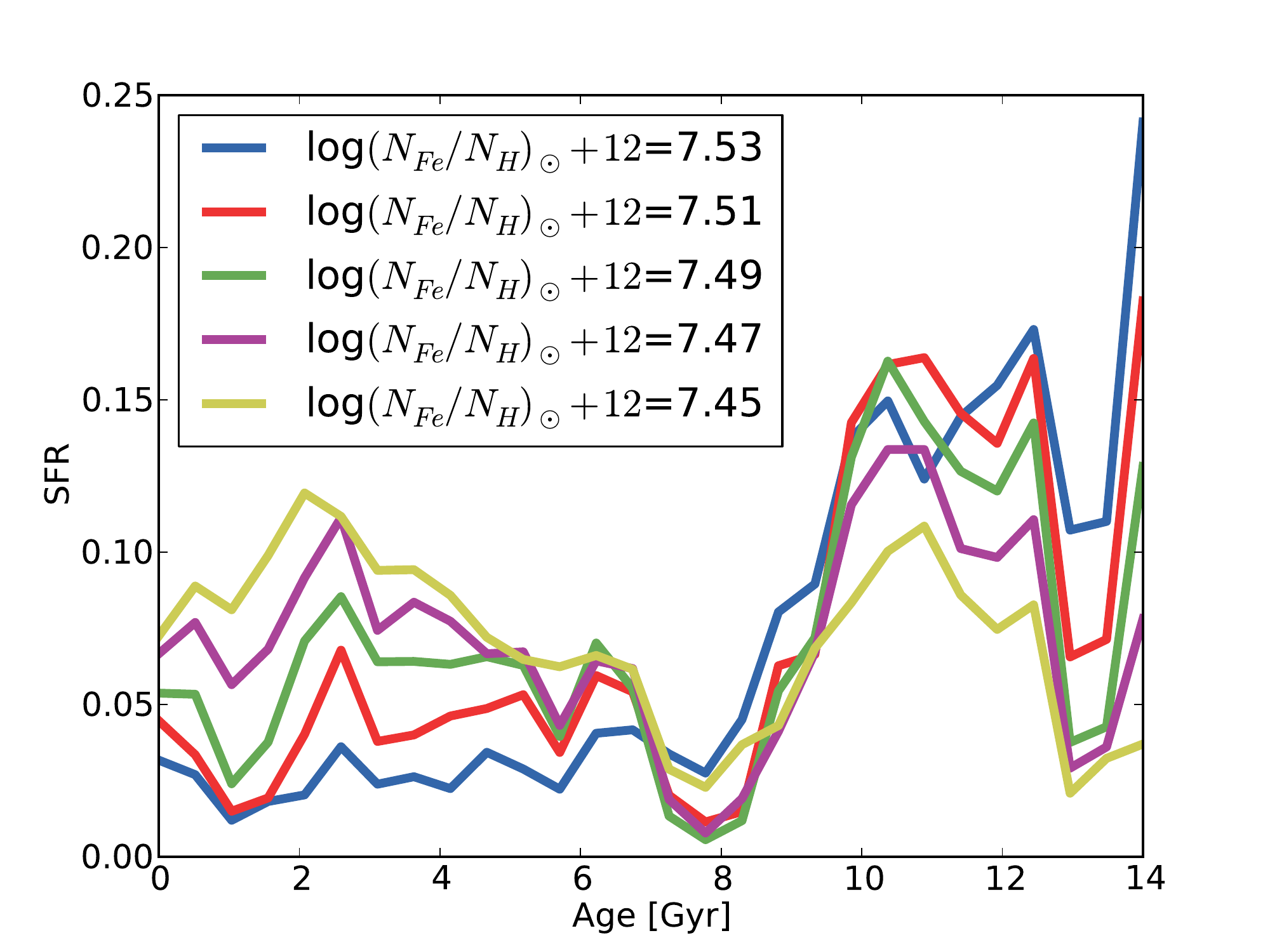} &
\includegraphics[width=3.0in]{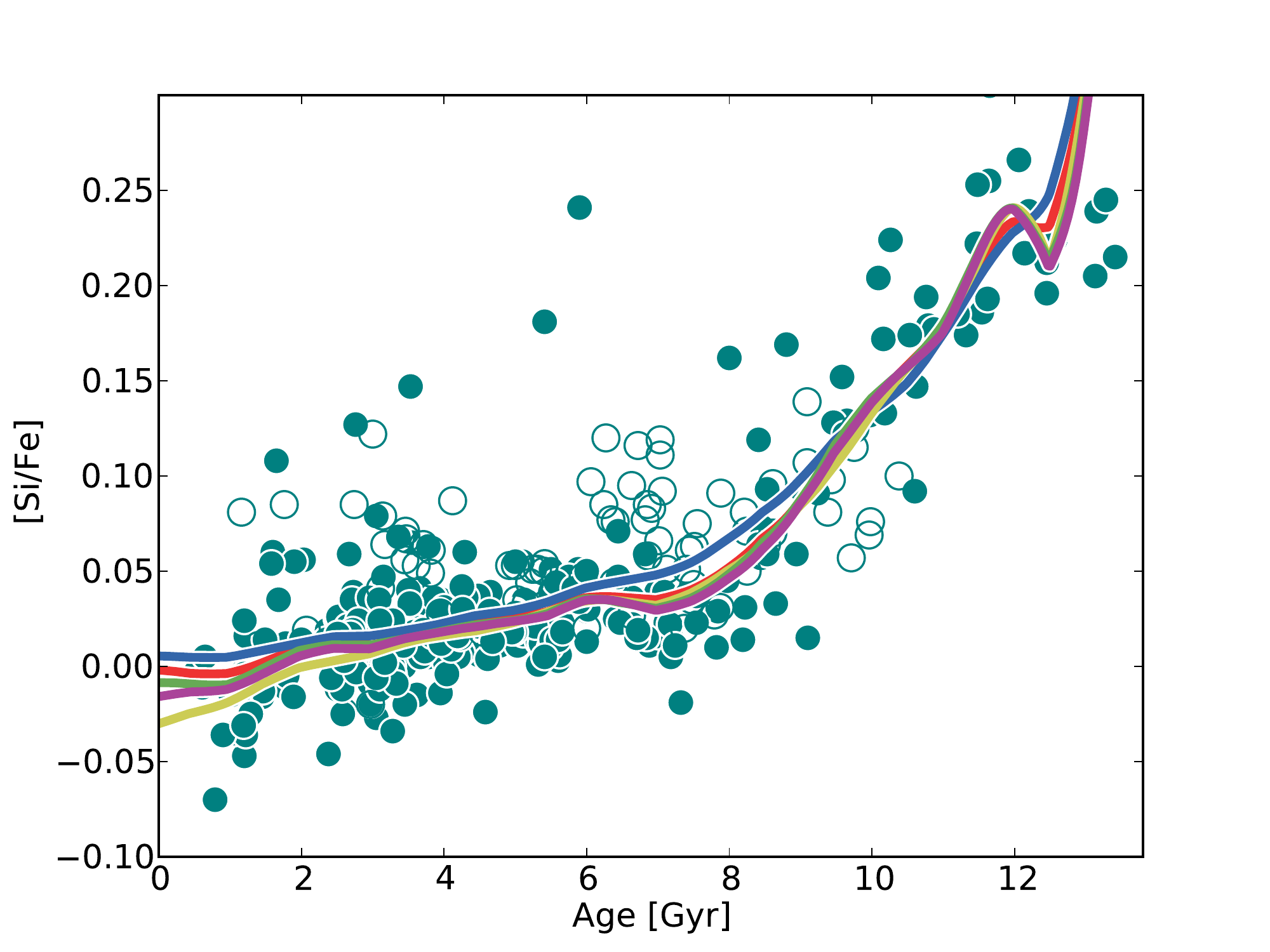}\\
(a)  & (b)\\
\includegraphics[width=3.0in]{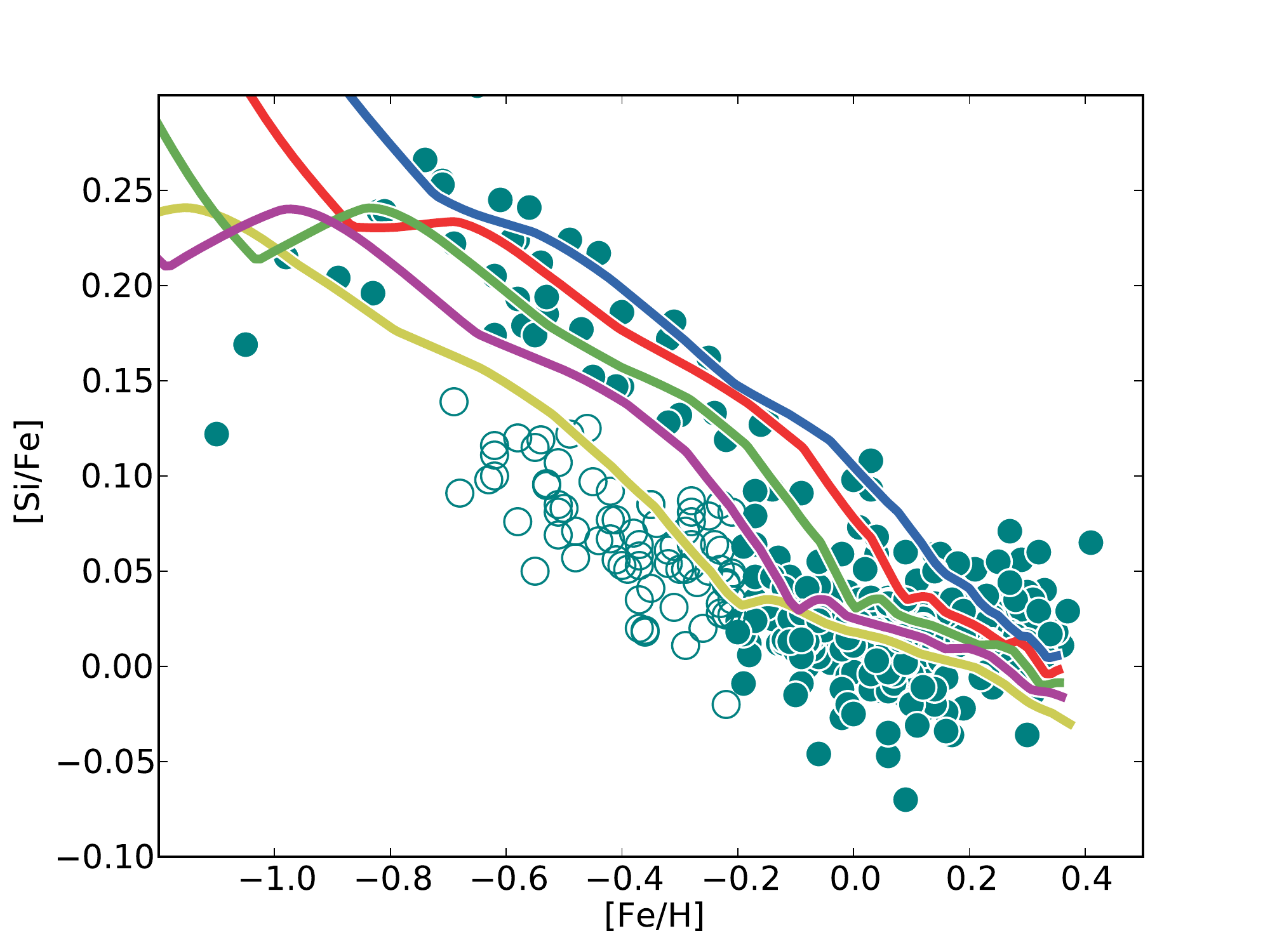} &
\includegraphics[width=3.0in]{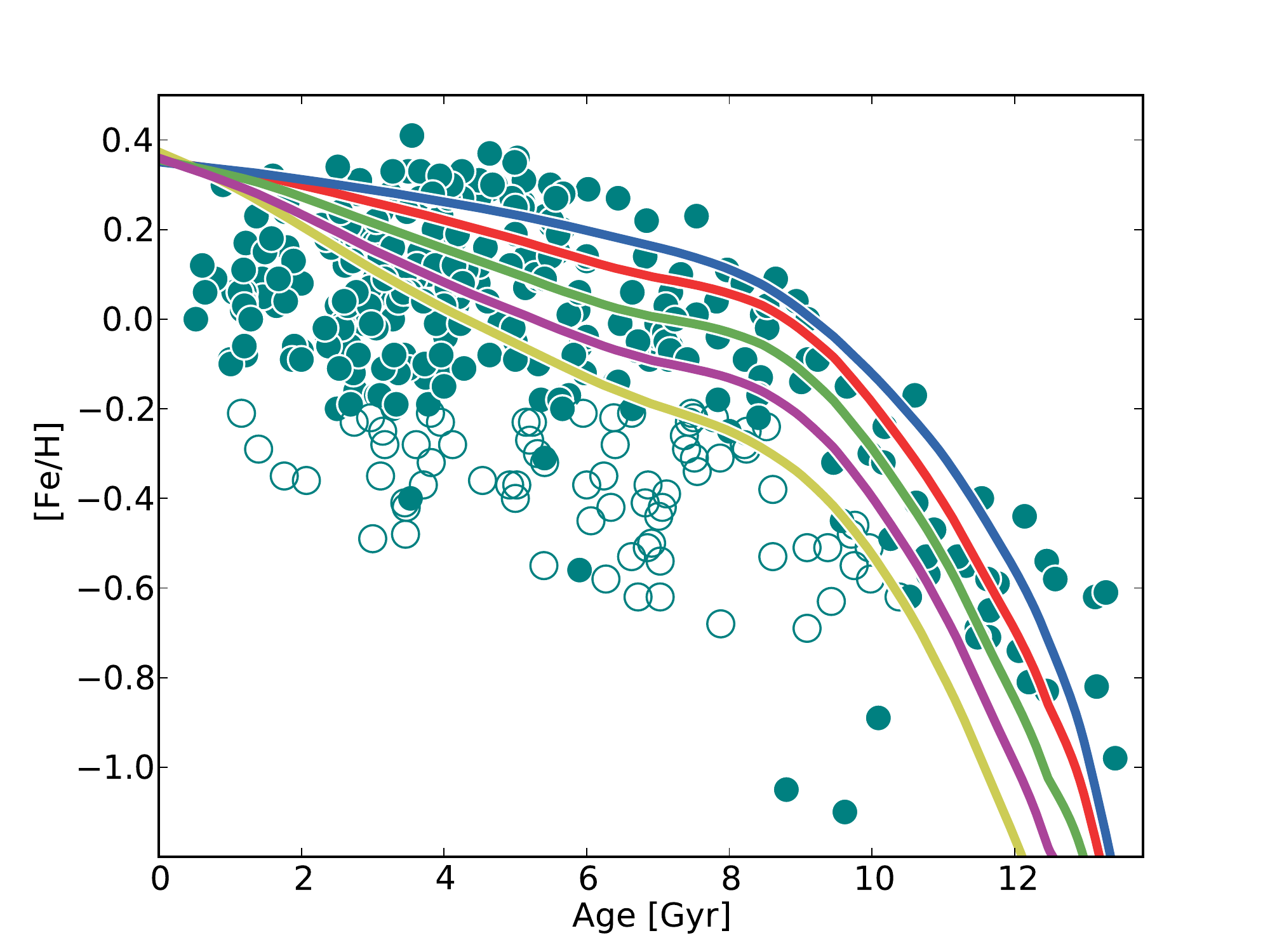}
\\
(c)  & (d)\\
\end{tabular}
\caption{The impact of changing the solar abundance normalisation on the recovered SFH (values given in the legend in panel (a)). The different SFHs give a thick disc mass fraction of 78\%, 68\%, 59\% and 48\% and 39\% respectively. The red line is the mean SFH recovered in Fig. \ref{Fig:bestfitsiandersmet}. The outer disc data is given as empty circles, while the inner disc is shown as filled circles.}

\label{Fig:bestfitmgandersmetsolarvary}
\end{figure*}

There is a distinct change in the form of the recovered SFH with changing A(Fe/H)$_\odot$, shown in Fig. \ref{Fig:bestfitmgandersmetsolarvary}.  Here, A(Fe/H)$_\odot$ indicates the value of log$(N_{Fe}/N_{H})+12$ in the sun. Lower A(Fe/H)$_\odot$ values require a lower early time SFR, and higher late time SFR, compared to the SFH recovered for high A(Fe/H)$_\odot$ values. Indeed, the fraction of stars in the thick disc part of the distribution (ages $>$ 8 Gyr) range from 78\% to 39\% for A(Fe/H)$_\odot$ = 7.53, 7.45 respectively, which is a considerable difference for a solar abundance change  of only 0.08 dex, (which corresponds to 20\% change).  

We also see that the metallicity-[Si/Fe] data (panel (c)) is best fit by the intermediate range, i.e. 7.49-7.53, with the best fit to this plot would be 7.51, the \citet{Anders1989} value used before. 

It is crucial to point out that even for A(Fe/H)$_\odot$=7.47 (the value used for the \citet{Adibekyan2012} data) over half the mass is in stars with ages greater than 8 Gyr. As previously discussed, we used the \citet{Anders1989} value for our solar iron normalization, {\it not the one used for the data}, in order to more accurately fit the [Si/Fe]-[Fe/H] distribution. 

Without changing the solar abundance of iron from the value of 7.47 used in the \citet{Adibekyan2012} to the \citet{Anders1989} value of 7.51, the track passes through the lower end of the thick disc envelope in the metallicity-[Si/Fe] plot rather than the centre.

The results remain qualitatively the same whether we utilise the solar normalisations used in the data, or our preferred value from \citet{Anders1989}. Even if the mass of the thick disc has been slightly reduced, it retains a mass comparable to the thin disc. This is justified by the uncertainties in the yields, and the lack of consensus on which set of theoretical yields accurately reflect the observations. If the stars in the yield tables produce slightly less iron, for example, it will have the same effect as a high A(Fe/H)$_\odot$. There is also no accord in the literature as to the abundance of elements in the sun, such that even recent solar values \citep[e.g.][]{Asplund2009} are not universally adopted  \citep[e.g.][]{Adibekyan2012}. 

 It is interesting that a shift of only 0.02 dex can result in considerable changes in the track in panels (c) and (d). It can be seen that amplifications up to 4 or 5 times the offset in the solar value occur. Due to the narrow range in final [Si/Fe] value discussed earlier, the star formation rate must change considerably with changing solar iron value in order to fit to panel (b). This has a much more significant effect on the total amount of metals ejected at any given time than on the abundance ratios.  The total metallicity is more sensitive to the SFH than is the [Si/Fe] evolution, as can be seen in Fig. \ref{Fig:SFRexponential}.   

It is also worth noting that the spread in end point metallicity is smaller than in the offset in solar iron value. The end point values vary from 0.350 to 0.373 for the  A(Fe/H)+12 values of 7.53 and 7.45 respectfully. This is because there is more iron locked up in stars which have not yet produced a SNIa in the low solar iron case than in the high solar iron case, because of the form of the recovered SFH. The actual value of the solar iron normalisation effectively compensates for this locked up fraction.

It is telling, however, that the transition time from the thick disc to the thin disc does not change considerably, and that, except for the extremely high solar iron value, the dip is clearly visible. It is noted, however, that where the dip does occur it is always in the same place irrespective of solar value. The knee feature still appears robust.

\subsection{IMF variations}\label{imfvar}

An important consideration in a chemical evolution model is the effect of the IMF. The functional form of the IMF controls the fraction of stars with a particular mass which forms from a given amount of gas. This has important consequences for the chemical evolution, as the amount of a given metal released when a star dies depends on the  mass of that star. There is a considerable history of IMFs beginning with \citet{Salpeter1955}, and continuing to the present day. Further, because the IMF must be normalised, and some IMFs, like the \citet{Salpeter1955} IMF, are divergent, a suitable range of masses must be chosen over which to integrate.  For each IMF we will explore, we normalise the function by integrating between a mass of 0.1 and 100, as is traditional. We select a range of common IMFs from the literature an implement them in our model.  

\begin{figure*}
\centering
\begin{tabular}{cc} 
\includegraphics[width=3.0in]{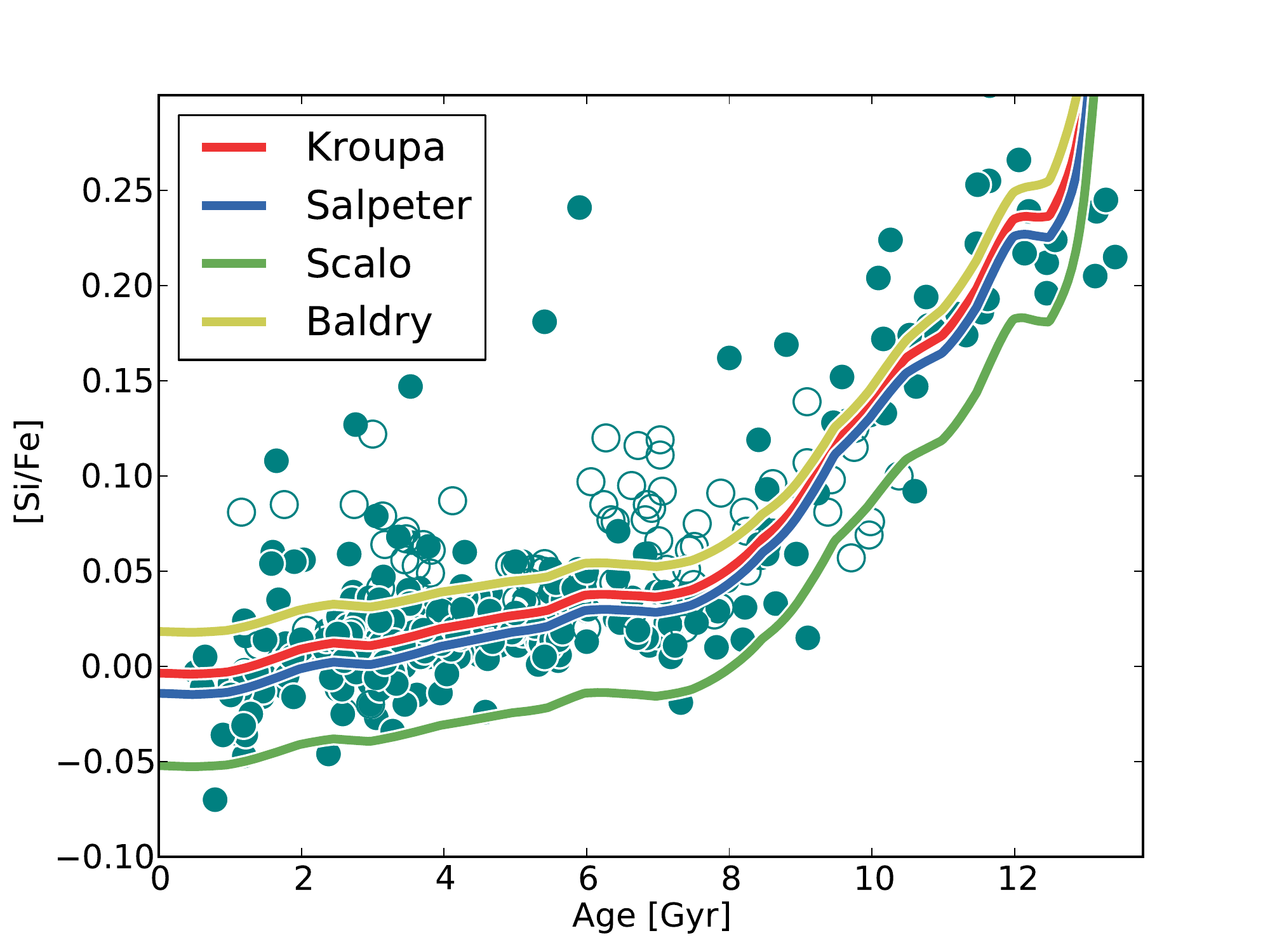} &  
\includegraphics[width=3.0in]{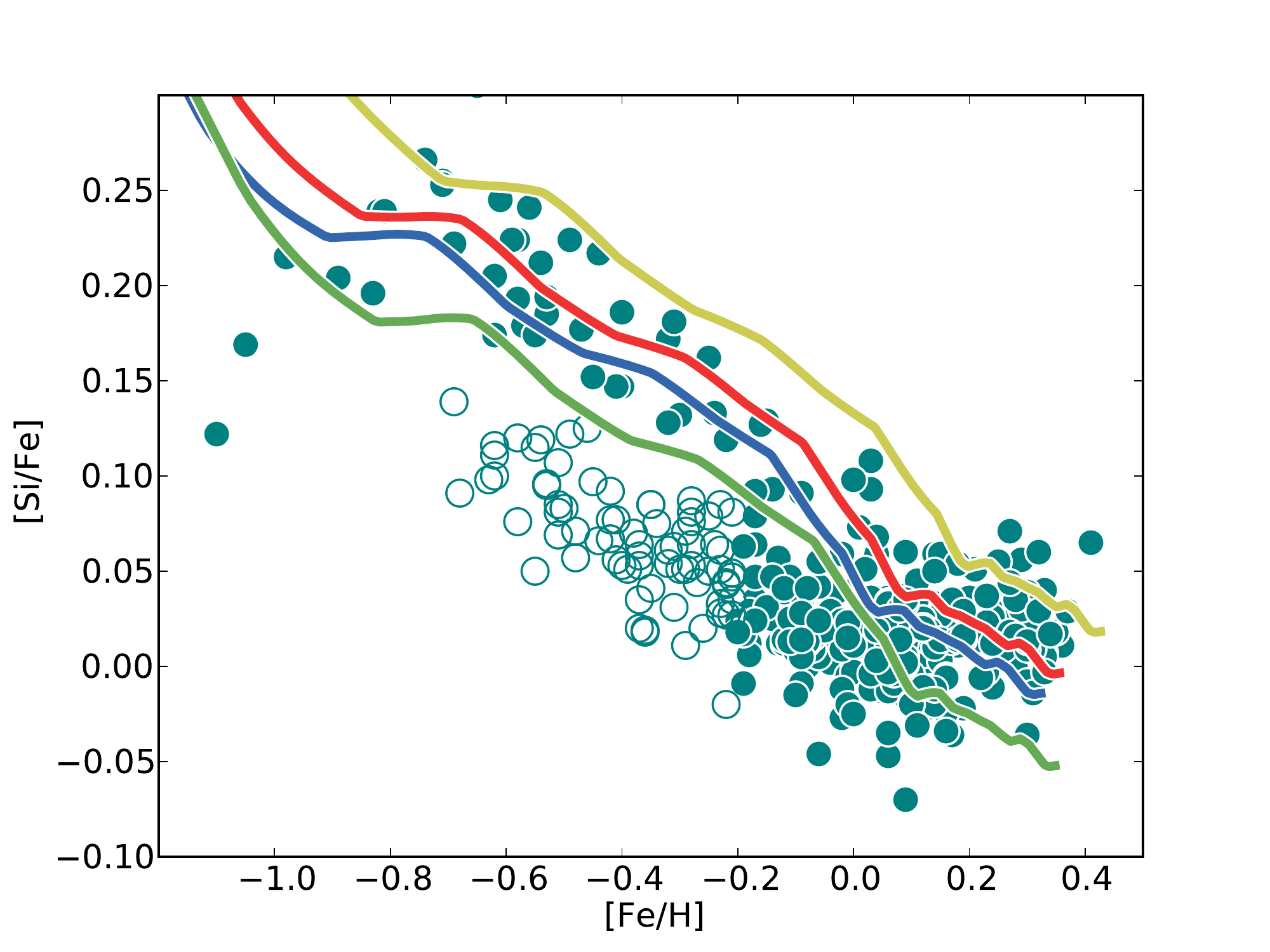}\\
(a) & (b) \\
\includegraphics[width=3.0in]{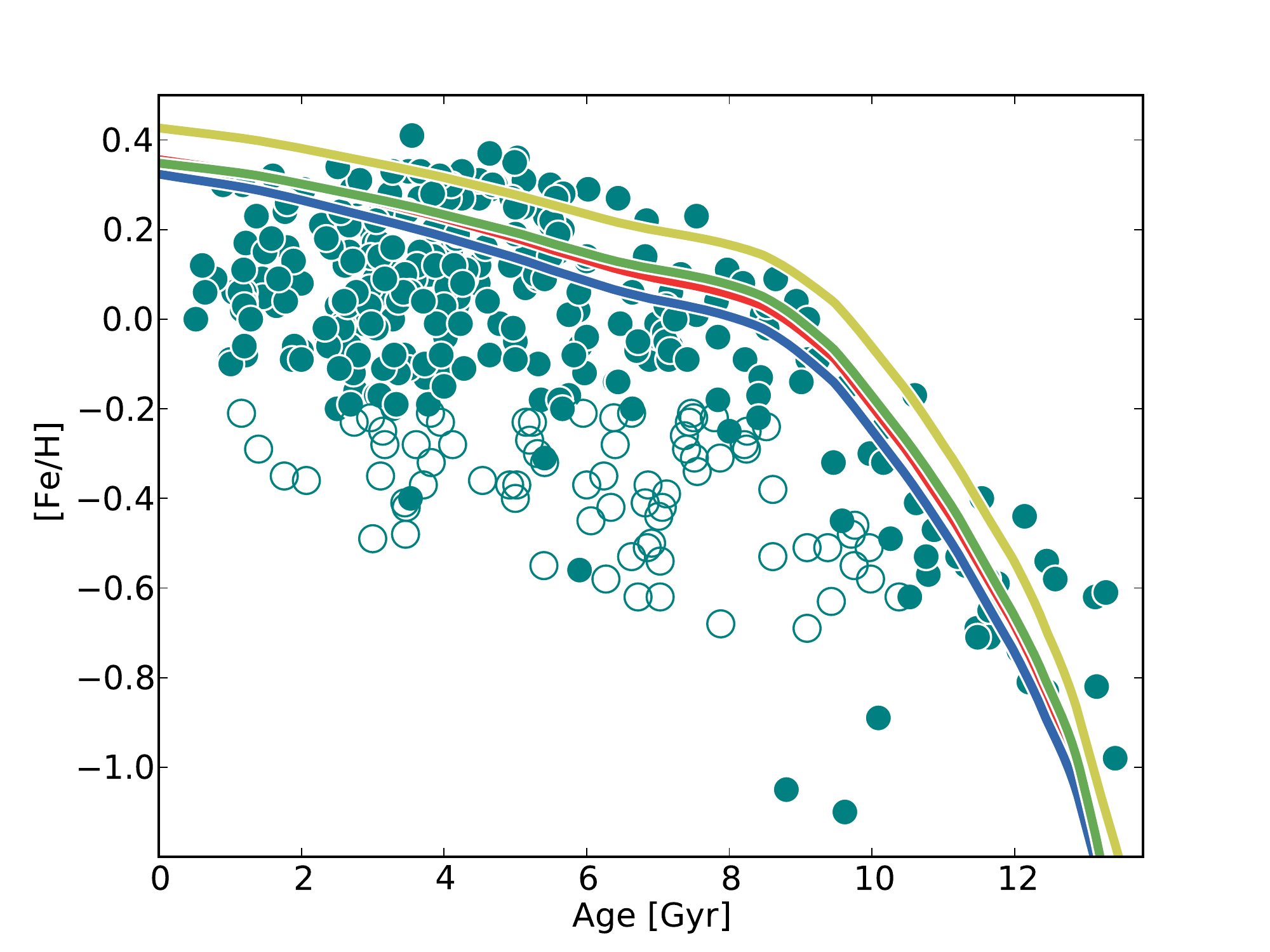} &  
\includegraphics[width=3.0in]{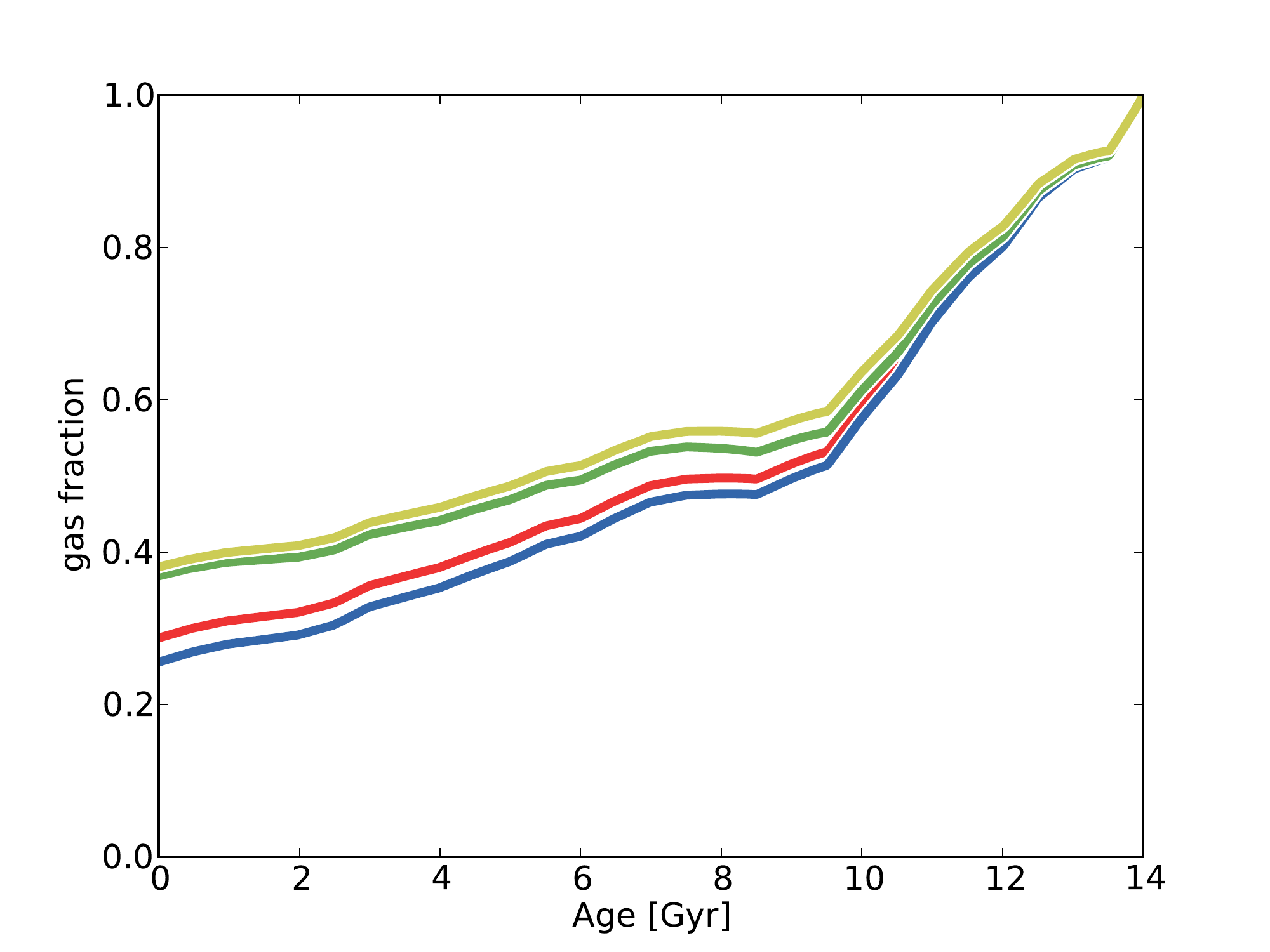}\\
(c) &(d)

\end{tabular}
\caption{The chemical tracks using different IMFs (shown in Table \ref{Tab:imflit}) using the SFH derived from fitting the inner disc stars in Section \ref{SFR} (Fig. \ref{Fig:bestfitsiandersmet}, panel (a)) . Only the IMFs are changed, the SFH is the same on each run. The chosen IMFs are given in the legend in panel (a). Panel (a) shows the evolution of [Si/Fe] with stellar age.  Panel (b) shows the evolution of [Si/Fe] with metallicity, panel (c) shows the evolution of metallicity with age and panel (d) shows the evolution of the gas fraction.  The points are \citet{Haywood2013} data and open circles depict outer thin disc stars, while solid circles are thick disc and inner thin disc stars.}
\label{Fig:differentimfs}
\end{figure*}

We show, in Fig. \ref{Fig:differentimfs}, how various IMFs affect the chemical tracks, while keeping the same SFH. By using the same SFH we can explore how the different IMFs change the contributions of silicon and iron and the amount of gas released, in a controlled way. We chose the popular IMFs of \citet{Salpeter1955} and \citet{Kroupa2001}, the historically important \citet{Scalo1998} IMF,  and for illustrative purposes the IMF of \citet{Baldry2003}. These are representative of the range of IMF gradients that have been proposed, and are useful to compare with other work. 

The functional form of our chosen IMFs are summarised in Table  \ref{Tab:imflit}. We see some interesting behaviour in the IMFs, in that the different panels of Fig. \ref{Fig:differentimfs} provide different details. The simplest form is the old Salpeter IMF, which has no change in slope, while the Kroupa IMF has a similar slope (-1.3 compared to -1.35) but a break at 1 solar mass. The Baldry IMF has a shallower slope, meaning proportionally more high mass stars, while the Scalo IMF has a steeper slope, meaning proportionally fewer high mass stars. 

Figure \ref{Fig:differentimfs} shows that, as expected, the Kroupa and Salpeter IMFs produce similar chemical tracks, although the Kroupa IMF is  slightly more iron abundant {than the Salpeter IMF}. This is due to the break in the IMF at 1 M$_\odot$. SNIa and AGB stars involve a component of the IMF where M $<$ 1 M$_\odot$, which occurs at later times  but mainly because it also changes the normalization of the IMF which is calculated by the integral of the IMF between 0.1 and 100 solar masses.  Panel (b), however, shows that for a given metallicity, the Kroupa IMF has more silicon than the Salpeter. This is also due to the shape of the IMF, which produces more high mass stars in the Kroupa IMF because of the (slightly) shallower slope.

Panel (a) shows that the [Si/Fe]-age evolution of the Kroupa and Salpeter IMFs are almost identical, while the Scalo IMF is both the  less iron abundant  relative to the silicon, and is the more different to the data than the other IMFs. The end time difference of the [Si/Fe] of each IMF from the standard Kroupa IMF (used hitherto) is -0.01, -0.05 and +0.02 for the Salpeter, Scalo and Baldry IMFs respectively. The end point metallicity, the differences are -0.03, -0.006 and 0.07  at age=0, but with a maximum difference of 0.07, 0.05 and -0.17 respectively. We see that the Kroupa, Scalo and Salpeter IMFs give fairly similar metallicity tracks with a much more divergent Baldry.

The explanation for these differences can be explained by the shape of the IMFs and the mass fraction of the stellar population which releases gas. 

\begin{enumerate}
\item The Baldry IMF produces more SNII compared to SNIa, and so the ISM is more silicon rich while producing more gas overall (panel d).  The ISM is, therefore, enriched more than with the other IMFs. This is why the Baldry IMF track is the rightmost track in panel (b).
\item A considerable fraction of the iron is made in SNIa  and is lacking in the Baldry IMF (compared to the Kroupa IMF) due to the  shallower slope. The vast (relative) overabundance of SNII means, however, that overall, the Baldry IMF produces more iron than the other models. This high level of SNII is also why the gas fraction is higher. 
\item For the Scalo IMF, the lack of SNII relative to the Baldry IMF is compensated for by the higher proportion of low mass stars (and SNIa and AGB stars). This means that the gas is returned to the ISM more gradually  and explains the lack of [Si/Fe] in panels (b)  and (a)), i.e. the iron has not yet been returned to the ISM. 
\item The Salpeter IMF produces the least gas because of the higher proportion of stars with M $<$ 0.9 M$_\odot$ that have not yet released their gas and metals. 
\end{enumerate}

Other popular IMFs have not been shown because of the similarity between them and either the Salpeter or Kroupa IMFs. The popular Chabrier \citep{Chabrier2003} IMF , for example, has a slope identical to Salpeter but a locked in fraction (stars which have not yet ended their lives) similar to Kroupa. 
 
\begin{figure*}
\centering
\begin{tabular}{cc} 
\includegraphics[width=3.0in,]{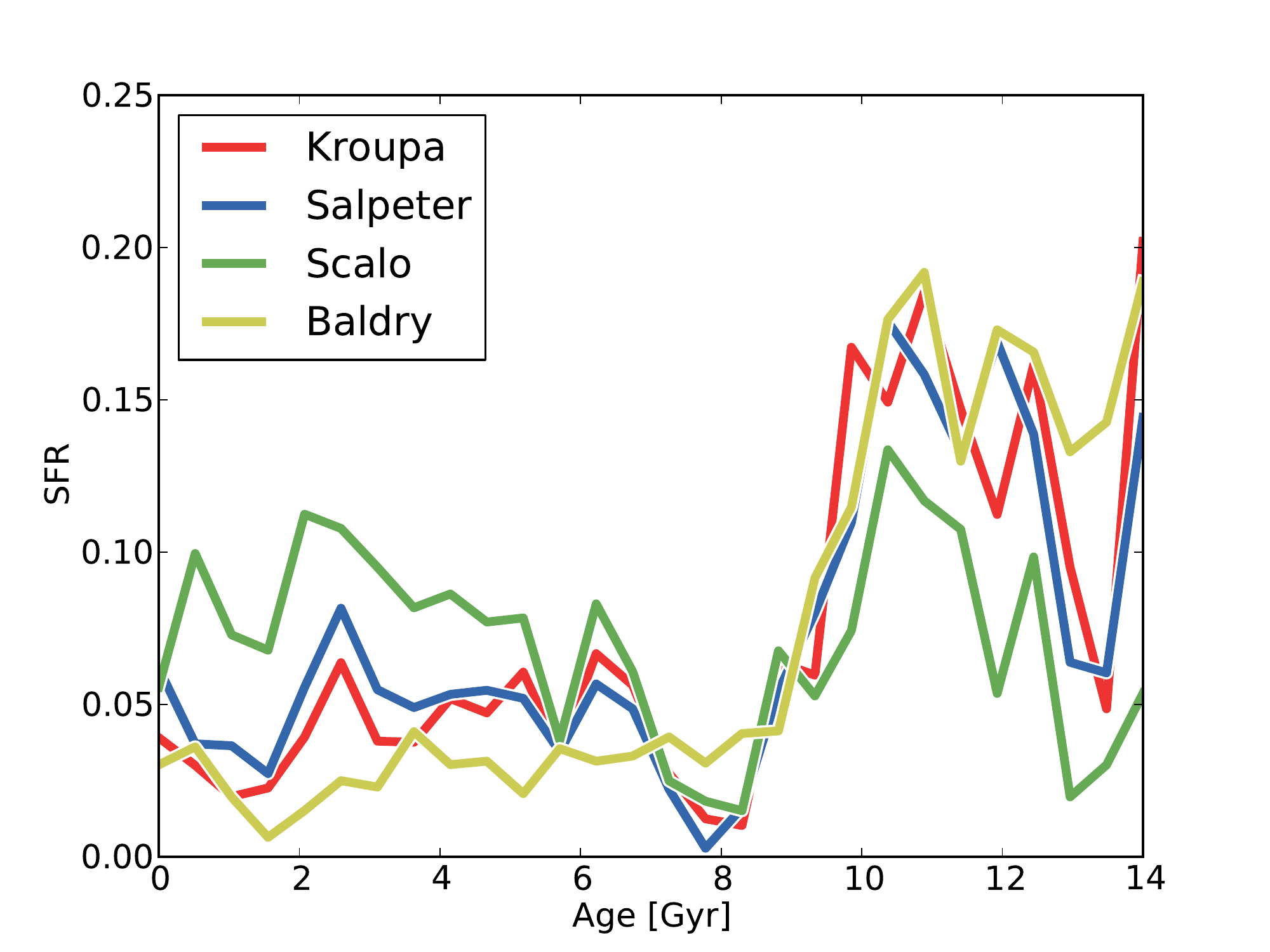}\\
(a) \\
\includegraphics[width=3.0in]{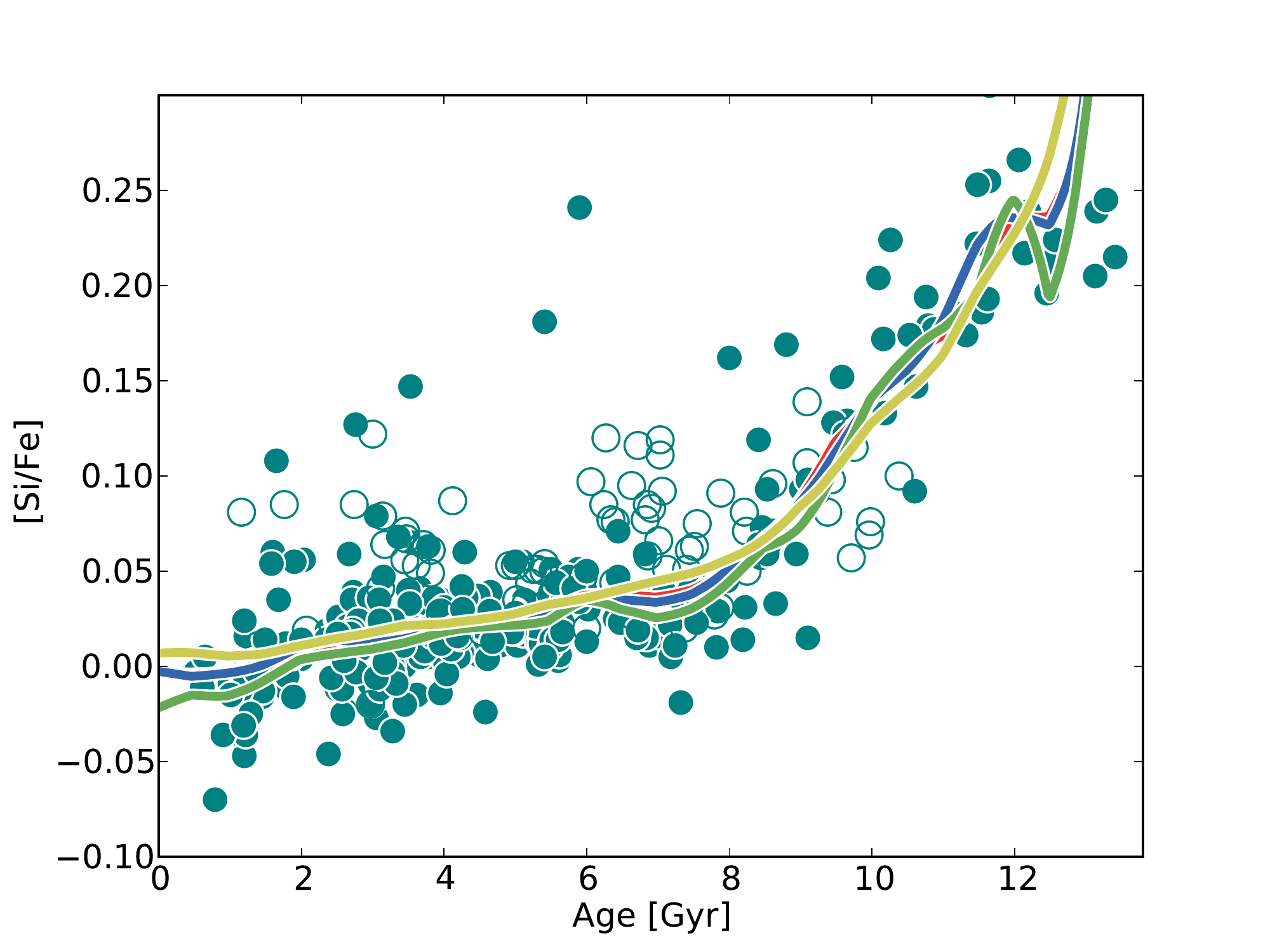} &  
\includegraphics[width=3.0in]{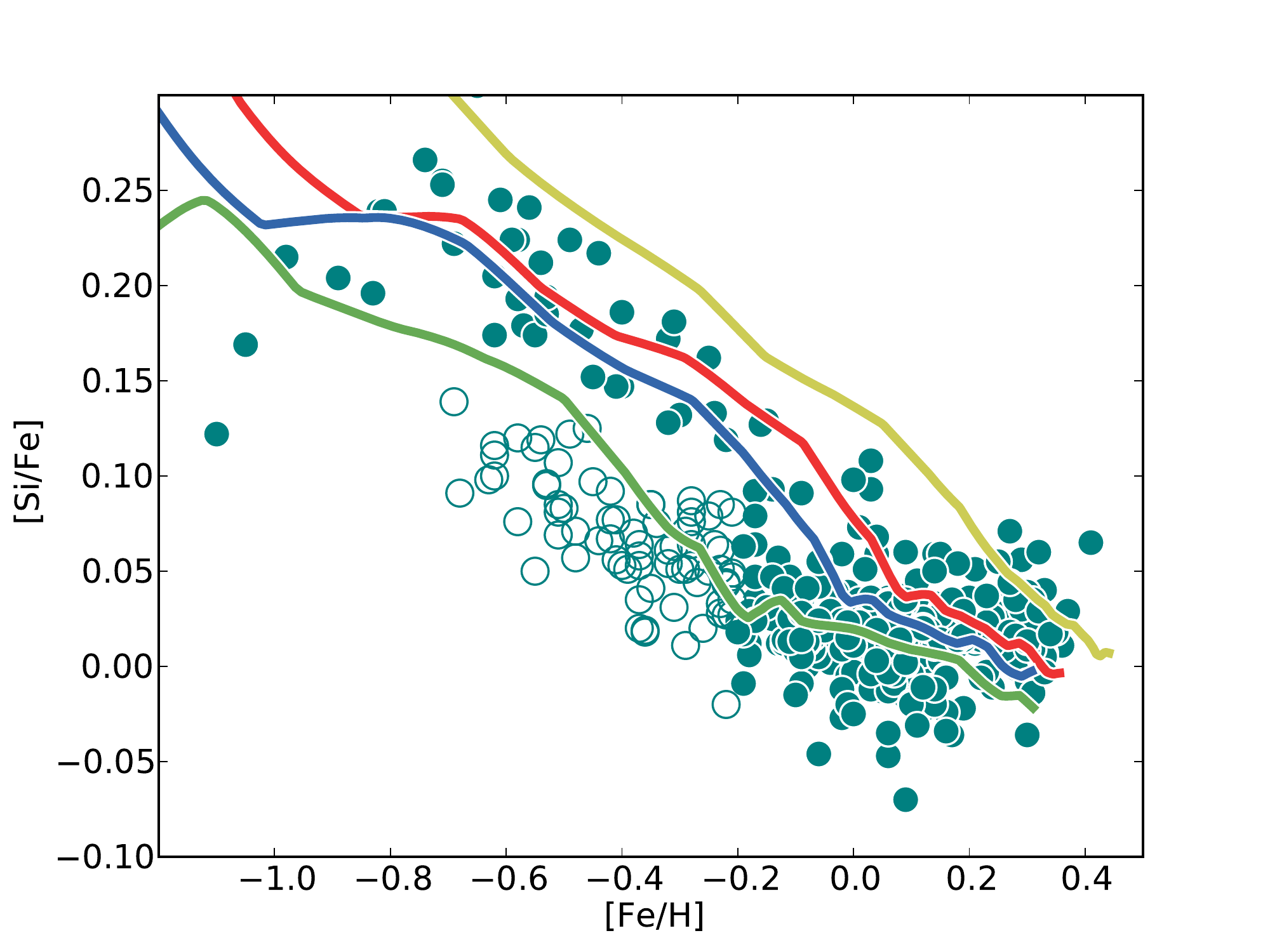}\\
(b) & (c) \\
\includegraphics[width=3.0in]{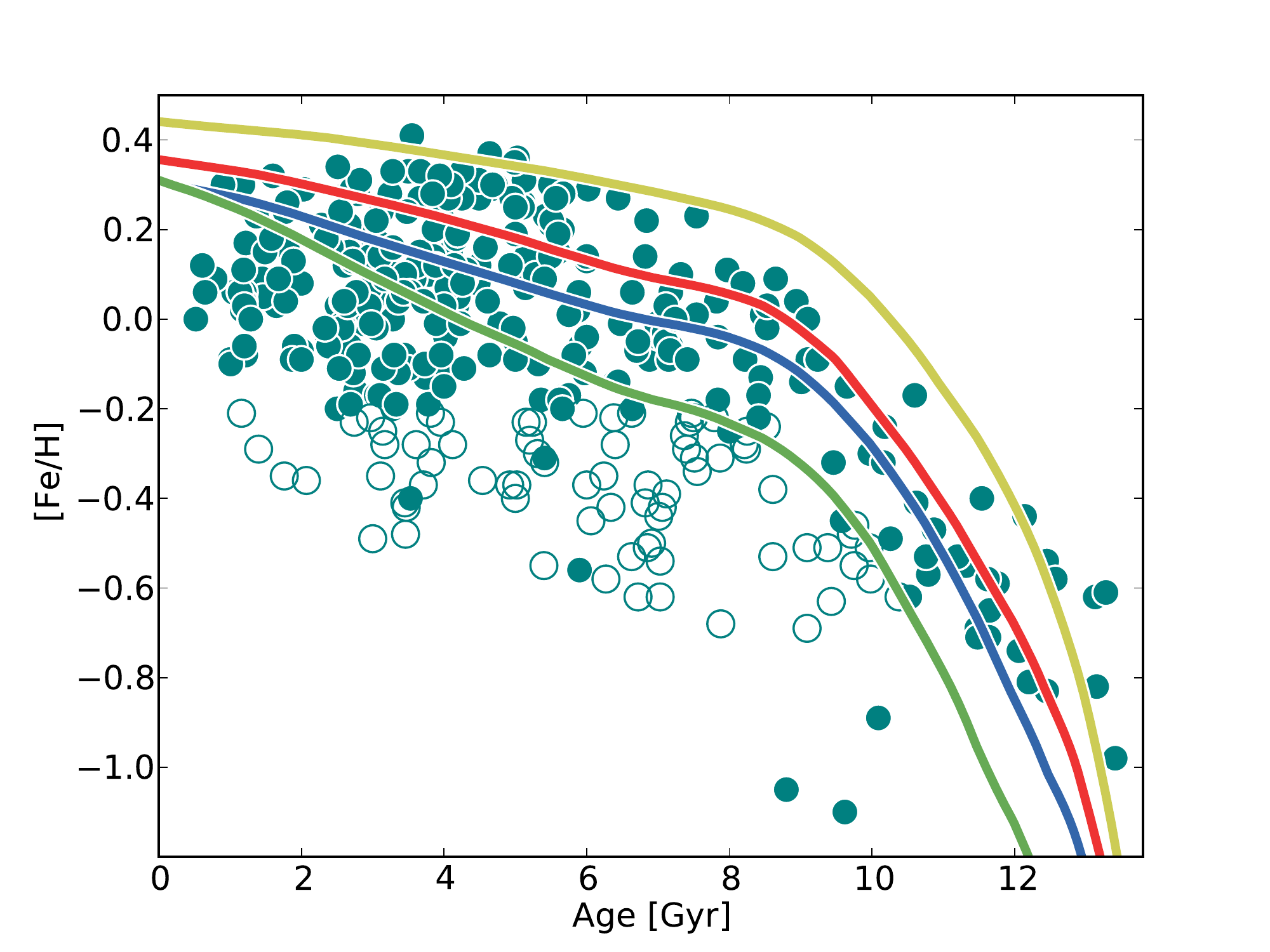} &  
\includegraphics[width=3.0in]{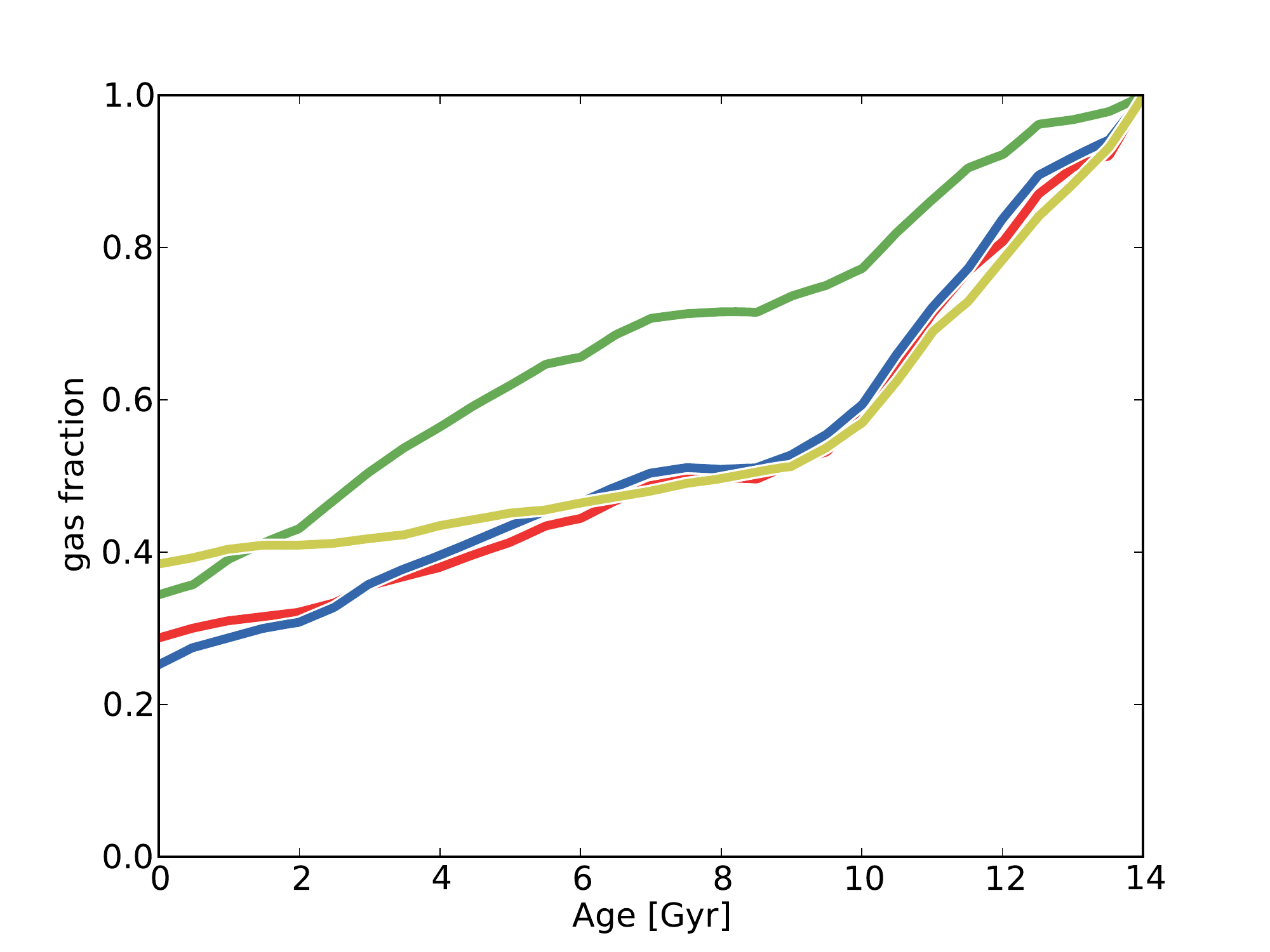}\\
(d) &(e) \\

\end{tabular}
\caption{The chemical tracks of the best fit SFHs using the different IMFs described in Table \ref{Tab:imflit}. The chosen IMFs are given in the legend in panel (a). Panel (a) shows the different fitted SFHs, panel (b) shows the evolution of [Si/Fe] with stellar age and is the distribution used to fit the data.  Panel (c) shows the evolution of [Si/Fe] with metallicity, panel (d) shows the metallicity evolution with stellar age, and panel (e) shows the evolution of the gas fraction with time. The points are from \citet{Haywood2013}. Open circles are outer thin disc stars, solid circles are thick disc and inner thin disc stars.}
\label{Fig:differentimfsfit}
\end{figure*}

Figure \ref{Fig:differentimfsfit} shows the different SFHs (panel (a)) and chemical tracks recovered by using the bootstrap procedure on each of the four selected IMFs.  We have once again attempted to fit the inner thin and thick discs. Each IMF can be closely fit to the age-[Si/Fe] distribution  (panel (b)) but show markedly different tracks through the metallicity-[Si/Fe] and age-metallicity distributions. The Baldry IMF overproduces iron, while the Scalo IMF is under-abundant in iron. The Kroupa and Salpeter IMFs are fairly similar. This suggests that we should discard the Scalo and Baldry IMFs based on this data. The different IMFs also imply a different fraction of stars in the thick disc, ranging from 37\% (Scalo) to 61\% (Baldry). There is considerable degeneracy between varying the solar abundance and choosing the IMF. We also see that the SFH derived for the Scalo IMF produces a much more gas rich ISM (panel (e)), and that the Baldry IMF does not recover the `dip' at 8 Gyr. 

\begin{table}
\centering
\begin{tabular}{c|cccc|c}
\hline
	Name & 0.1-0.5 & 0.5-1.0 & 1.0-10. & 10.-120 & C\\
\hline
	Kr01     & -0.8 & -1.7 & \multicolumn{2}{|c|}{-1.3}&  0.331\\
	S55              &\multicolumn{4}{|c|}{-1.35} & 0.171\\
	Sc98                  & \multicolumn{2}{|c|}{-0.2} & -1.7 & -1.3 &0.395\\
	B03    & -0.5 & \multicolumn{3}{|c|}{-1.2} &0.323\\
\hline
\hline
	\multicolumn{6}{c}{SSP mass fraction} \\
\hline
\hline
Name & SNII & AGB  & 8-3. & 2.6-1.8 &1.5-0.9  \\
\hline
Kr01 & 0.363 & 0.630 &  0.233 & 0.111& 0.187\\
S55  & 0.337 & 0.655 & 0.235 &  0.117 &0.202\\
Sc98 & 0.240 &  0.750 & 0.207 & 0.135 & 0.286\\
B03 & 0.792  &  0.207 & 0.121 &  0.030 & 0.030\\
\hline
\end{tabular} 
\caption{Various IMFs taken from the literature as defined in Table 2 of \citet{Baldry2003}. IMFs are normalised between 0.1-100 M$_\odot$. Kr01 is from \citet{Kroupa2001}, S55 is \citet{Salpeter1955}, Sc98 is from \citet{Scalo1998}, B03 is from \citet{Baldry2003}. C is the constant of normalisation between 0.1 and 100 M$_\odot$. The lower section of the table shows the integrated mass fraction of stars involved in SNII, AGB and the different components of the SNIa. }
\label{Tab:imflit}
\end{table}

\subsection{Normalisation}
At this point it is useful to discuss our choice of normalisation, $A$. In the rest of the paper we have required that the integral of the SFH is equal to unity. Essentially, this sets the star formation efficiency of the system, with lower values having less efficient star formation. 

\begin{figure*}
\centering
\begin{tabular}{cc} 
\includegraphics[width=3.0in]{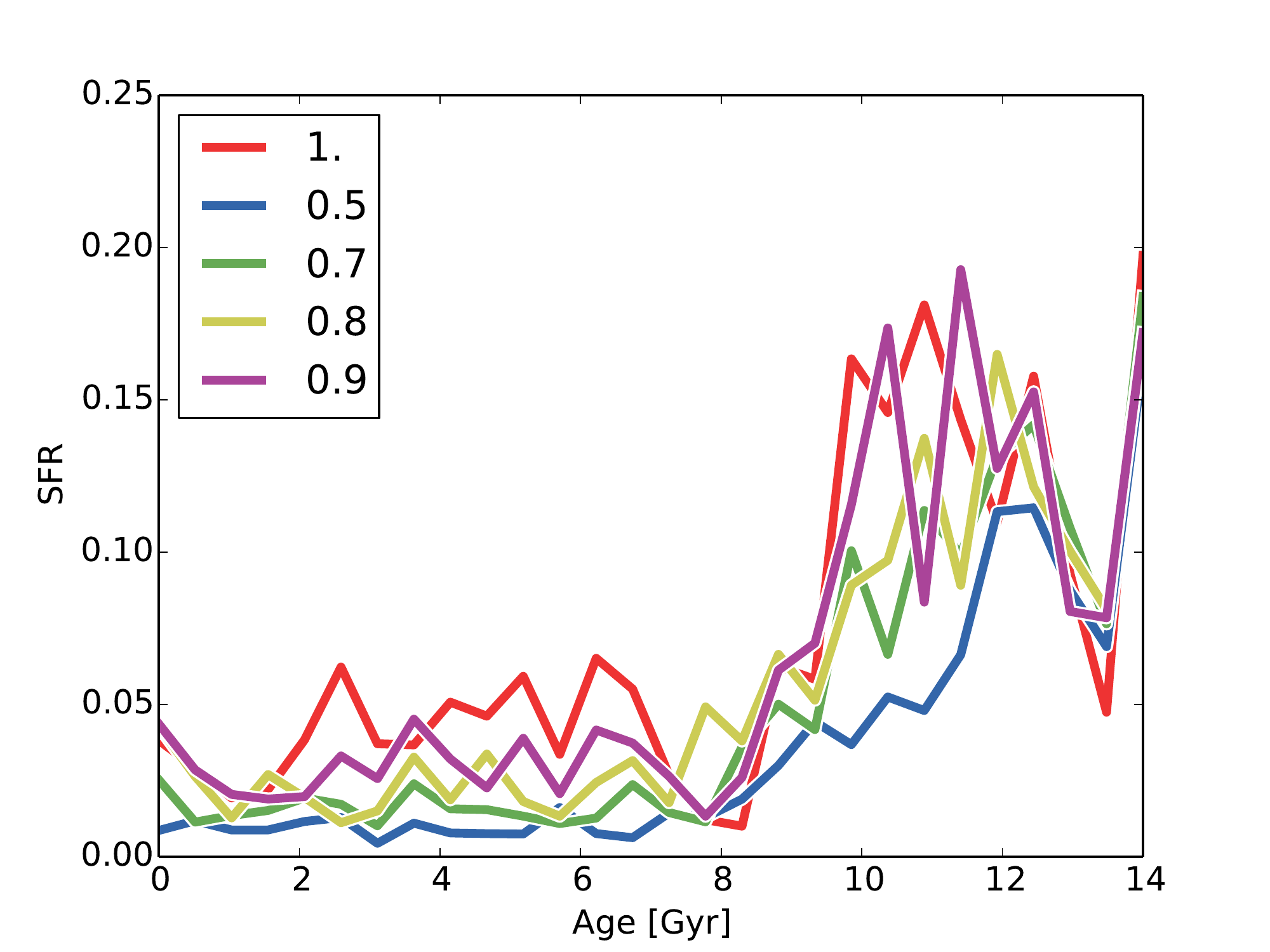} &
\includegraphics[width=3.0in]{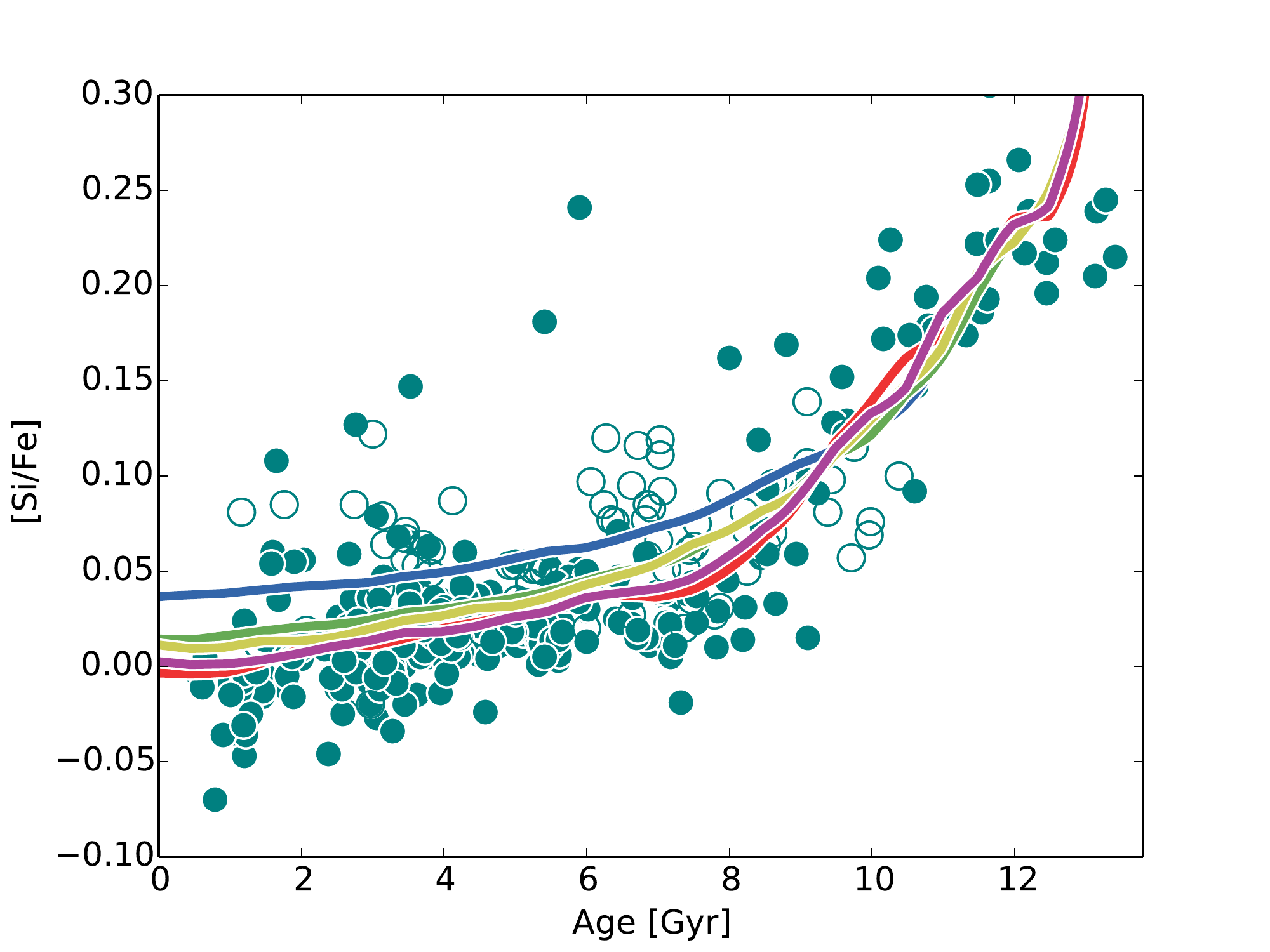} \\
(a) & (b)  \\
\includegraphics[width=3.0in]{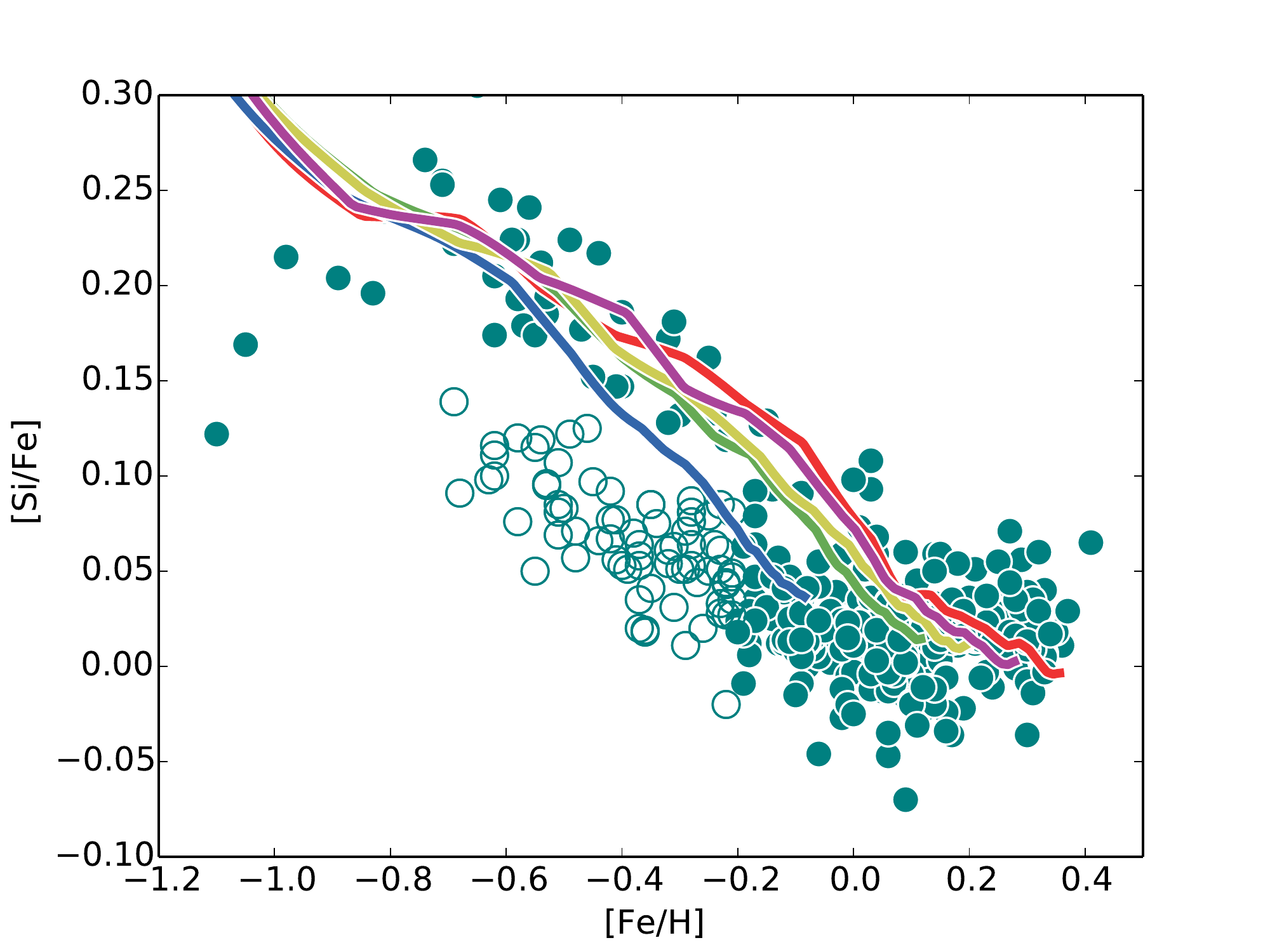} &
\includegraphics[width=3.0in]{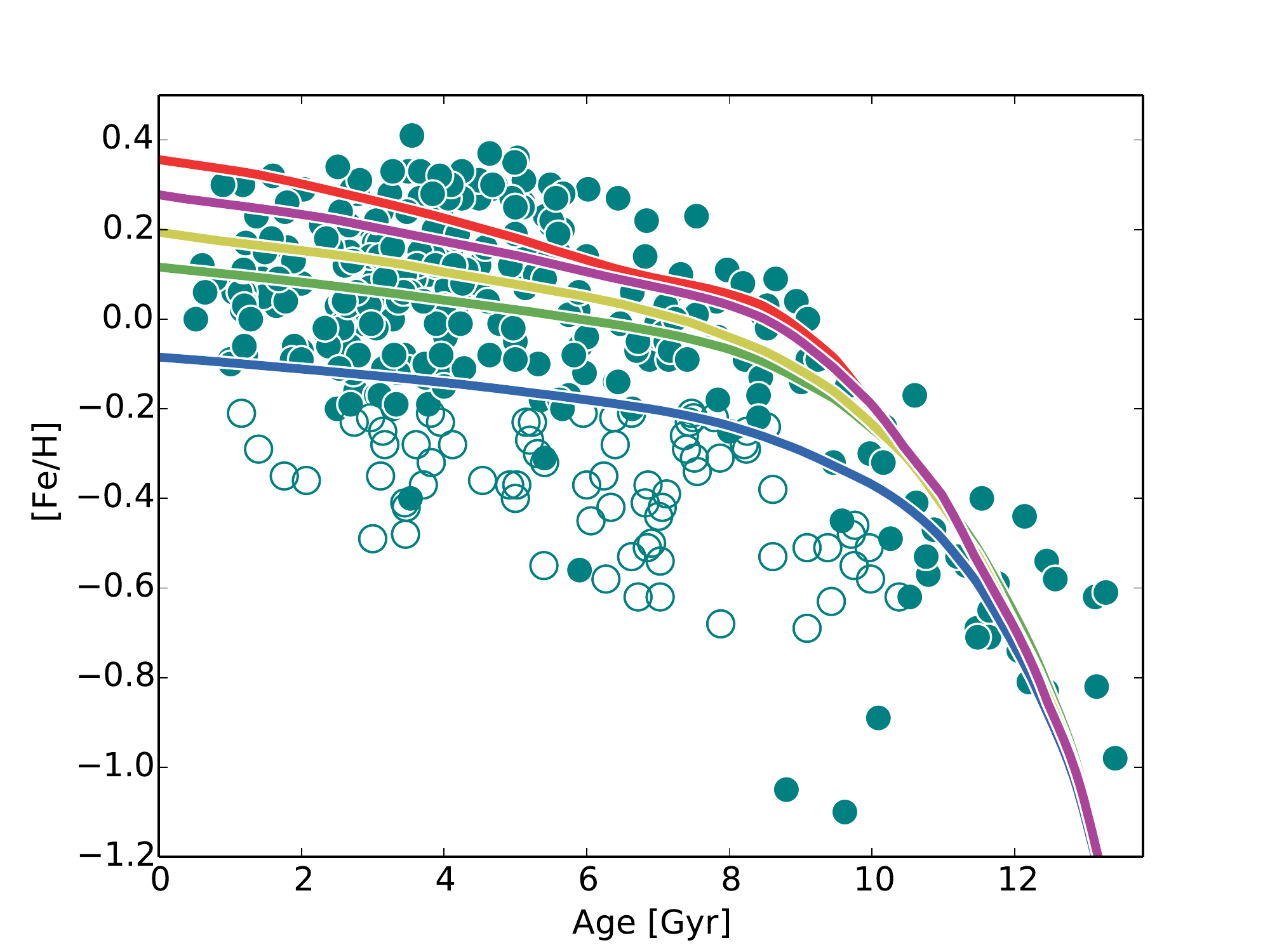} \\
(c) & (d)  \\
\end{tabular}
\caption{Comparison between the best fitted (to thick and inner thin disc stars) chemical tracks for different total SFH normalisation. The values of the integral of the SFHs are given in the legend in panel (a). Panel (a) shows the different fitted SFHs, panel (b) shows the evolution of [Si/Fe] with stellar age and is the distribution used to fit the data.  Panel (c) shows the evolution of [Si/Fe] with metallicity and panel (d) shows the metallicity evolution with stellar age. The points are from \citet{Haywood2013}. Open circles are outer thin disc stars, solid circles are thick disc and inner thin disc stars. }

\label{Fig:withdeltanorm}
\end{figure*}

The first clear result of lowering $A$ is that it quickly destroys the otherwise robust `dip' feature (Fig. \ref{Fig:withdeltanorm}). Although by $A=0.5$ the age-[Si/Fe] track does not evolve to a iron rich enough end state, where $0.7<A$ the chemical tracks match the data fairly well due to the intrinsic scatter. Nevertheless, it is only in the normalisation equals 1 case that the dip and `knee' is very sharp (panel b).  

One potential advantage of these lower ($A>0.7$) normalisations is that they have a low [Fe/H] end state, and so identify high metallicity stars which are assumed to originate from subsolar galactic radii (as suggested in Section \ref{featuresSFR}). 

Naively, one would assume that changing $A$ would have only a minimal impact on the chemical evolution, other than to shift the plot to lower [Fe/H] values. However, this simplistic view is complicated by the fact that the yields from stars depend on the metallicity at which the star is formed. This is why, in the outer disc, introducing a dilution changes the track in [Si/Fe]. The physical interpretation of the normalization is simply that the ISM is more massive compared to the stellar mass of the galaxy, and so ejected metals are diluted into a greater quantity of primordial gas. Thus, the star formation efficiency is less intense. 

A further impact of lowering the normalisation is that there is insufficient contrast between the top and bottom of the dip at 8 Gyr, and so this feature no longer has an effect on the chemical evolution.  

\subsection{Using different Elements}\label{diffel}

Here we discuss the differences in the tracks of magnesium and oxygen using \citet{Nomoto2006} SNII yields and \citet{Iwamoto1999} SNIa yields. We chose to follow these elements in addition to silicon because magnesium is also used in \citet{Haywood2013}, and oxygen is both the dominant $\alpha$ element, and very common in numerical simulations, \citep[e.g.][]{Brook2012, Gibson2013}. These elements differ from silicon because they are produced in high amounts by  SNII alone, while silicon has contributions from both types of SN. We use the \citet{Anders1989} solar normalisation for magnesium, and the solar normalisations derived by \citet{Ramirez2013} for oxygen and iron when dealing with oxygen data\footnote{These values are given as 8.64 and 7.45 for oxygen and iron from \citet{Ramirez2013}. In \citet{Anders1989} the Si, Mg and Fe values are 7.55, 7.53 and 7.51 respectively. } We will discuss the oxygen tracks and data first, and then explore the differences between the three different alpha species. 

\begin{figure*}
\centering
\begin{tabular}{cc} 
\includegraphics[width=3.0in]{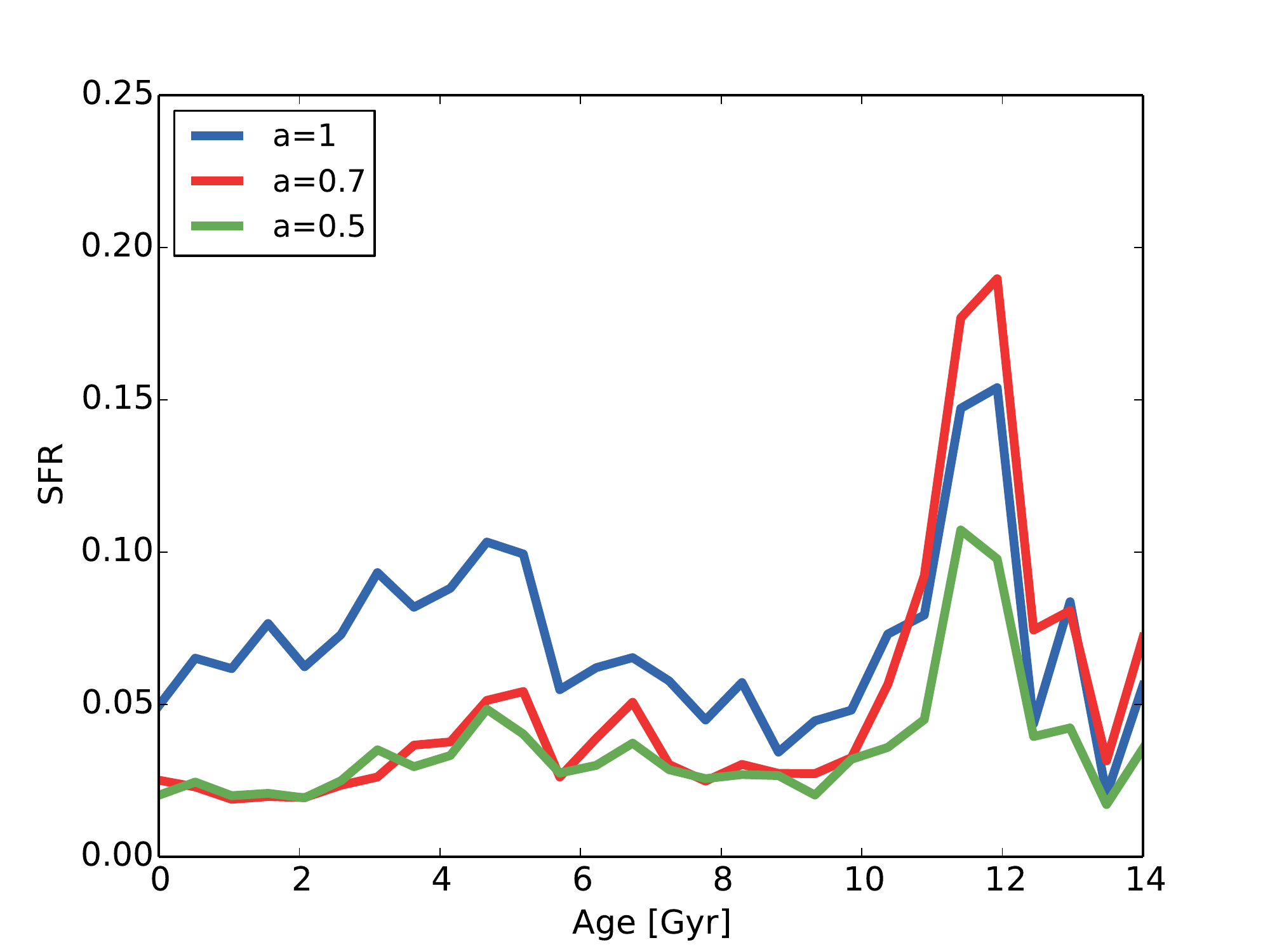} &
\includegraphics[width=3.0in]{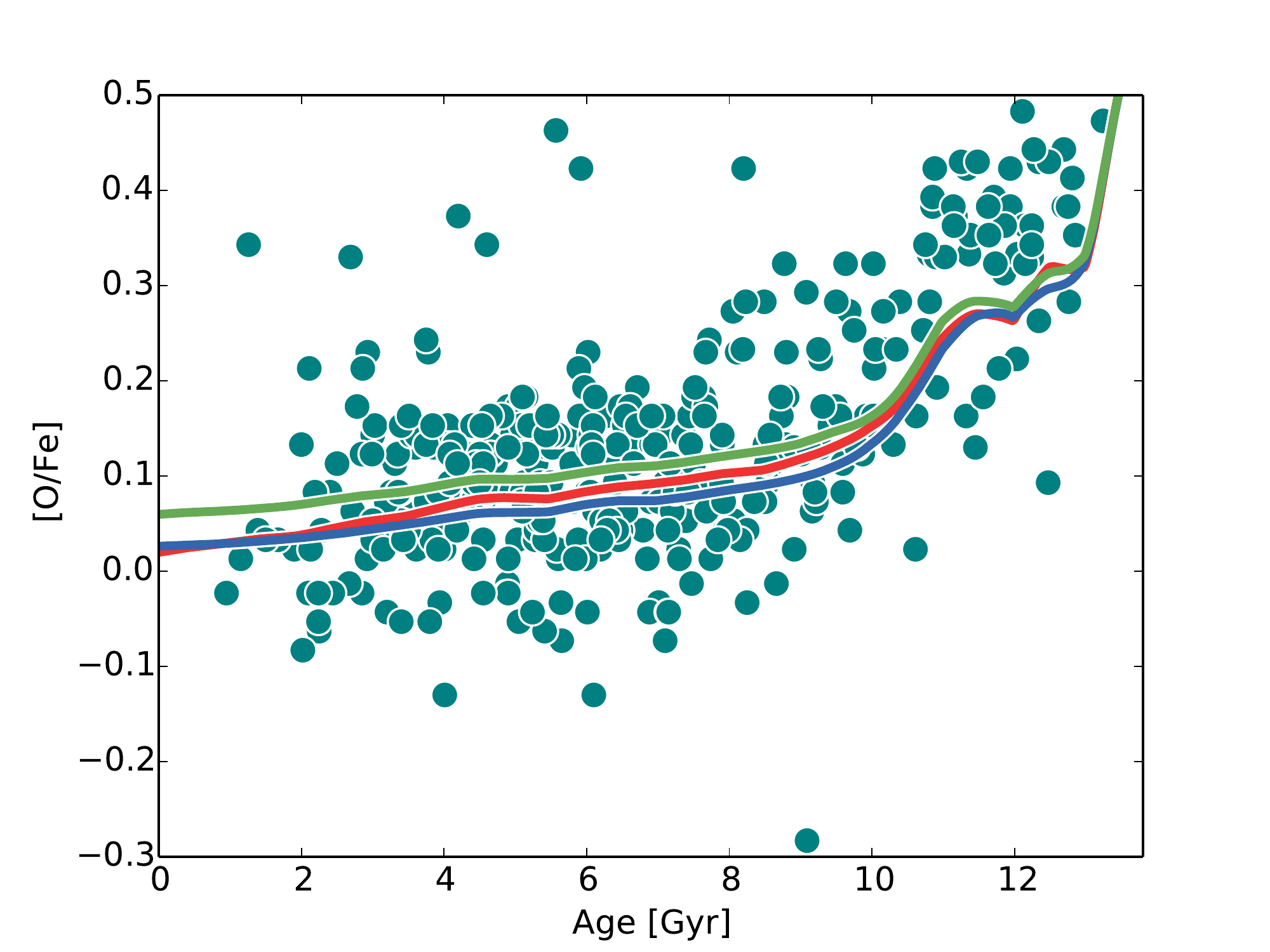} \\
(a) & (b)  \\
\includegraphics[width=3.0in]{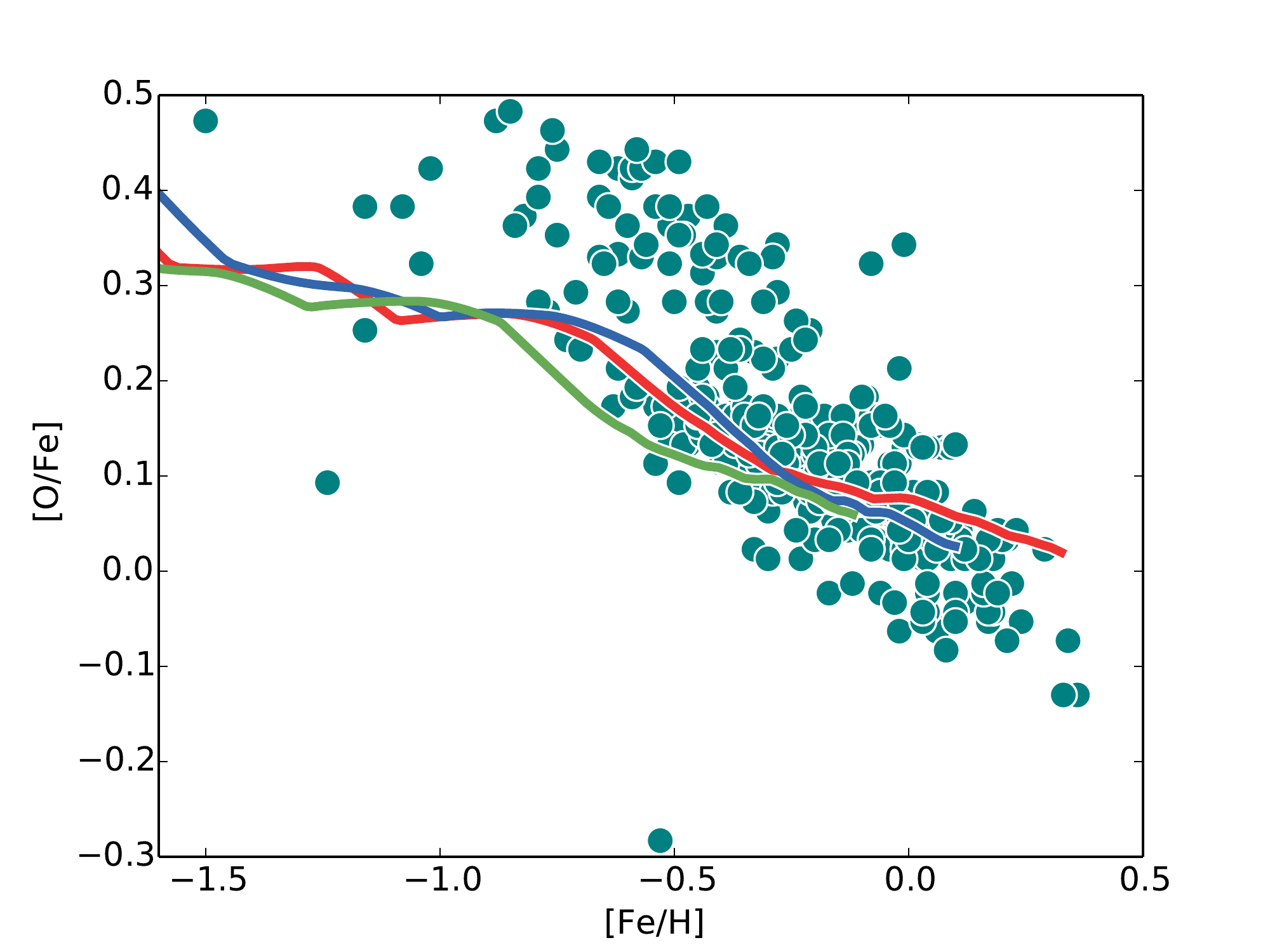} &
\includegraphics[width=3.0in]{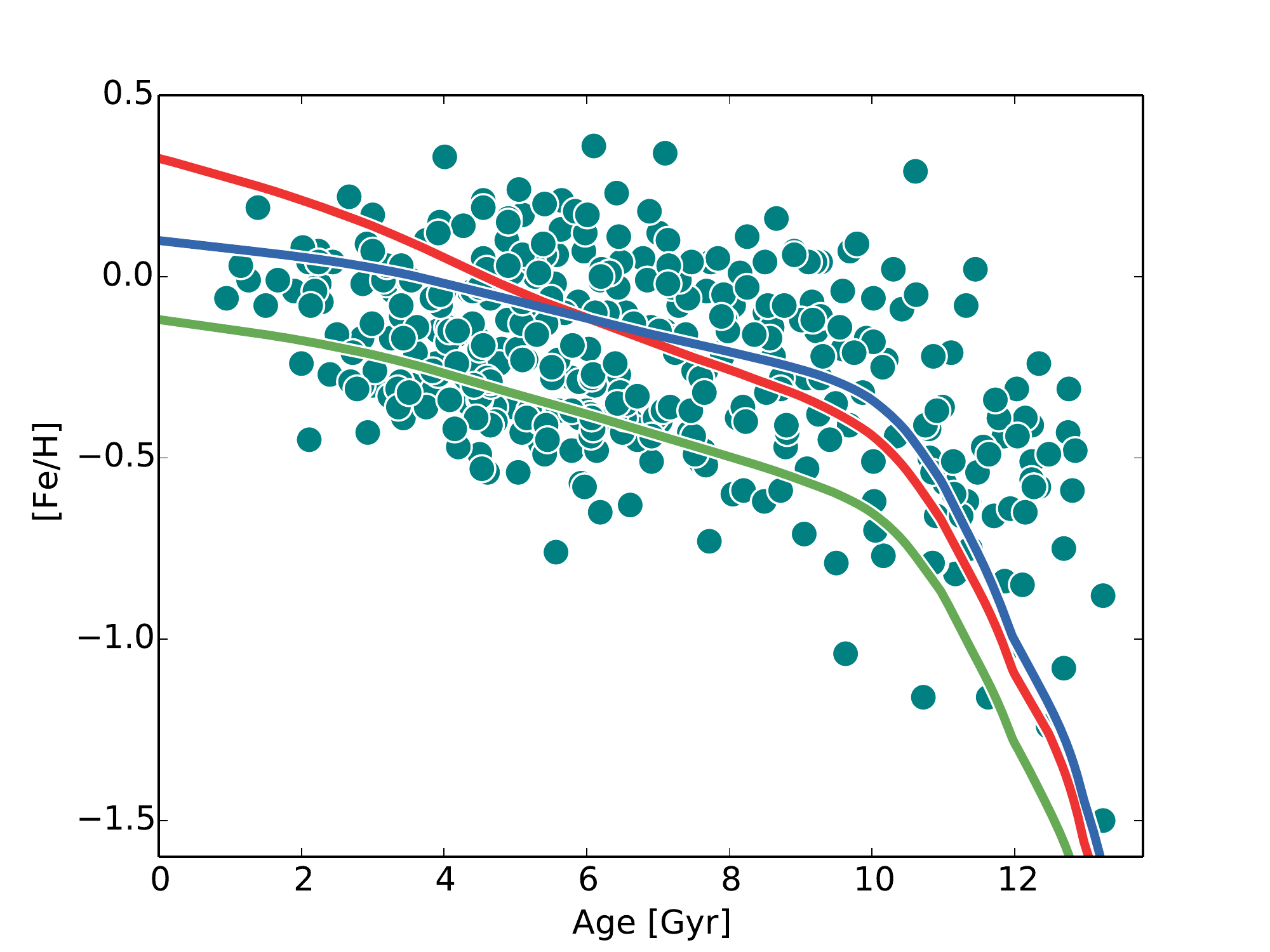} \\
(c) & (d)  \\
\end{tabular}
\caption{The best fitted SFR for oxygen abundances for different total normalisations of the SFH. The values used in the normalisation are given in the legend in panel (a). The points are \citet{Ramirez2013} data. }

\label{Fig:bestfitoxygen}
\end{figure*}

\begin{figure*}
\centering
\begin{tabular}{ccc} 
\includegraphics[width=2.0in, trim={0cm 0cm 1.5cm 0cm}, clip]{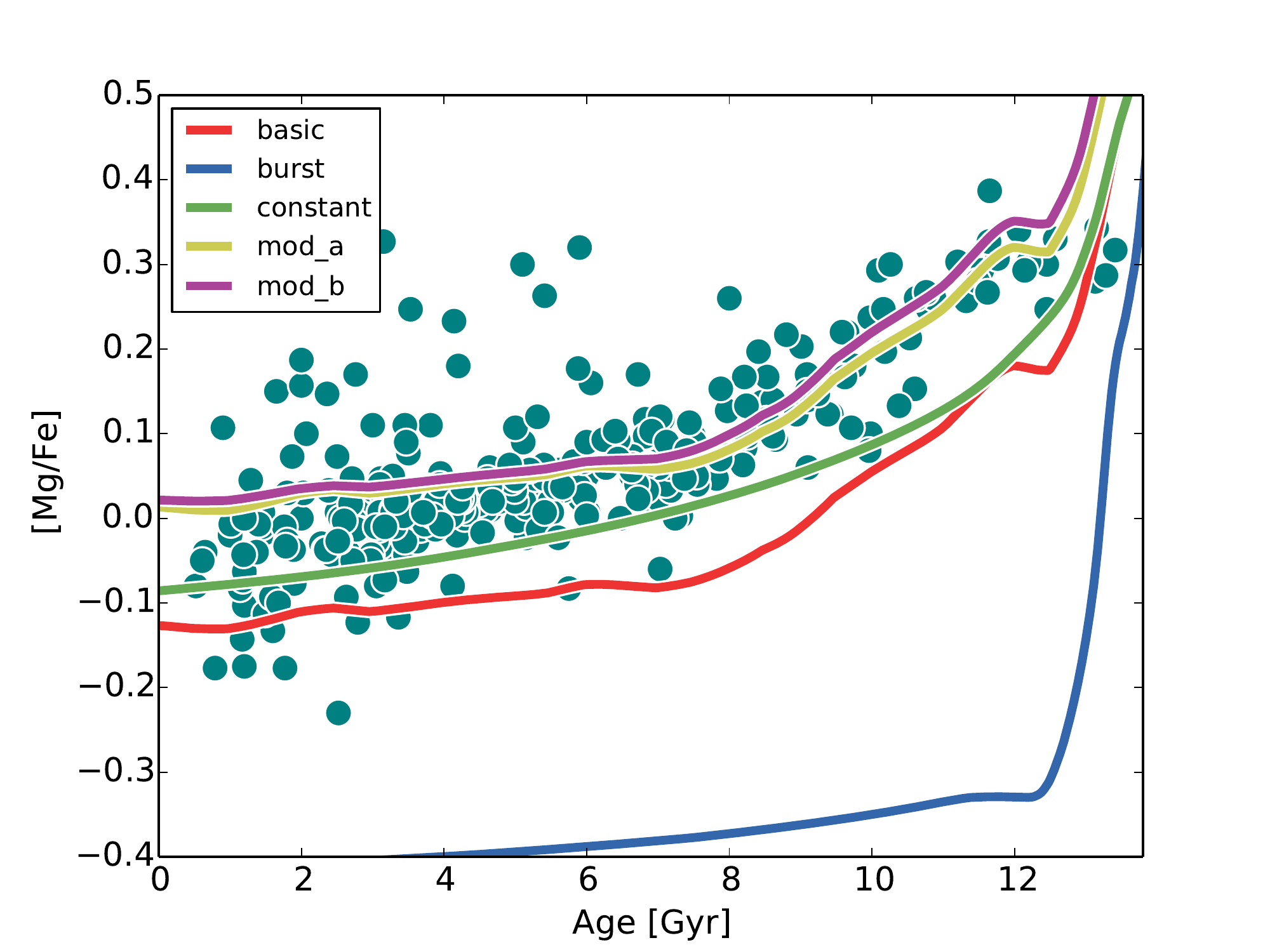} &
 \includegraphics[width=2.0in, trim={0cm 0cm 1.5cm 0cm}, clip]{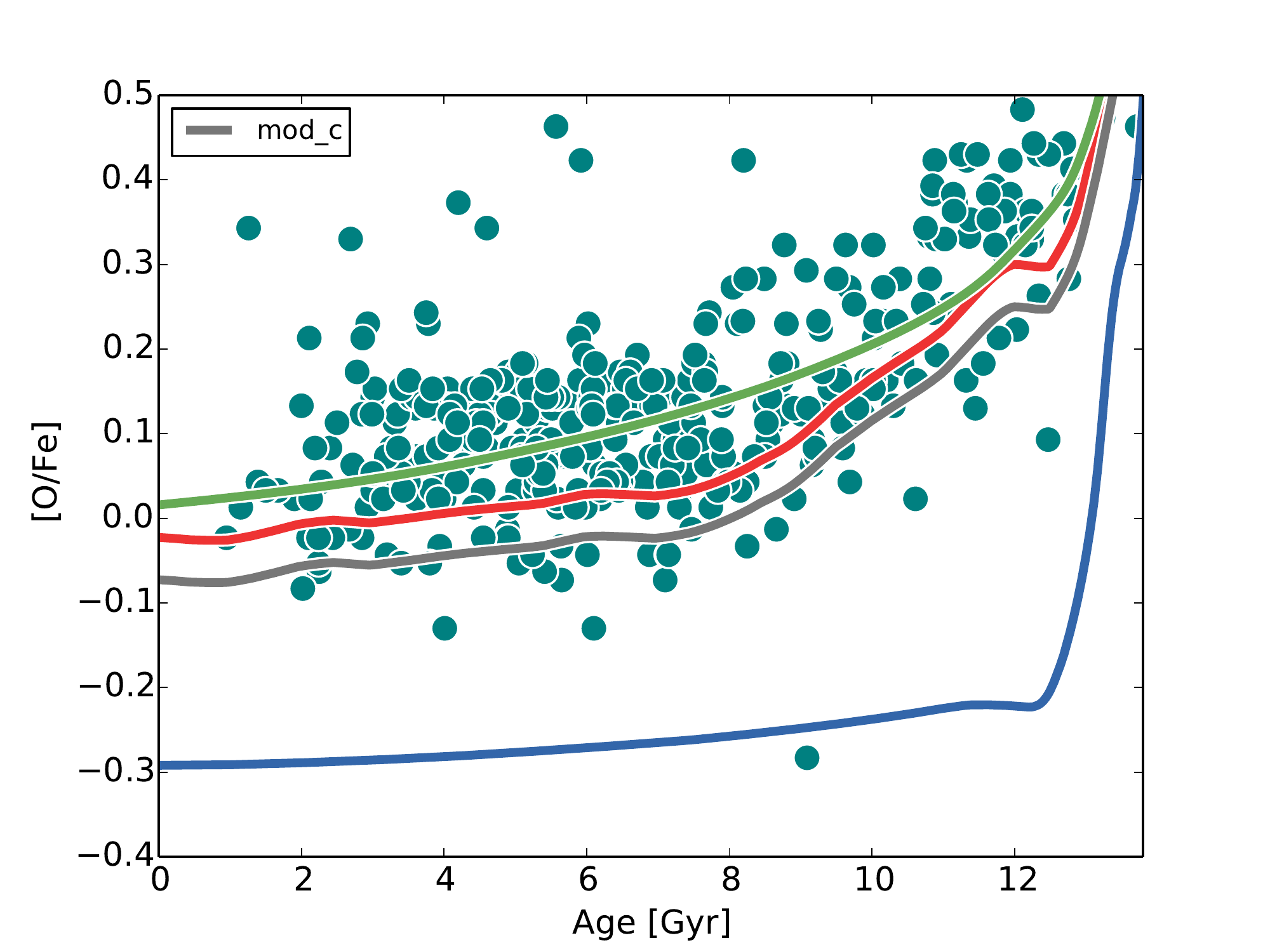} &
  \includegraphics[width=2.0in, trim={0cm 0cm 1.5cm 0cm}, clip]{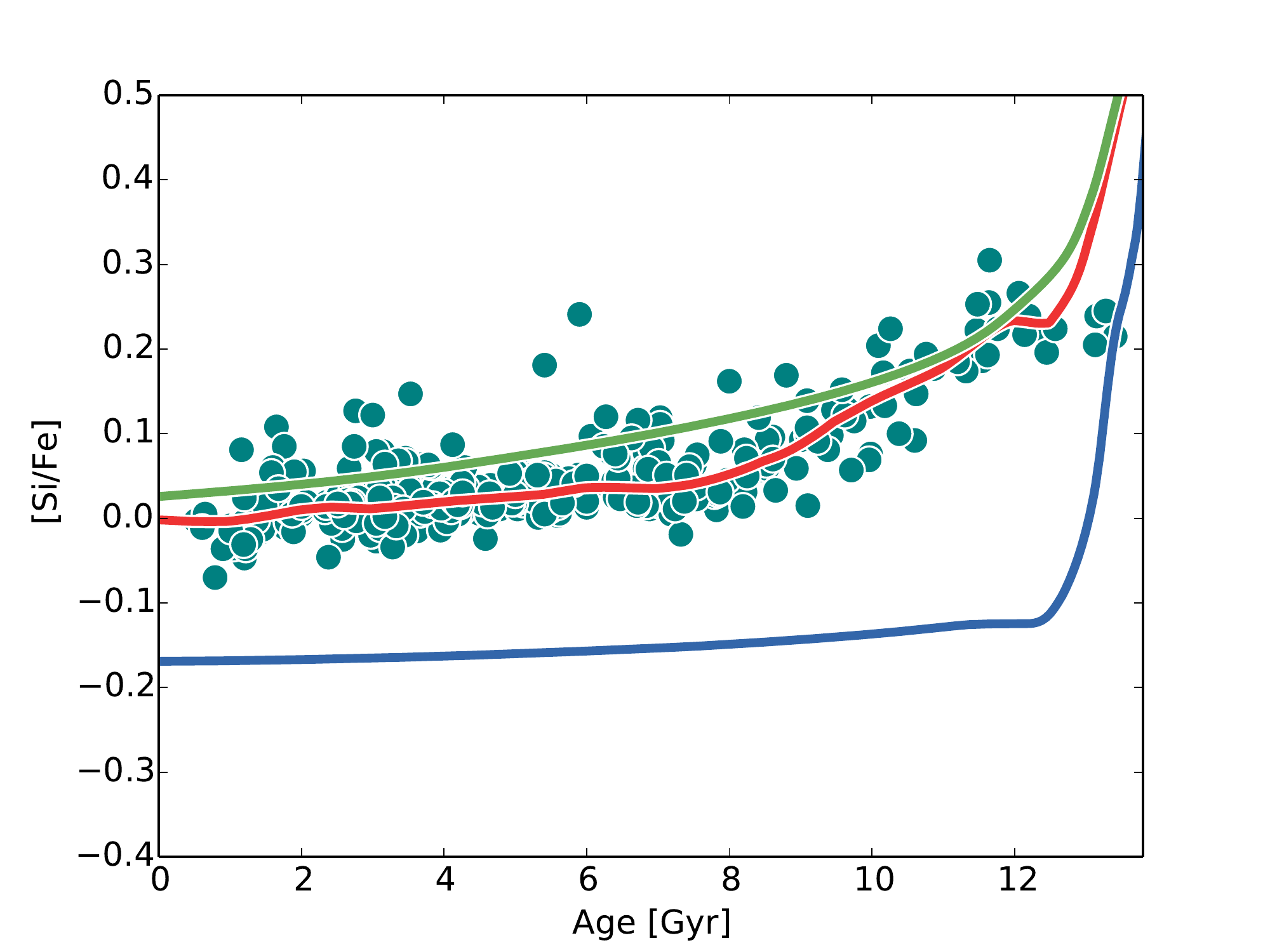}  \\
(a)  & (b) & (c)\\
\includegraphics[width=2.0in, trim={0cm 0cm 1.5cm 0cm}, clip]{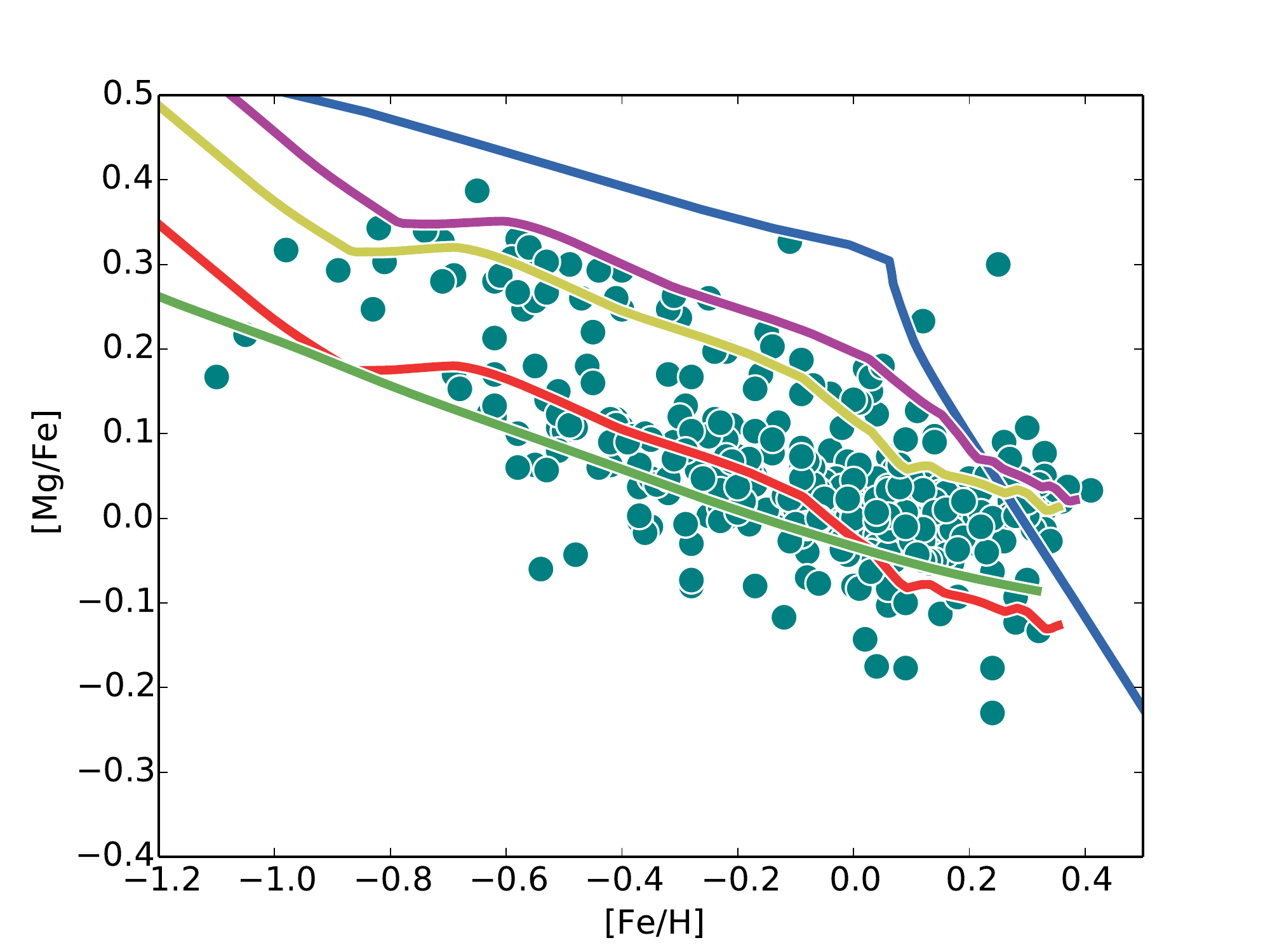} &
 \includegraphics[width=2.0in, trim={0cm 0cm 1.5cm 0cm}, clip]{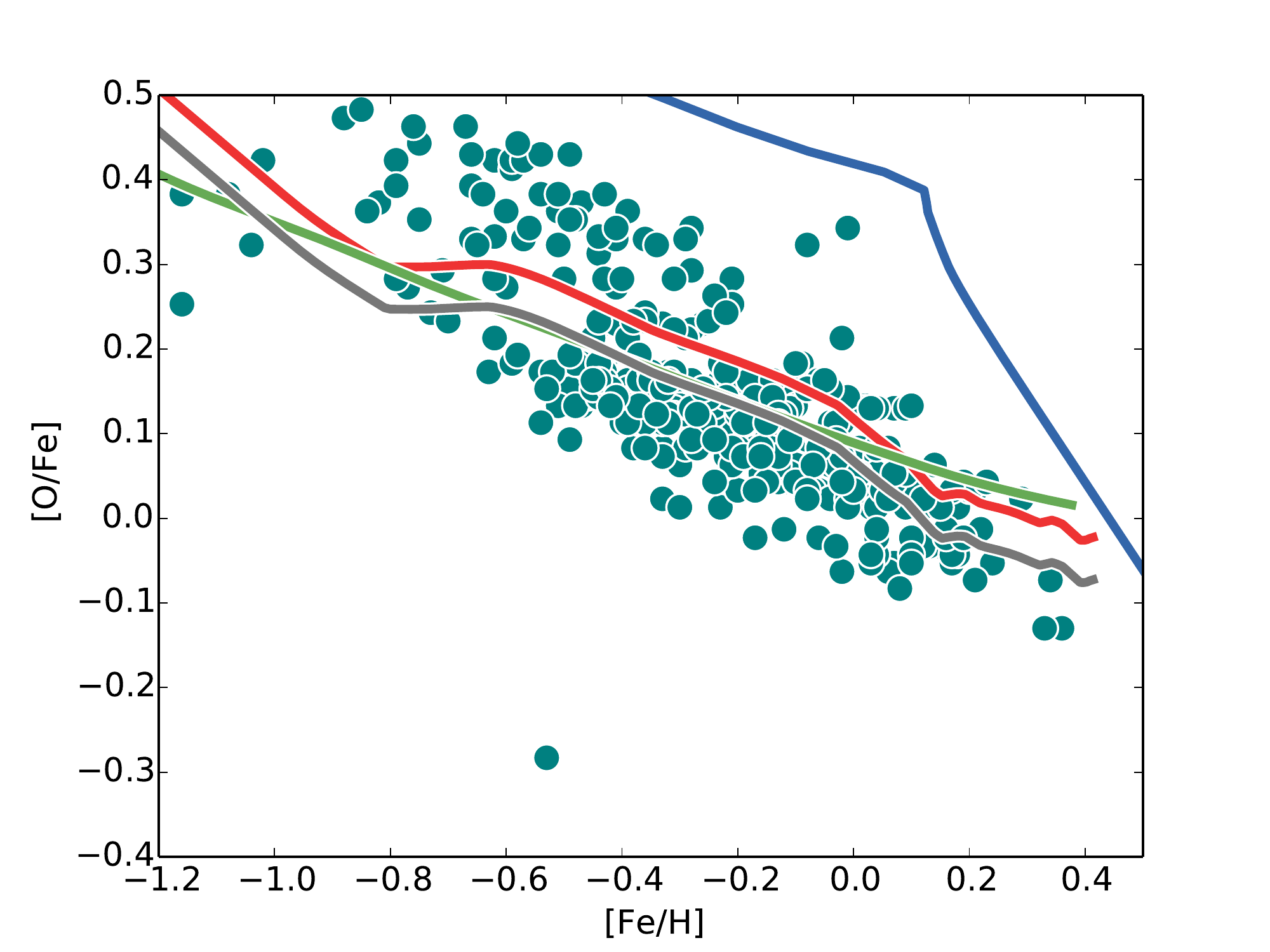} &
   \includegraphics[width=2.0in, trim={0cm 0cm 1.5cm 0cm}, clip]{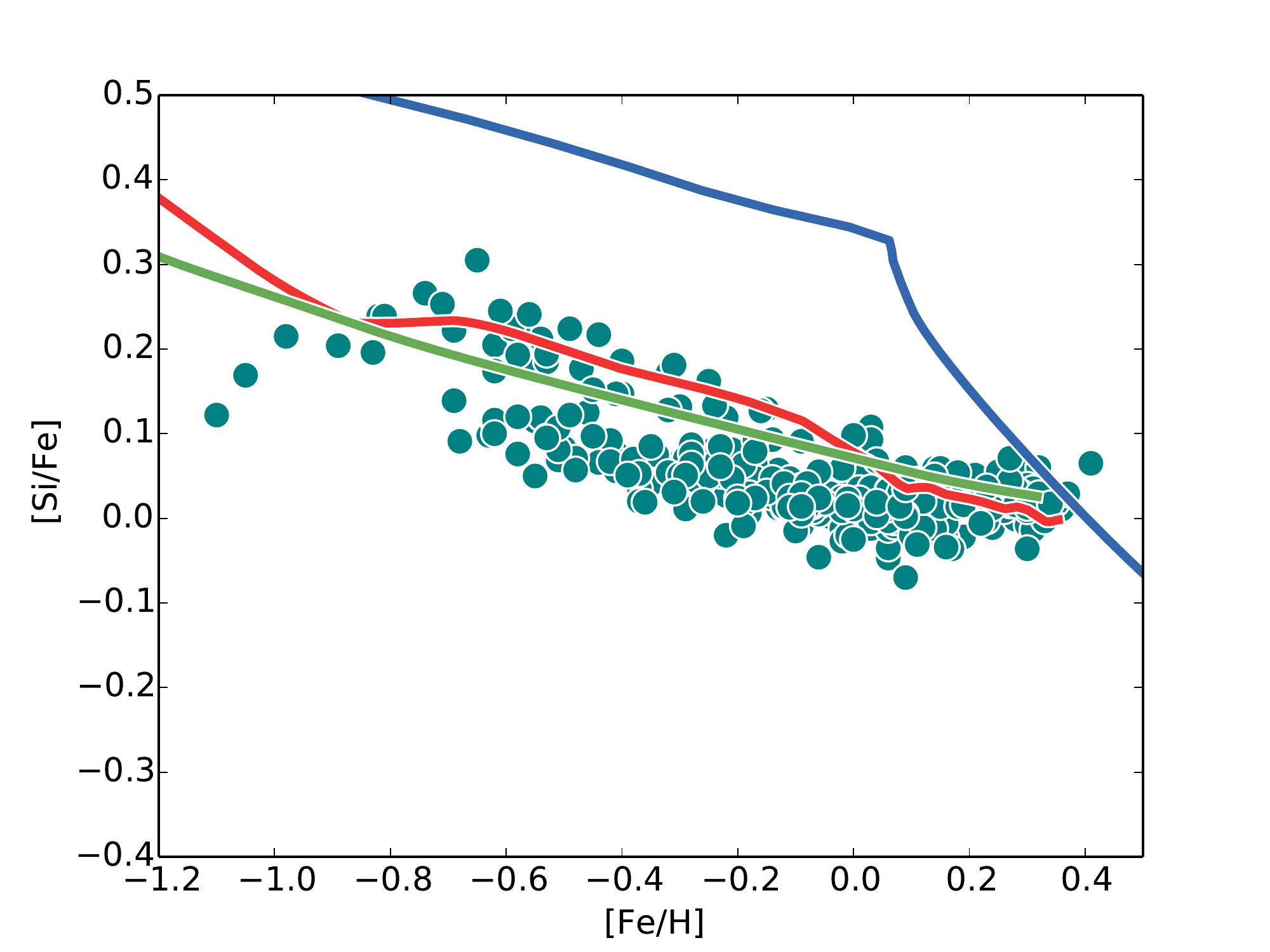}  \\
(d)  & (e) &  (f)\\

\end{tabular}
\caption{The chemical evolution of [$\alpha$/Fe] with age and metallicity for several SFHs. We used the  best fitted SFH for silicon (Fig. \ref{Fig:bestfitsiandersmet}) using \citet{Anders1989} meteoric solar abundances ('basic', red line) as well as a sharply initially bursting SFH ('burst', blue line) and a constant SFR ('constant', green line). The left hand column shows the magnesium evolution, the centre column shows the oxygen evolution, and the right hand column shows the silicon evolution.  The additional lines illustrate various modifications to the best fitted SFH.  `mod\_a' (yellow line) is the track, based on the `basic' SFH where a constant of 0.14 dex has been added to the solar magnesium. `mod\_b' (purple line) is the track where a constant of 0.18 dex has been added to the 'basic' track, but the late time (stars older than 8 Gyr) SFR has been reduced by 2.  `mod\_c' (grey) is the 'basic' track where the solar oxygen value is shifted by -0.05 dex. }
\label{Fig:bestfitsiandersmet_o_mg}
\end{figure*}

The best fit SFH for oxygen has a significantly different form.  There is a low initial SFR, which is needed so that the [O/Fe] track is close to the thick disc sequence in panel (b). Even so, the oxygen track lies below the thick disc data. This low initial SFR has an impact on the initial rise of the [Fe/H]-age track (panel (c)). The initial rise in [Fe/H] is slower than the data suggests, and so does not reach the [Fe/H] values of the oldest stars. The low initial SFR is followed by a peak in SFR for each of the three normalisation values. The ratio of the early to late time SFR is lowest for the case where $A=1$ (where $A$ is the normalisation), but has a similar form to the SFH recovered for silicon where $A<$1. However there is no `dip' at 8 Gyrs in any of the oxygen SFHs. This may be due to either the solar normalisation (discussed in the previous subsection) or the larger scatter in the dataset. 

Figure \ref{Fig:bestfitoxygen} shows the best fit SFH for the \citet{Ramirez2013} oxygen yields and ages. There is considerably more scatter in the data in the age-[O/Fe] plot than in the \citet{Haywood2013} sample, and it was not possible to identify a specific inner and outer disc using oxygen abundances. This is because the \citet{Ramirez2013} data have a significantly lower S/N ratio than the \citet{Adibekyan2012} data. We can see that the best fit SFHs give a poor fit to the metallicity-[O/Fe] distribution in panel (c), and shift the tracks to the lower left hand side of the plot. If we assume that the \citet{Nomoto2006} + \citet{Karakas2010} +\citet{Iwamoto1999} yields  and our tracks for SSPs (Section \ref{partA}) are correct, then these plots suggest that there has been considerable dilution of the metals due to either gas inflow or some other  process. Considering the uncertainties in the theoretical yields, and the choices we have made in our model, however, it is impossible to know whether it is not just a matter of inaccurate theoretical stellar yields.  A solution to this issue may involve a modification to our truncation of the stellar yields at 40 solar masses. This works well for silicon, but is, perhaps, less appropriate for oxygen and/or magnesium (see below). Including the yields from stars with M$_*>$40 M$_\odot$ is apparently required to produce enough oxygen to match the data, but a straight extrapolation to higher masses overestimates the amount of oxygen. This would produce a much more massive thick disc, in order to lower the [O/Fe] value to match the data, see Fig. \ref{Fig:SFRexponential}.

Due to the scatter in the data, the separation between the inner and outer thin discs, seen in the \citet{Adibekyan2012} sample, is not apparent  (see panel c). This, as well as the uncertainties in the model, may be hiding the details of the SFH, such as the dip. 

What it noticeable in the recovered SFHs is that  a high SFR during the  `thick disc' era is required, followed by a long period of lower star formation in order to fit the data. This is qualitatively the same as for silicon, and confirms  that the thick disc is a substantial component of the Galaxy, formed during an intense phase of star formation.

When we fit the age-[O/Fe] distribution in the same manner as the age-[Si/Fe] track we find that we cannot fit the [Fe/H]-[O/Fe] plot at the same time. Additionally, using the \citet{Ramirez2013} dataset there is no clear way of splitting the data between the thick disc/inner thin disc and the outer thick disc like in the \citet{Haywood2013} data. The slope of the data in [O/Fe]-[Fe/H] is too  shallow for our model to recover. The best fit SFH also fits the [O/Fe]-age distribution fairly well, if we shift it by 0.05 dex (panels (b) and (e)), except at high [Fe/H] in panel (e).

The data for oxygen (see Fig. \ref{Fig:Combes} and \ref{Fig:bestfitoxygen}) show the same tightly correlated thick disc component, and a change at around 8 Gyr. {\it The recovered SFHs are different for silicon and oxygen, but in the details rather than in the general case.} Either way, the thick disc is very massive  and contains $\sim$38\%, $\sim$56\% and $\sim$48\% of the stars for $A$=1.0, 0.7 and 0.5 respectively. In each case this would give the thick disc a mass of around 2.1$\times$10$^{10}$M$_{\odot}$, 3.2$\times$10$^{10}$M$_{\odot}$, and 2.7$\times$10$^{10}$M$_{\odot}$ respectively, assuming a total Milky Way stellar mass of 5$\times$10$^{10}$M$_{\odot}$. 

%===========

Using a selection of three simple SFHs, the oxygen and magnesium abundance tracks can be compared to the silicon tracks and the data. Our model under-produces magnesium relative to the iron to an extent of 0.14 dex or 40\% (Fig. \ref{Fig:bestfitsiandersmet_o_mg}, panel(a)). The late time slope is flatter than in the observations for this element, and the evolution of [Mg/Fe] is not bracketed by burst and constant SFHs, suggesting that, if it is to be fit at all, the SFH would be increasing. 

The [O/Fe] evolution is an  equally bad match to the data  as magnesium for any SFH (panels (a) and (b)  Fig. \ref{Fig:bestfitsiandersmet_o_mg}), this is  equally true in the metallicity-[$\alpha$/Fe] tracks. These tracks show that the data lie between the burst and constant SFHs for the magnesium  and oxygen. Silicon is by far the best at fitting both the panels.  

Using the model in table \ref{Tab:summarymodel}, the magnesium evolution cannot be converged to the data. If the data is only just covered by the increasing SFR (i.e. where y=kx and k is the constant of normalisation),  the features of the plots cannot be recovered using our best fitting procedure. An alternative approach would be to release the requirement that the integral of the SFR is equal to one, allowing $A<1$ and thus producing a greater variation in the end point of the age-[Mg/Fe] evolution. This, however, influences the [Fe/H] distribution considerably, shifting the tracks to lower [Fe/H]. This means that the [Fe/H]-[Mg/Fe] cannot be recovered simultaneously with age-[Mg/Fe].  Alternatively, this difference would be a matter of using the correct IMF, improved yields, etc.  Although we have offset the solar iron value in the last section in order to fit the data well, the magnesium would require a much more significant offset in order to fit the distribution of points.

This demonstrates that the theoretical magnesium yields do not correspond well to the observed chemical evolution of the Galaxy. This is a further reason why we consider the [Si/Fe] fit most appropriate for this paper. 

Having said this, if we shift the magnesium value so that the theoretical ratios  between magnesium and silicon are the same as in the data (a shift of +0.14 dex) we found that the early time canonical silicon SFH is reproduces the [Mg/H]-age fairly well (Fig. \ref{Fig:bestfitsiandersmet_o_mg}, panels (a) and (d)). At later times a reduction of 50\% to the SFR improves the fit further (with a shift to +0.18 dex). The [Fe/H]-[Mg/H] is also recovered well. 

Conversely, we can interpret this more negatively.  The model under-produces magnesium relative to the iron to an extent of 0.14 dex or 40\% (Fig. \ref{Fig:bestfitsiandersmet_o_mg}).  (Again, a possible solution involves the extrapolation to high mass stars, i.e. M$_*>40$). The late time slope is flatter than in the observations of magnesium, and the evolution of [Mg/Fe] is not bracketed by a flat SFR. This suggests that, if it is to be fit at all, the SFH would be {\it increasing with time}. 

As previously discussed there is considerable variation between the available theoretical yields, each of which return different chemical tracks for a given element. With this uncertainty in mind, it is unsurprising that for any given set of yields the various species return varying SFHs. It is encouraging that silicon and oxygen and magnesium return SFHs which are similar.

\subsection{Yields and metallicity threshold for SNeIa}

\begin{figure*}
\centering
\begin{tabular}{cc}
 \includegraphics[width=2.2in, trim={0cm 0cm 0cm 0cm}, clip]{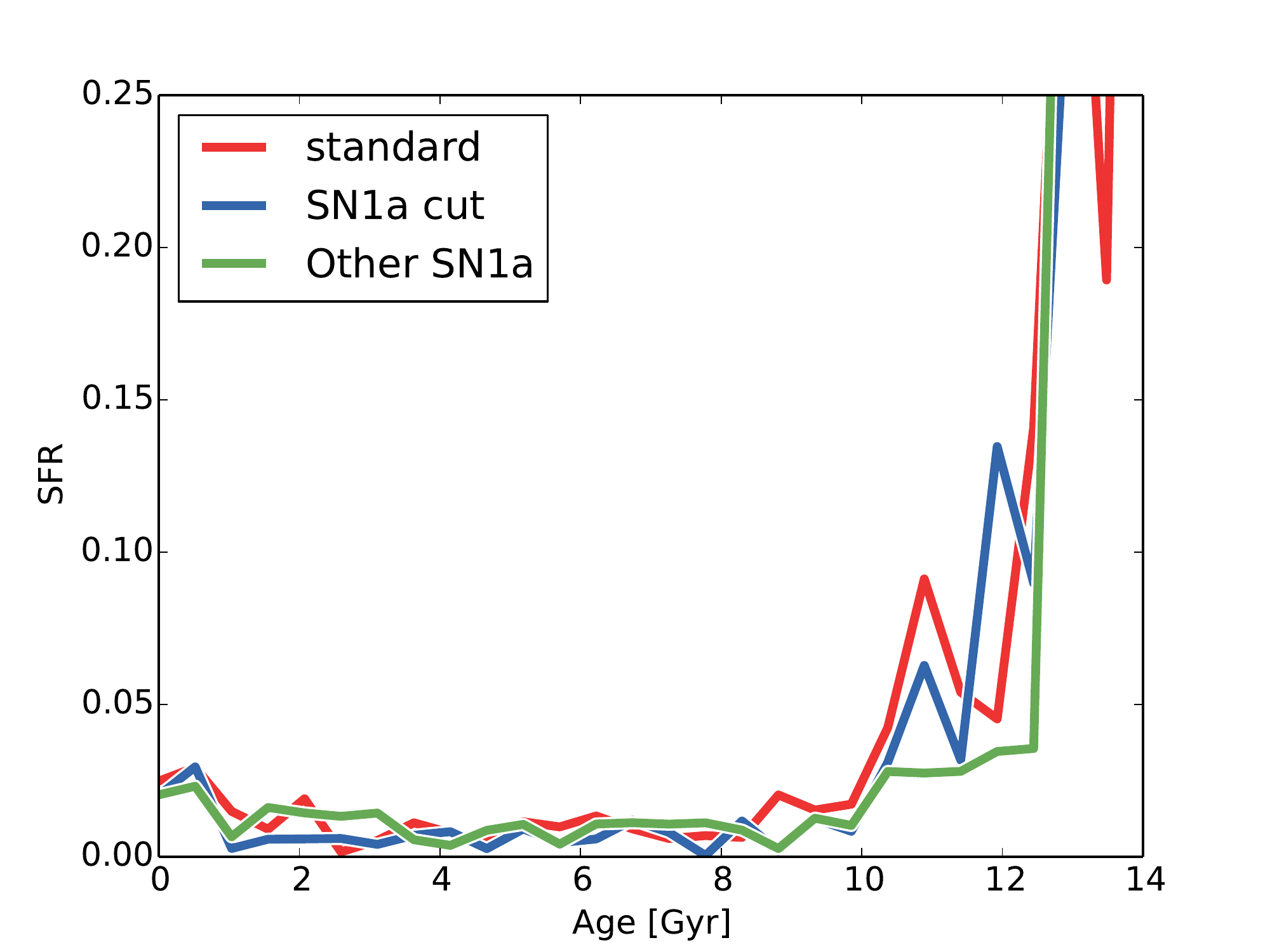} &
 \includegraphics[width=2.2in, trim={0cm 0cm 0cm 0cm}, clip]{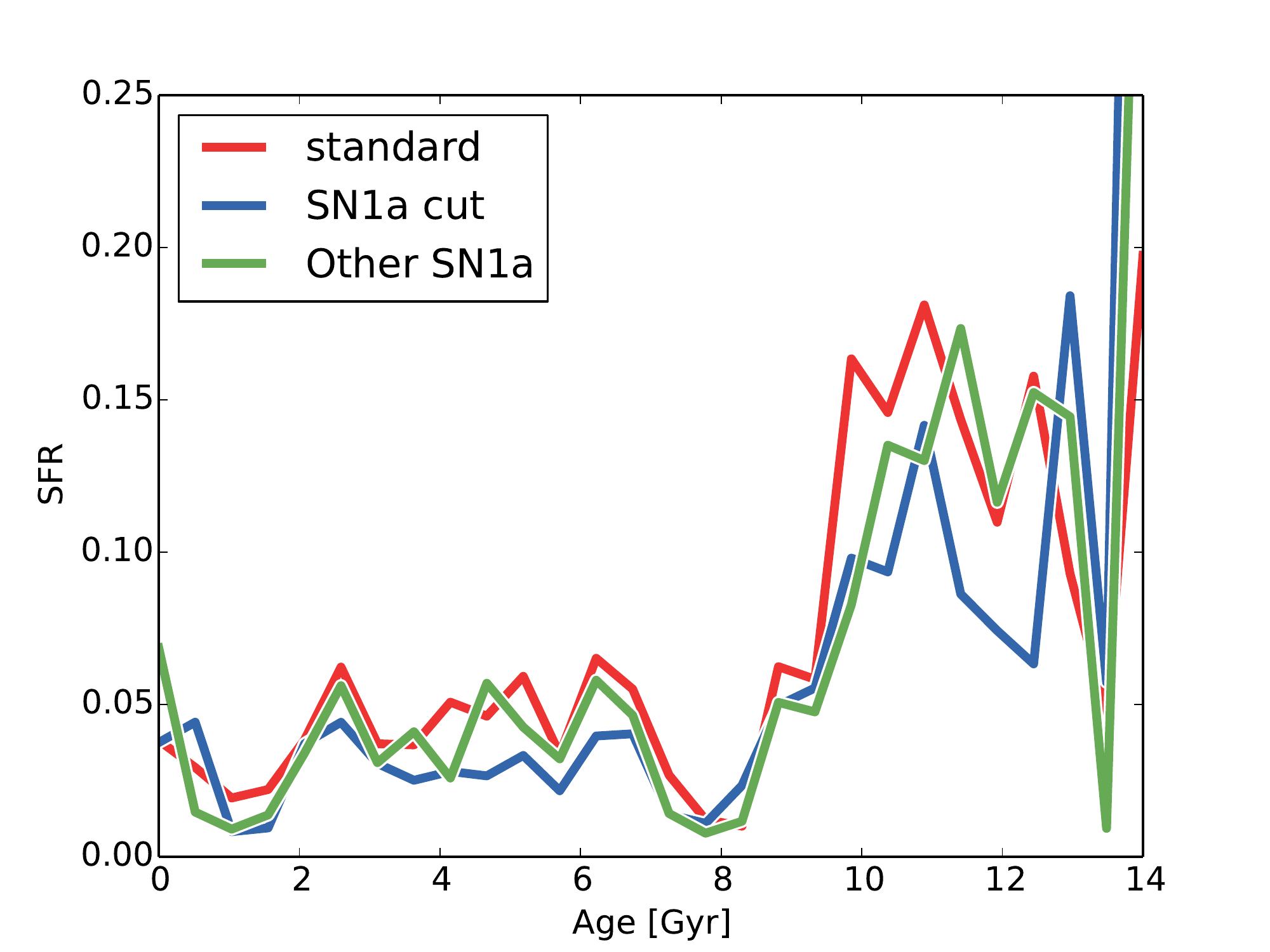} \\
 (a) & (b) \\
\includegraphics[width=2.2in, trim={0cm 0cm 0cm 0cm}, clip]{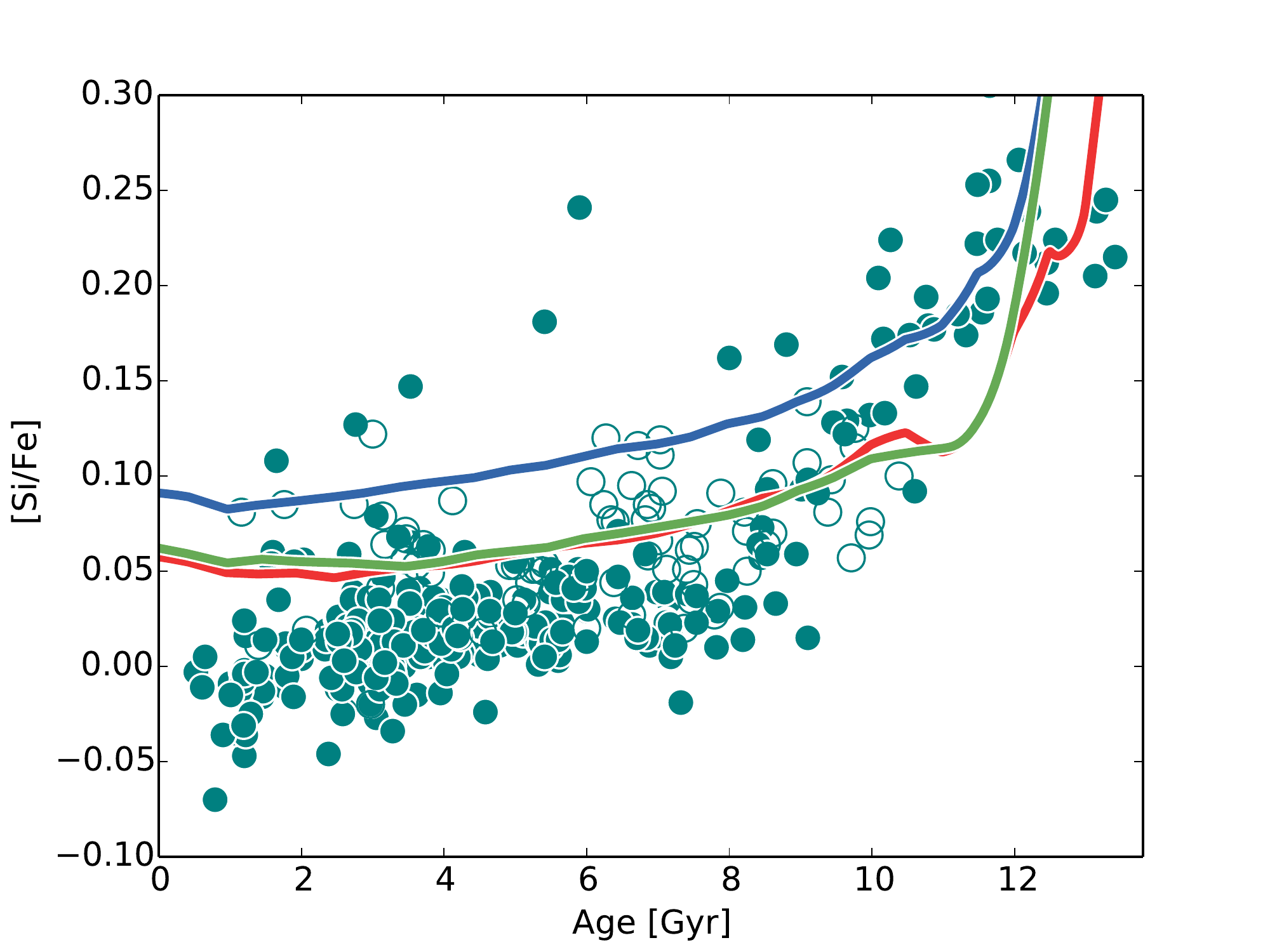}&  
\includegraphics[width=2.2in, trim={0cm 0cm 0cm 0cm}, clip]{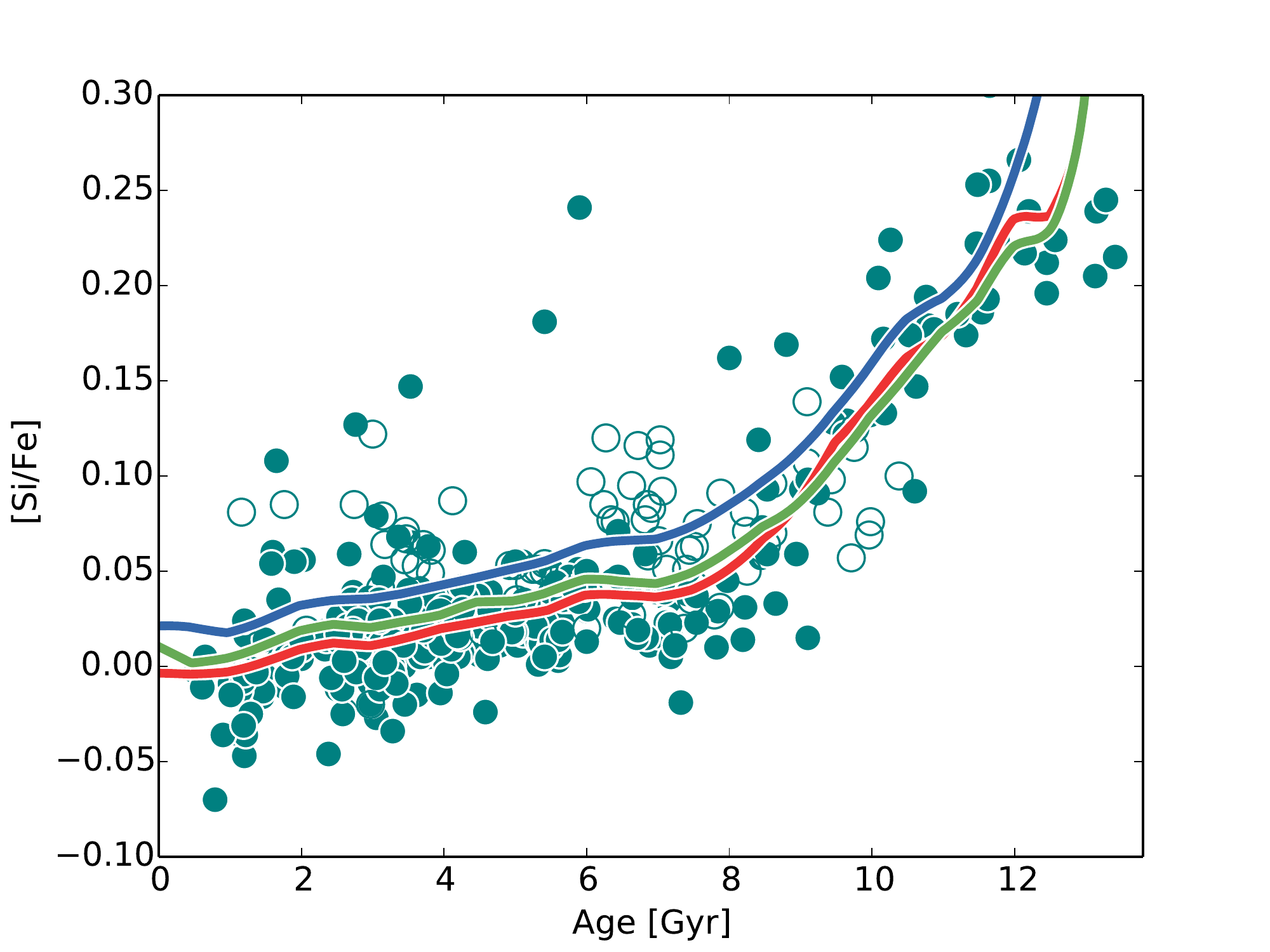} \\ 
 (c) & (d) \\  
 \includegraphics[width=2.2in, trim={0cm 0cm 0cm 0cm}, clip]{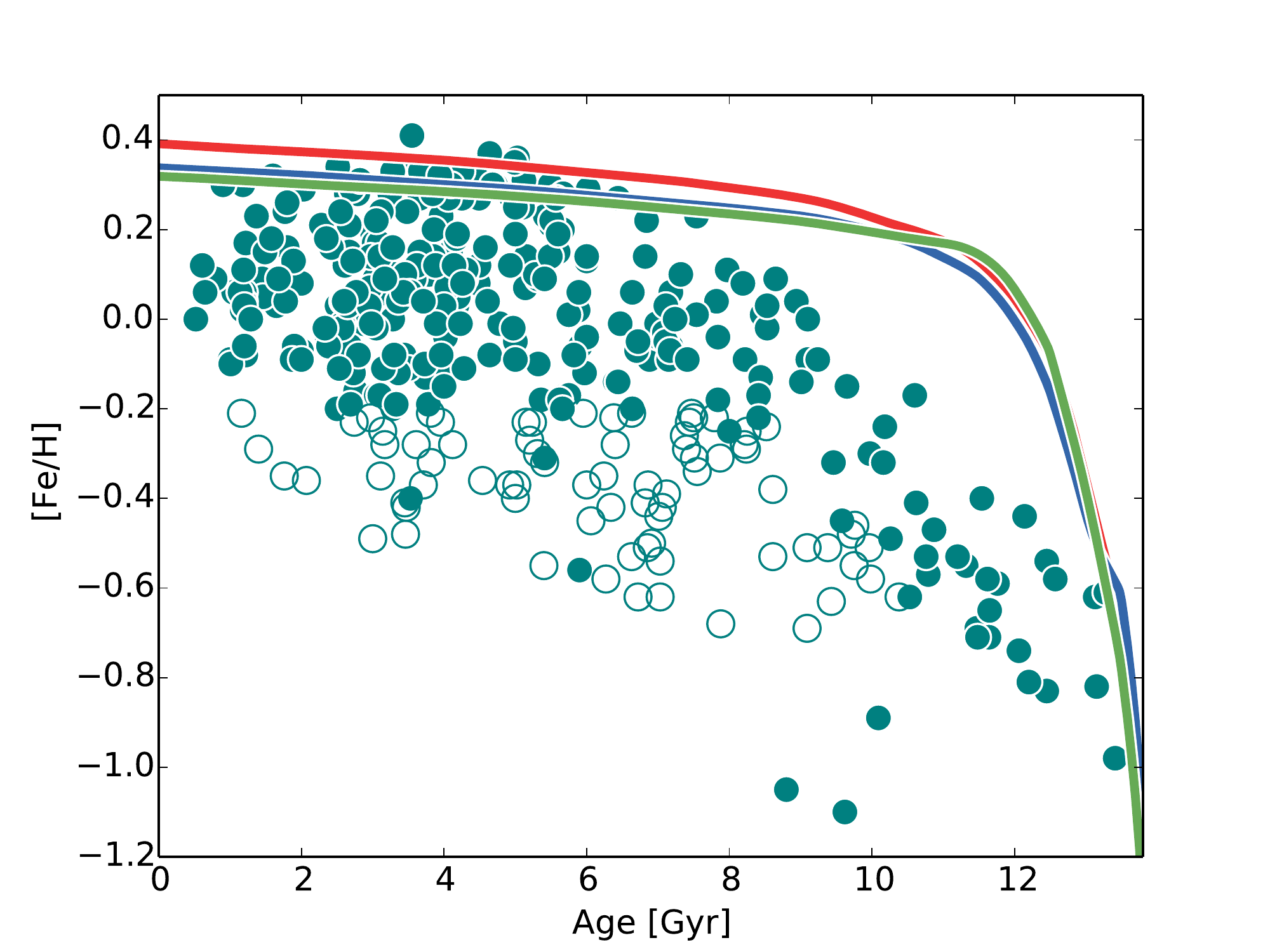}&
 \includegraphics[width=2.2in, trim={0cm 0cm 0cm 0cm}, clip]{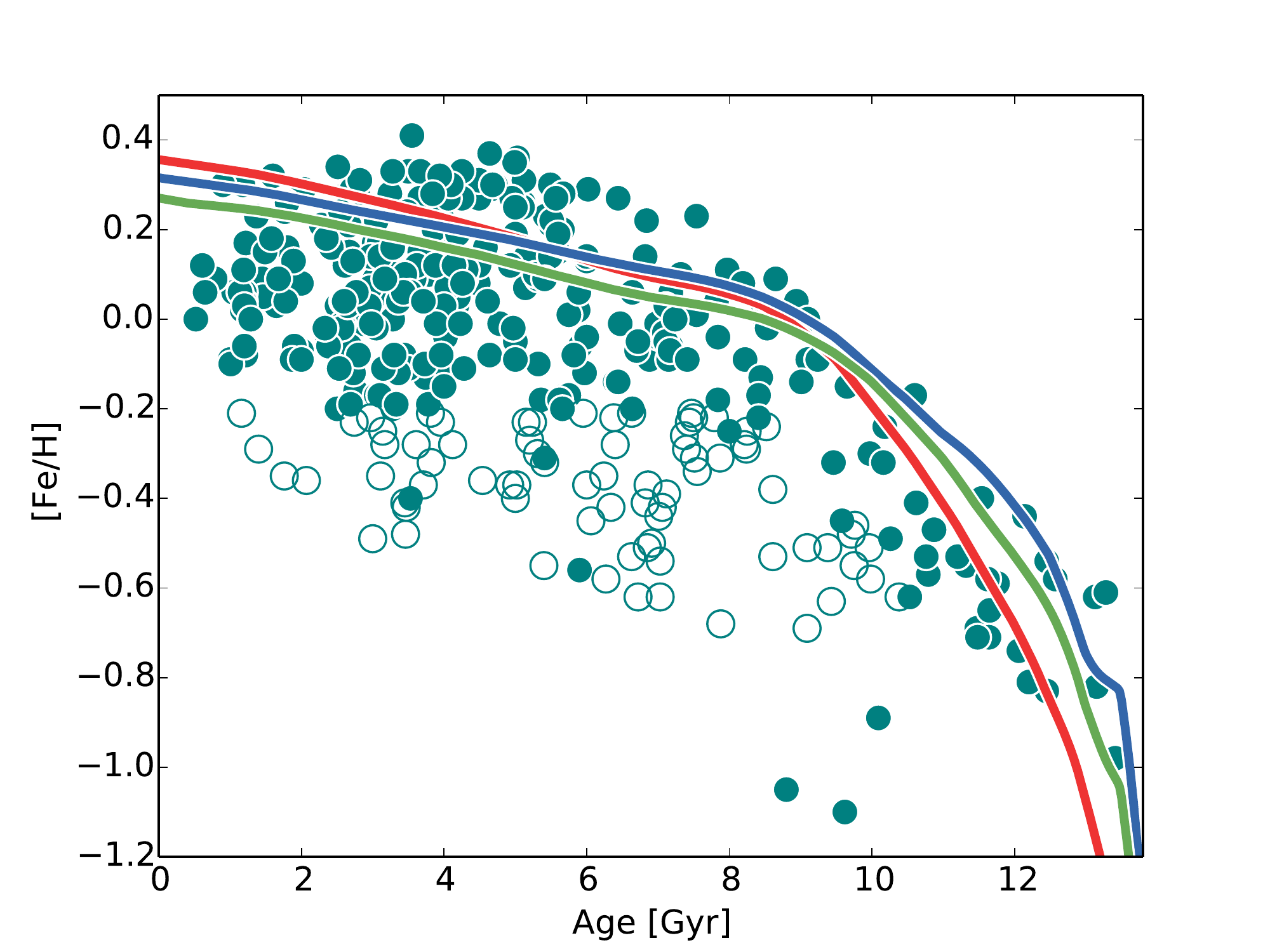}\\
 (e) & (f) \\
 \includegraphics[width=2.2in, trim={0cm 0cm 0cm 0cm}, clip]{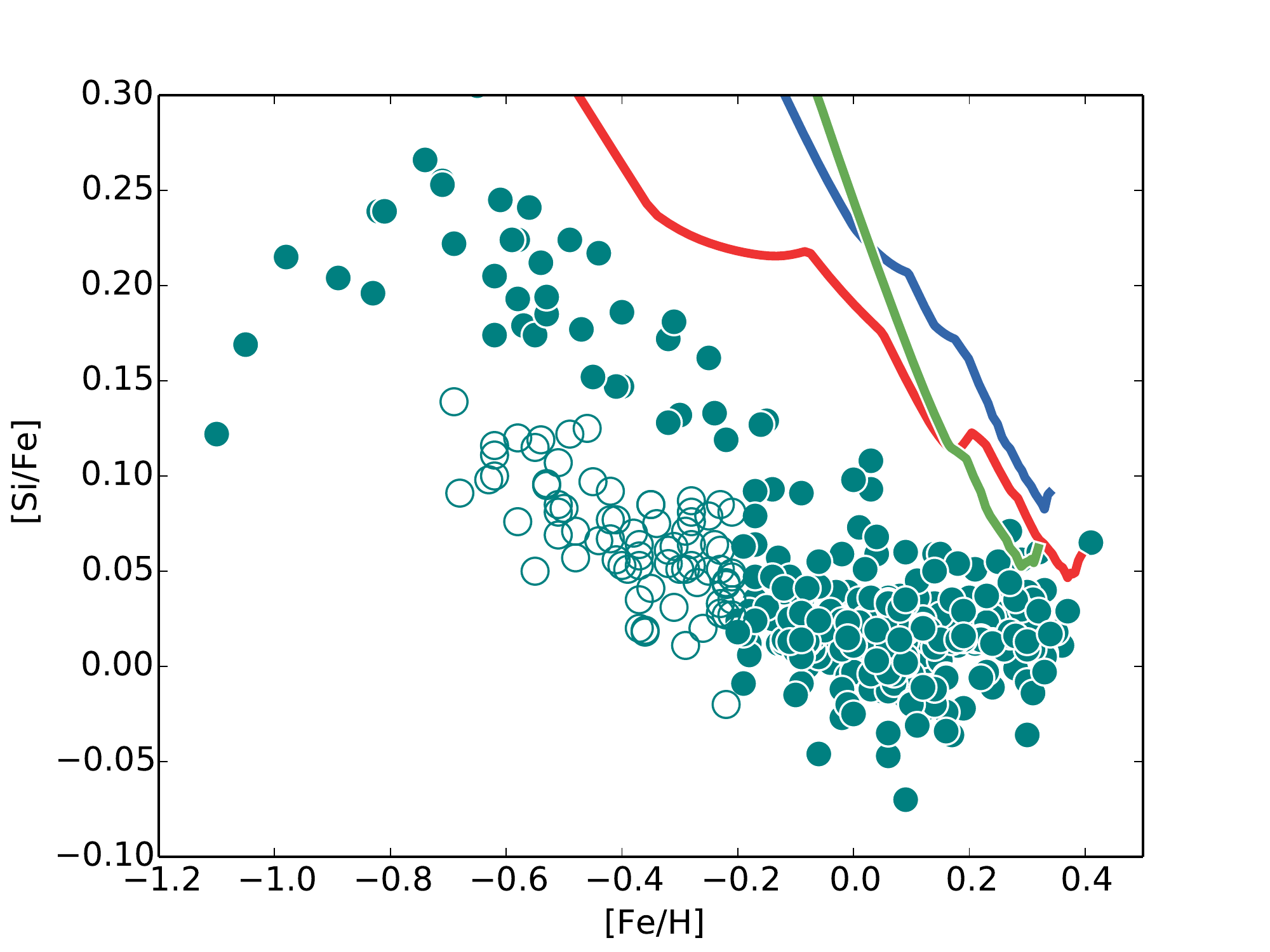} &
 \includegraphics[width=2.2in, trim={0cm 0cm 0cm 0cm}, clip]{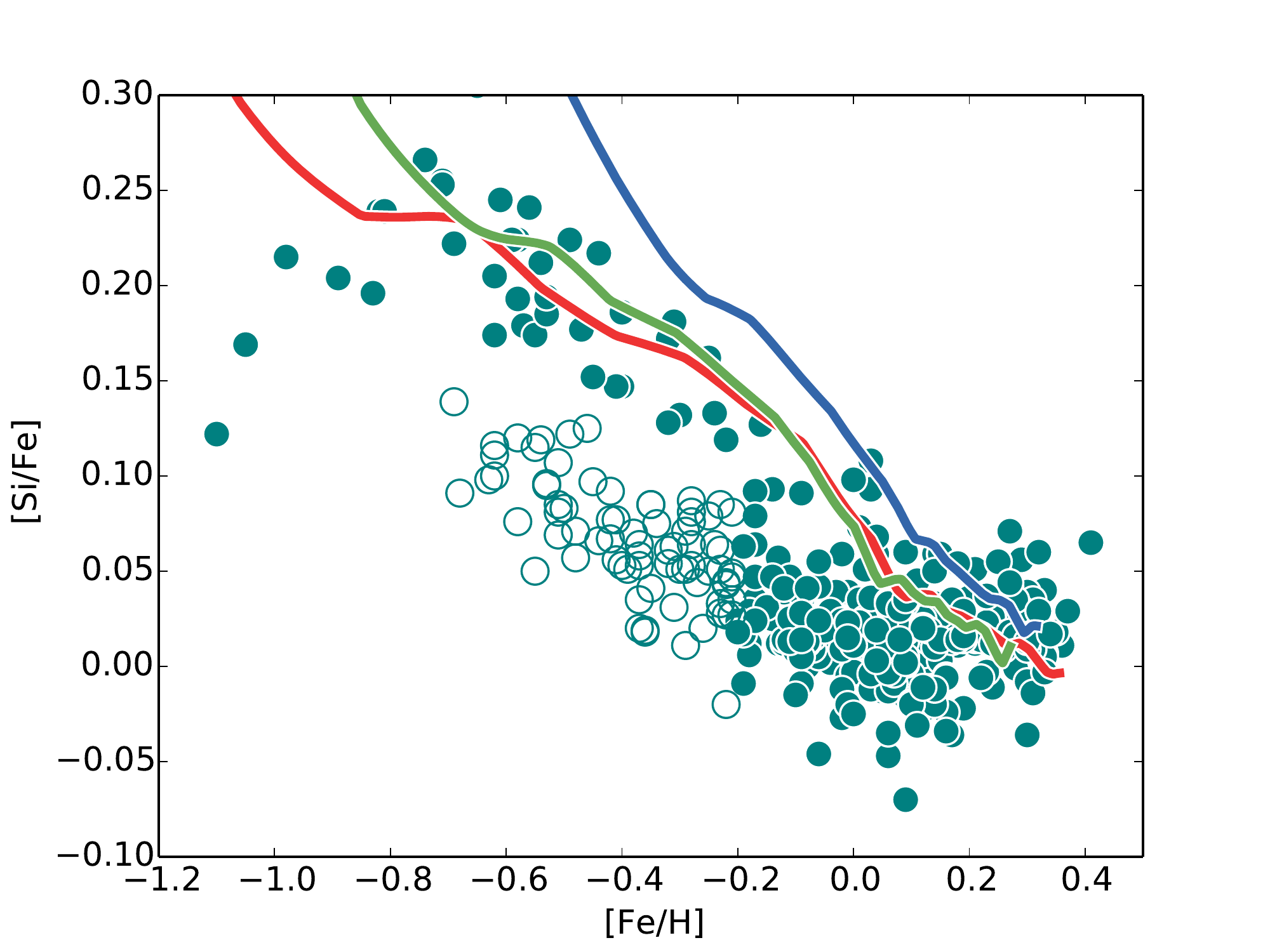}\\
 (g) & (h) \\ 
\end{tabular}
\caption{The comparison of three different models and yields. The left hand column shows the best fitted chemical tracks for \citet{WW1995} SNII yields, while the right hand column shows the tracks for \citet{Nomoto2006} SNII yields. The best fitted chemical track, and recovered SFH, for the standard GCE model summarized in Table \ref{Tab:summarymodel} is the `standard' track (red line).  The blue and green lines are the chemical tracks where a threshold  (Z = 0.1Z$_\odot$) for activating SNIa is included (`the SN1a cut' ), and  an alternative set of SNIa yields from \citet{Travaglio2004} are used (`Other SN1a') respectively. The recovered SFHs are given in panels (a) \& (b), and the standard distributions: age-[Si/Fe], age-[Fe/H], [Fe/H]-[Si/Fe] are given in rows two to four respectively. The points are from \citet{Haywood2013}. Open circles are outer thin disc stars, solid circles are thick disc and inner thin disc stars.}

\label{Fig:otheryieldsmodels}
\end{figure*}

The results presented in Sects.~\ref{Patterns} and \ref{SFR} are based on the adoption of \citet{ Nomoto2006}  yields for SNII,  \citet{Karakas2010}  yields for AGB stars and  \citet{Iwamoto1999}  yields for SNIa. Here, we  discuss the extent to which the recovered SFH of the Galaxy is sensitive to this choice, and how much the results would change if we adopted another set of yields for SNII and SNIa. This topic has been extensively studied by other authors  \citep[e.g.][]{Romano2010} but we include it here in order to be comprehensive.\\

We will also show the impact on the SFH of adopting a metallicity threshold of Z = 0.1Z$_\odot$, as in \citet{Matteucci2008, Kawata2003},  for the onset of SNIa.  This choice is justified in the work of \citet{Kobayashi1998}. The origin of this concept is that only gas with [Fe/H] $>-1$ can produce the optically thick winds required to stabilise gas accretion onto the white dwarf. However, more recent observations suggest that SNIa can occur very early in the history of the Universe, throwing doubt on the existence of any threshold \citep{Mannucci2005,Mannucci2006}, which is why we do not implement it in the main model used hitherto.

Figure \ref{Fig:otheryieldsmodels} shows the huge impact of using different SNII yields on the chemical evolution model. 
We choose \citet{WW1995} yields, as an alternative to \citet{Nomoto2006}, because they are the most common in the literature. We include, however, the modification to the \citet{WW1995} yields by \citet{Timmes1995} where the iron abundance is divided by two. This is required in order to have a high enough initial [Si/Fe] value at $t$= 14 Gyr (the oldest stars). 
The alternative SNIa yields are from \citet{Travaglio2004}. 
For all the choices adopted in this section, we run our best fitting procedure on the inner disc stars, in order to recover the corresponding SFH.

It is clear from Fig. \ref{Fig:otheryieldsmodels} that the recovered SFH using \citet{WW1995} yields produces chemical tracks which are very poor fits to the data. We find that:

\begin{enumerate}
\item The end point in the age-[Si/Fe] is  too high in all cases, with the worst fit being the model where a metallicity threshold for the onset of SNIa is included (panel (c)). 
\item The metallicity-[Si/Fe] track shows that there is too much iron for the given [Si/Fe] ratio as well (panel g).
\item The age-[Fe/H] track (panel (e)) shows that most of the iron forms too early. 
\item The SFH for these yields all burst very early (panel a) and has an almost exponential form.
\end{enumerate}
We would expect to evolve in age-[Si/Fe] most quickly for the sharp early burst (see Sect. \ref{exp}). This implies that there exists no possible SFH which can fit the data using these yields (in this model). 

For the \citet{Nomoto2006} yields the model with the SNIa threshold is again the poorest match of the three models employing these yields (right column, Fig.~\ref{Fig:otheryieldsmodels}). Using the different SNIa yields, however, produces tracks indistinguishable in either the chemical tracks or the recovered SFH. Thus, we do not require a SNIa metallicity threshold to fit the data, and find that such a threshold actually hinders fitting. 

\section{Discussion}\label{Discussion}
\subsection{Summary of results}

We found that fitting our model to the chemical trends as a function of age 
is a very effective way to recover the SFH of the Galaxy. We emphasize that 
as long as the stars originate from different parts of the disc, but 
nevertheless show homogeneous chemical properties, the recovered SFH is 
representative of the whole population, without the need for specific 
corrections due to volume effects. 
While this is  less correct for the youngest part of the disc, because the stars
of this population visible at the solar vicinity explore a restricted range of 
galactic radii, it is certainly correct for the thick disc population.

This method therefore, provides a unique and novel way to investigate the 
history of star formation in the first Gyrs of the Galaxy formation, 
for which there is very little hope of significant progress from counting stars
as a function of age, as least before Gaia parallaxes become available.

Fitting [Fe/H]-[$\alpha$/Fe] distributions is a much less 
effective way to distinguish between models because the chemical tracks
are degenerate with time. Without age information it is difficult to
constrain the chemical evolution and the SFH.
Fig. \ref{Fig:SFRexponential} illustrates that the thick disc  [Fe/H]-[$\alpha$/Fe] data is compatible 
with SFHs having a broad range of timescales (2-8 Gyr), only a very sharp burst shows a
significantly deviating  track. Age information is vital in
order to distinguish the SFR from the chemistry alone.
Before summarising our results, a number of limitations and cautionary remarks are worth noting, 
mainly due to our incomplete knowledge and inconsistency of the stellar yields:

\begin{itemize}

\item We require \citet{Nomoto2006} SNII yields in order to fit the Galaxy
chemical evolution. The \citet{WW1995}+\citet{Timmes1995} yields give a
too high final [Si/Fe] ratio.

\item We cannot converge the Galaxy chemical evolution using the solar
abundances used in \citet{Adibekyan2012}  (the iron used is from \citet{Gonzalez2000} and has a value of 7.47) and recover a track matching all three projections of the data, but must shift the iron abundance
by 0.04 dex (or 10\%) in order to match the age-[Si/Fe] and
metallicity-[Si/Fe] plots simultaneously. In this we follow 
\citet{Timmes1995} who divided the \citet{WW1995} by two in order to fit
the data (though to a lesser degree).
The recovered SFH changes strikingly if the solar abundance shifts,
changing the relative contributions of the thick and thin disc.
Higher solar iron increases the mass of the thick
disc , while a low solar iron emphasises the importance of late time
star formation. However, only solar iron values between 7.49 and 7.53
allow us to fit the [Fe/H]-[Si/Fe] data for the inner disc.

\item We cannot fit the upper  track using magnesium and silicon at the same time.
This is because the abundance ratios between these elements are much lower
in the observations than in the model.

\item Allowing the normalisation to change (the total fraction of gas used to make stars)
introduces some degeneracy in the fitting procedure. However, all normalisations 
$\ge$0.5 give similar results within the errors, but we note that lower 
normalisations produce larger thick disc components.
Normalisations $\le$0.5 are too low to fit the age-[Si/Fe] distribution. 
\end{itemize}

Having mention these caveats, our results are:

\begin{itemize}

\item Our best fitted SFH for the Milky Way using silicon, shows that the Galaxy 
passed through a phase of intense stars formation between $\sim$ 9 and 13 Gyr
ago. 

\item This phase was followed by a rapid drop (a decrease of 80\% in 1 Gyr). For stars of age 9.75 Gyr the SFR has a value of 0.14 Gyr$^{-1}$ 
 but at 8.75 Gyr it has dropped to 0.06 Gyr$^{-1}$, and by 8.25 it has fallen to 0.015 Gyr$^{-1}$ at
the bottom of the dip (where the integral of the SFH is 1).
The SFR apparently stalls for about 1 Gyr. The measured dip is
highly robust
to all our tests (where A $>$ 0.8).

\item The SFR recovers after 1 Gyr, at around 7 Gyr, and continues at a lower intensity 
up to the present. 

\item The chemical evolution of the outer disc is parallel to that of the thin disc, but starting
from different initial conditions: stars of the outer disc start to form $\sim$ 2 Gyr before the inner disc 
at a higher alpha abundance and lower metallicities. We require
 a dilution of metals by 3$\times$ as much hydrogen in order to fit the outer disc distribution, which
otherwise shows a flat SFH similar to that of the inner disc.

\item  Standard IMF such as \citep{Kroupa2001}'s allow us to fit all 
projections of the chemical tracks at the same time. 
The \citet{Salpeter1955} IMF gives similar results and
the SFH recovered by these two IMFs is very
similar within the error.
\item The changes to the model discussed in Section \ref{Robust} change the recovered SFH only slightly. The same form is nearly always required, with an early period of vigorous star formation, followed by a dip, and then a quasi-linear era. However, changes to the canonical model presented in Section \ref{SFR}, the yields or IMF etc, usually result in our being unable to fit the chemical tracks to the data in all three distributions (age-[$\alpha$/H], age-[Fe/H] or [Fe/H]-[$\alpha$/Fe]). In general we either succeed or fail without much of a grey area between.

\end{itemize}

\subsection{Interpretation}

\subsubsection{Life with no (G-dwarf) problem}

The results presented above introduce a significant change of perspective regarding chemical evolution modelling 
of the Milky Way disc. As mentioned in the introduction, most (in fact, to our knowledge, all recent) chemical evolution models 
are based on the need to suppress the formation of too many stars at low metallicity
to be in accord with the solar neighbourhood MDF. 
Several studies in the early 1980s \citep{Chiosi1980,Lacey1985,Matteucci1986,Matteucci1989}, also motivated by the idea of an inside-out 
building of discs \citep{Larson1976}, suggested that long term infall was an elegant solution to this problem. 
 The results presented in the previous section have shown that, in apparent contradiction to the solar 
vicinity, the Galaxy has formed a huge number of dwarfs at intermediate metallicity. While we postpone 
a detailed discussion of the MDF, and the G-dwarf problem, to another paper (Haywood et al., in preparation), we briefly 
comment on this apparent contradiction.

The thick disc scale length measurement by \citet{Bovy2012c} shows that, at 1.8~kpc, this population 
has a much shorter scale length than previously thought. Assuming 250 pc, 3.0 kpc and 90\% for the thin disc scale height, scale length 
and local density and 1000 pc, 10\% and 1.8 kpc for the thick disc, the thin and thick disc would represent each 50\% of
the stellar mass. 
This is to be compared with fractions resulting from chemical models assuming infall, wherein the thick disc
usually amounts to, at most, 15 to 25\%.
Thus, the  lack of metal-poor stars is, in this context, a purely local bias\footnote{The vertical scale height of the thick disc is considerably longer than the scale height of the thin disc, 
which further enhances the idea that purely local observations do not sample the average density distribution of low metallicity stars well. This is discussed in detail in \citet{Haywood2001}.}. The low metallicity stars do exist, 
but they are mostly inside the solar radius. When the Galaxy is taken as a whole there is no under-abundance of low metallicity stars. 
It implies that a near solar metallicity was attained relatively early in the chemical evolution of the disc, not 
because of the low dilution permitted by long term infall, but because the Galaxy formed huge quantities 
of stars. This permitted the metallicity of the ISM to rise from [Fe/H]$<$-1.0 dex to about -0.2 dex. 
Of these stars, a large majority inhabit the inner Galaxy.

The case of the outer disc may be different, because it is possible that infall may have contributed on a longer time scale.
Moreover, as already said, the initial metallicity and abundance ratios may have been set by the combination of both outflows
from the inner disc, and accretion of gas. So it is possible that infall models would represent the outer disc evolution, 
although a clear assessment must await new data.

\subsubsection{The stellar mass evolution of the Galaxy}

The best fit SFH obtained in section \ref{sec:inner} (Fig. \ref{Fig:bestfitsiandersmet}) shows that 52\% of the stellar mass formed in the Milky Way disc(s)
between 13 and 9 Gyr. This percentage confirms the estimate, based on the density parameters given in the previous 
section, that the thick disc represents half the stellar mass of the disc. 
The only other component that may have contributed significantly to the general mass growth of the Milky Way at early epochs is 
the bulge.
Most recent results, however, suggest that the contribution from this population is minor. 
\citet{Shen2010} proposed, from adjusting a model to the bulge kinematics data, that the Milky Way 
was possibly a pure disc galaxy, and, that a classical bulge, if present, could  not represent 
more than 10\% of the disc stellar mass. This estimate has been confirmed by \citet{Kunder2012}. From the Argos
survey, \citet{Ness2012} argue that the components they detected in the decomposition of 
the MDF of the bulge are attributable to the thick and thin discs. \citet{DiMatteo2013} 
suggested that component B could be due to the young thick disc, and confirmed that if a spheroidal
component exist in the bulge, it must be small ($<$10\%).
These conclusions make the Milky Way fit its local environment: \citet{Kormendy2010} and \citet{Fisher2012}
showed that within 10 and 12 Mpc, the dominant type among giant discs are either pure discs or disc with
pseudo bulges. 
Hence, we deduce from these studies, and from our results, that the main component to have formed in 
the Galaxy at these epochs is the galactic thick disc. 
This is in accord with the result by \citet{vanDokkum2013}, which shows that 
progenitors of the Milky Way type galaxies did not grow a disc starting from a naked bulge.

Moreover, \citet{vanDokkum2013} found that progenitors of the MW have assembled $\sim$ 50\% of their
mass at z$\sim$ 1.5 (or 9.5 Gyr), well in accord with our findings. 
Our SFH follows closely the description given by \citet{vanDokkum2013} about MW progenitors: ``the implied star formation
rate is approximately constant at 10-15 M$_{\odot}$ yr$^{-1}$ from z$\sim$ 2.5 to z $\sim$ 1 then decreases rapidly
to $\leq$  2 M$_{\odot}$ yr$^{-1}$  at z=0.''

\subsection{Comparison with other models}
\label{infall}

We now discuss our findings in the more general context of chemical evolution models. 
As already mentioned, analytical chemical evolution models of the Milky Way disc 
are based on two basic ingredients, the infall of gas on long time scales (5 Gyr 
or larger),  and a Schmidt-Kennicutt law. 
The fit to the solar vicinity MDF, a constraint that we discard,  requires that 
the production of stars of intermediate metallicities is limited, something that models
achieve through infall, i.e the metallicity rises in the system by limiting the dilution. 
We, however, increase the metals in the ISM by allowing a large number of stars of 
intermediate metallicity to be created. Hence, the two radically different views: the one
developed here, and the one presented in previous models, makes direct comparison with previous models somewhat 
hazardous, particularly since in many studies, the thick 
disc is not considered as a component of the modelling. 
Although the possible importance of the Galactic thick disk was recognized as early as 1986 \citep{Gilmore1986},
this population is rarely taken into account explicitly in models.
For instance, there is no explicit thick disc phase in the models of \citet{Fenner2003} or \citet{Naab2006}.
For infall models the estimated stellar mass generated before their model reaches
a metallicity of -0.2 dex is usually less than 10\% of the total stellar mass. 

In \citet{Chiappini1997}, a halo-thick disc phase is explicitly considered and has a formation
time scale of 1 Gyr, but its characteristics ([Fe/H]$<$-1 dex) are difficult to compare with the 
known properties of the observed thick disc.  This follows the low mass model of the stellar halo of \citet{Carney1990}.
 \citet{Fenner2003} use a
similar two-phase infall model, based on \citet{Chiappini1997} but
consider the first infall phase to be the halo, the thick disc then has
thin disc-like properties and the two are not distinguished.
In \citet{Micali2013}, where a thick disc phase is also explicitly considered, the thick 
disc is of the order of 25\%. 
In our case, as we have no infall, we do not define the
history in terms of gas accretion, but in terms of the properties of the
SFR. In the case of the infall based models, the thick disc phase is
relatively short, while the data of \citet{Haywood2013} implies it lasts
approximately 4-6 Gyr, in keeping with our result. \citet{Chiappini1997,Chiappini2001} therefore, favor a rapidly formed thick disc, while  \citet{Fenner2003} interpret their thick disc as a thickened thin disc \citep{Wyse2001}.
Both these scenarios are at odds with the \citet{Haywood2013}
interpretation, which our model favors.
It is important to emphasize  that the length of the different phases etc.
were not implemented in the model directly, but are the result of using a
closed box model and fitting the \citet{Haywood2013} data.
The dip at ages of 8 Gyr (Section \ref{SFR}) is superficially reminiscent of the
dip at ages of 11 Gyr in the \citet{Chiappini1997} dual infall model, but
has a different effect. In our model the dip is required, coupled with the
large drop in SFR between the inner thin and thick disc sequences, to
produce a sharp transition at the knee of the age-[Si/Fe] distribution. In
\citet{Chiappini1997} the dip serves to allow rapid dilution of [Fe/H],
and allows the production of the lower sequence in the
[Fe/H]-[$\alpha$/Fe] distribution once the thick disc and halo have been
formed. Thus, in effect, the thick disc sequence is constructed separately
in the traditional GCE code, with its own infall rate etc, and then the
lower sequence is a continuous evolution occurring subsequently. This does
not entirely match the data of \citet{Haywood2013} as the earliest stars
in the ``outer disc'' sequence are formed concurrently with
later stars of the thick disc. Also, our thick disc stars continue to be
formed until a lot later than in \citet{Chiappini1997} and its
descendants.

In order to accurately capture the physics of the gas in the Galaxy a number of different phases of gas should be considered. The most well known of which are: the low temperature star forming gas (HI) in the disc which is eligible to form stars; and the hot gas reservoir which lies in the halo. A third form of gas often invoked to explain the `missing baryons' is the warm enriched circumgalactic medium, which comprises gas ejected from the disc by feedback \citep{Sommer2006}. This accounts for much of the missing mass in a galaxy, i.e. that gas which should be present according to cosmology, and which is not seen in the baryonic mass observed. In traditional closed box models this warm component is not considered part of the `box', and is considered an outflow \citep{Hartwick1976,MatteucciC1983,Prantzos2003}. We do not distinguish between different forms of gas, and refer to all different forms of gas in the same manner. As we do not need the HI surface density to control star formation, the gas reservoir is simply both somewhere to dilute the metals produced by the stars, and something which at {\it some point} forms stars. In effect, our `closed box' is referred to as such because no gas or metals escape from the system as a whole, but gas and metals are free to change from cold gas to warm gas to hot gas etc. This is no worse an approximation than any, as the way different modes of gas interact is still largely unknown. Thus, when we say that no gas escapes, we mean, in effect, that it does not escape permanently.

A dramatic difference is the sensitivity of the chemical evolution to
features in the SFH. In traditional, infall based, models a change  to the system can
have a very strong effect on the chemical history \citep{Colavitti2008}. The lack of a large reservoir of gas means that metals are
not well diluted in the short term. That gas is then cleared out of the
system by star formation and pristine gas falls in. This has two effects
on the chemical evolution. It means that the initial increase of
metallicity is more rapid (making infall a solution to the G-Dwarf problem
as previously discussed), it also has the effect that a burst or hiatus in
star formation quickly changes the direction of the chemical track
(reducing the ``chemical inertia''). In our case, with a large
reservoir of gas, the chemical tracks are slow to change direction because
new metals are diluted into an ISM that is massive, and new metals are
added to a large quantity of other metals released from previously formed
stars. This can be seen by the effect of the dip in the SFR on the
[Si/Fe] evolution. Our pause of star formation lowers the [Si/Fe] value by
a small amount (Fig. \ref{Fig:sfhfeatures}). A similar pause in a traditional GCE code has a
much more significant effect \citep{Romano2010,Colavitti2008}.

In traditional GCE models there is apparently little consensus on
the ``best'' parameters. For example, \citet{Naab2006} uses a Salpeter IMF, the recent paper by \citet{Spitoni2014} or \citet{Micali2013} uses a \citet{Scalo1986} IMF while \citet{Fenner2003} use a \citet{Kroupa1993} IMF. The choice of yields remains
uncertain \citep[][etc.]{Romano2010, Kobayashi2011}.
Although we are sensitive to our chosen yields, and chosen elements, this
is an issue intrinsic to the theoretical yields. Other models similarly
provide different results for different elements \citep[e.g.][]{Romano2010}
and yields. There is no clear ``winner'' although some
elements and yields fit more poorly than others. For example, \citet{Chiappini1997} use
\citet{WW1995} yields while \citet{Fenner2003} use \citet{Limongi2000,Limongi2002}
yields etc.  There is also no consensus on the exact form of the gas
density to star formation rate (\citet{Chiosi1980}, \citet{Fenner2003}, see
\citet{Matteucci2008} for a review) or on the exact form infall should take
\citep{Colavitti2008}. Another issue with GCE codes is in
the choice of parameters for which there seems to be little justification, 
other than just permitting more flexibility to fit the data. For example, \citet{Chiappini1997}
change the star formation efficiencies between the halo/thick disc phase and
the thin disc phase, while three different star formation efficiency are 
adopted in \citet{Micali2013} for the halo, thick disc and thin disc phase. 

Many GCE codes attempt to replicate the [Fe/H]-[Si/Fe] distribution only \citep[e.g.][]{Romano2010}, without any age information. While this is not a difference in the models as such, it is important when classifying the model as one which works well or not. As we showed in Fig. \ref{Fig:SFRfavouries}, without age information the fitting of SFHs is degenerate. Further, the age-metallicity relation shows considerable scatter, with an almost 1 dex spread in metallicity at a given age, particularly at more recent times. It is the age-metallicity relation which is used when age information is used at all in traditional models \citep[e.g.][]{Chiappini1997}. Thus, we would argue that age information and alpha abundances are essential if a GCE model is to be considered a good match to the data. 

Our model is much simpler and can  fit the data of
\citet{Haywood2013} (for silicon) very well. We have interpreted the
model in the context laid out in \citet{Haywood2013}, which is somewhat
different to that in previous models, and what's more, the parameters of the 
model were not tuned, except in so far as we required that the chemical
tracks actually fit the data  in order to derive a star formation history.

\section{Conclusions}

We have presented a new method to derive the SFH from stellar abundances, with a full assessment of the
possible uncertainties. 
We derive a SFH for the inner disc (R$<$7-8kpc),
which shows two distinct phases that correspond to the formation of the thick and the thin discs. 
The thick disc formation lasts 4-5 Gyr during which the SFR reaches 10-15 M$_{\odot}$.yr$^{-1}$. 
After the thick disc phase, star formation stalls for about 1 Gyr, and then resumes for 
the thin disc phase at a level of  2 M$_{\odot}$.yr$^{-1}$ for the remaining 7 Gyr. 
The SFH of the outer disc, as derived from metal-poor thin disc stars seen at the solar vicinity 
shows a similarly flat history, but preceded the thin disc by 2 to 3 Gyr.

The derived SFH implies that the thick disc represents about half the stellar mass formed in the Milky Way, 
something that has been largely unrecognized. 
As discussed in \citet{Snaith2014}, the mass growth of the Milky Way seems to be well in accord with 
the evolution of galaxies of its class. 

Overall, the present study proposes a change of paradigm in the way we see the first Gyrs of the evolution 
of the Galactic disc.
Up to now, chemical evolution models have described the early building of the Milky Way disc as a 
slow process. They have been built on the assumption that the thick disc is a minor (in terms of
stellar mass) component, in order to fit with the 'G-dwarf' constraint, but underestimating the role
of this population in the chemical evolution of our Galaxy.
The closed-box model used here is clearly only a zero order approximation description of the chemical evolution
of the Milky Way.  It does not pretend to capture the full complexity of the processes that regulate discs at high redshift, 
but it is thought to be more suitable than models which use infall on long characteristic time scales to represent the first Gyrs of the chemical 
evolution of the Milky Way.

\begin{acknowledgements}
The authors acknowledge support from the French 
Agence Nationale de la Recherche (ANR) under contract ANR-10-BLAN-0508 (Galhis project).
Support for ONS was partially provided by NASA through the Hubble
Space Telescope Archival Research grant HST-AR-12837.01-A from the Space
Telescope Science Institute, which is operated by the Association of
Universities for Research in Astronomy, Incorporated, under NASA contract
NAS5-26555. ONS thanks J.~Bailin for interesting discussions which improved the paper. The authors also thank F.~Arenou, F.~Royer and C.Babusiaux for their comments. We also thank the anonymous referee for their comments, which improved the quality of the paper. 
\end{acknowledgements}

\bibliographystyle{aa}
\bibliography{biblio}

\begin{thebibliography}{110}
\expandafter\ifx\csname natexlab\endcsname\relax\def\natexlab#1{#1}\fi

\bibitem[{{Abadi} {et~al.}(2003){Abadi}, {Navarro}, {Steinmetz}, \&
  {Eke}}]{Abadi2003}
{Abadi}, M.~G., {Navarro}, J.~F., {Steinmetz}, M., \& {Eke}, V.~R. 2003, \apj,
  597, 21

\bibitem[{{Adibekyan} {et~al.}(2012){Adibekyan}, {Sousa}, {Santos}, {Delgado
  Mena}, {Gonz{\'a}lez Hern{\'a}ndez}, {Israelian}, {Mayor}, \&
  {Khachatryan}}]{Adibekyan2012}
{Adibekyan}, V.~Z., {Sousa}, S.~G., {Santos}, N.~C., {et~al.} 2012, \aap, 545,
  A32

\bibitem[{{Allen} \& {Santillan}(1991)}]{Allen1991}
{Allen}, C. \& {Santillan}, A. 1991, \rmxaa, 22, 255

\bibitem[{{Anders} \& {Grevesse}(1989)}]{Anders1989}
{Anders}, E. \& {Grevesse}, N. 1989, \gca, 53, 197

\bibitem[{{Anders} {et~al.}(2014){Anders}, {Chiappini}, {Santiago},
  {Rocha-Pinto}, {Girardi}, {da Costa}, {Maia}, {Steinmetz}, {Minchev},
  {Schultheis}, {Boeche}, {Miglio}, {Montalb{\'a}n}, \&
  {Schneider}}]{Anders2014}
{Anders}, F., {Chiappini}, C., {Santiago}, B.~X., {et~al.} 2014, \aap, 564,
  A115

\bibitem[{{Asplund} {et~al.}(2009){Asplund}, {Grevesse}, {Sauval}, \&
  {Scott}}]{Asplund2009}
{Asplund}, M., {Grevesse}, N., {Sauval}, A.~J., \& {Scott}, P. 2009, \araa, 47,
  481

\bibitem[{{Aumer} \& {White}(2013)}]{Aumer2013}
{Aumer}, M. \& {White}, S.~D.~M. 2013, \mnras, 428, 1055

\bibitem[{{Baldry} \& {Glazebrook}(2003)}]{Baldry2003}
{Baldry}, I.~K. \& {Glazebrook}, K. 2003, \apj, 593, 258

\bibitem[{{Bensby} {et~al.}(2011){Bensby}, {Alves-Brito}, {Oey}, {Yong}, \&
  {Mel{\'e}ndez}}]{Bensby2011}
{Bensby}, T., {Alves-Brito}, A., {Oey}, M.~S., {Yong}, D., \& {Mel{\'e}ndez},
  J. 2011, \apjl, 735, L46

\bibitem[{{Binney} \& {Merrifield}(1998)}]{Binney1998}
{Binney}, J. \& {Merrifield}, M. 1998, {Galactic Astronomy} (Princeton
  University Press)

\bibitem[{{Bovy} {et~al.}(2012){Bovy}, {Rix}, {Liu}, {Hogg}, {Beers}, \&
  {Lee}}]{Bovy2012c}
{Bovy}, J., {Rix}, H.-W., {Liu}, C., {et~al.} 2012, \apj, 753, 148

\bibitem[{{Brook} {et~al.}(2014){Brook}, {Stinson}, {Gibson}, {Shen},
  {Macci{\`o}}, {Obreja}, {Wadsley}, \& {Quinn}}]{Brook2013}
{Brook}, C.~B., {Stinson}, G., {Gibson}, B.~K., {et~al.} 2014, \mnras, 443,
  3809

\bibitem[{{Brook} {et~al.}(2012){Brook}, {Stinson}, {Gibson}, {Kawata},
  {House}, {Miranda}, {Macci{\`o}}, {Pilkington}, {Ro{\v s}kar}, {Wadsley}, \&
  {Quinn}}]{Brook2012}
{Brook}, C.~B., {Stinson}, G.~S., {Gibson}, B.~K., {et~al.} 2012, \mnras, 426,
  690

\bibitem[{{Brusadin} {et~al.}(2013){Brusadin}, {Matteucci}, \&
  {Romano}}]{Brusadin2013}
{Brusadin}, G., {Matteucci}, F., \& {Romano}, D. 2013, \aap, 554, A135

\bibitem[{{Carney} {et~al.}(1990){Carney}, {Latham}, \& {Laird}}]{Carney1990}
{Carney}, B.~W., {Latham}, D.~W., \& {Laird}, J.~B. 1990, \aj, 99, 572

\bibitem[{{Cartledge} {et~al.}(2006){Cartledge}, {Lauroesch}, {Meyer}, \&
  {Sofia}}]{Cartledge2006}
{Cartledge}, S.~I.~B., {Lauroesch}, J.~T., {Meyer}, D.~M., \& {Sofia}, U.~J.
  2006, \apj, 641, 327

\bibitem[{{Casagrande} {et~al.}(2010){Casagrande}, {Ram{\'{\i}}rez},
  {Mel{\'e}ndez}, {Bessell}, \& {Asplund}}]{Casagrande2010}
{Casagrande}, L., {Ram{\'{\i}}rez}, I., {Mel{\'e}ndez}, J., {Bessell}, M., \&
  {Asplund}, M. 2010, \aap, 512, A54

\bibitem[{{Chabrier}(2003)}]{Chabrier2003}
{Chabrier}, G. 2003, \pasp, 115, 763

\bibitem[{{Chiappini} {et~al.}(1997){Chiappini}, {Matteucci}, \&
  {Gratton}}]{Chiappini1997}
{Chiappini}, C., {Matteucci}, F., \& {Gratton}, R. 1997, \apj, 477, 765

\bibitem[{{Chiappini} {et~al.}(2001){Chiappini}, {Matteucci}, \&
  {Romano}}]{Chiappini2001}
{Chiappini}, C., {Matteucci}, F., \& {Romano}, D. 2001, \apj, 554, 1044

\bibitem[{{Chiosi}(1980)}]{Chiosi1980}
{Chiosi}, C. 1980, \aap, 83, 206

\bibitem[{{Colavitti} {et~al.}(2008){Colavitti}, {Matteucci}, \&
  {Murante}}]{Colavitti2008}
{Colavitti}, E., {Matteucci}, F., \& {Murante}, G. 2008, \aap, 483, 401

\bibitem[{{Demarque} {et~al.}(2004){Demarque}, {Woo}, {Kim}, \&
  {Yi}}]{Demarque2004}
{Demarque}, P., {Woo}, J.-H., {Kim}, Y.-C., \& {Yi}, S.~K. 2004, \apjs, 155,
  667

\bibitem[{{Di Matteo} {et~al.}(2014){Di Matteo}, {Haywood}, {G{\'o}mez}, {van
  Damme}, {Combes}, {Hall{\'e}}, {Semelin}, {Lehnert}, \&
  {Katz}}]{DiMatteo2013}
{Di Matteo}, P., {Haywood}, M., {G{\'o}mez}, A., {et~al.} 2014, \aap, 567, A122

\bibitem[{{Diehl} {et~al.}(2006){Diehl}, {Halloin}, {Kretschmer}, {Lichti},
  {Sch{\"o}nfelder}, {Strong}, {von Kienlin}, {Wang}, {Jean}, {Kn{\"o}dlseder},
  {Roques}, {Weidenspointner}, {Schanne}, {Hartmann}, {Winkler}, \&
  {Wunderer}}]{Diehl2006}
{Diehl}, R., {Halloin}, H., {Kretschmer}, K., {et~al.} 2006, \nat, 439, 45

\bibitem[{{Fenner} \& {Gibson}(2003)}]{Fenner2003}
{Fenner}, Y. \& {Gibson}, B.~K. 2003, \pasa, 20, 189

\bibitem[{{Fenner} {et~al.}(2002){Fenner}, {Gibson}, \& {Limongi}}]{Fenner2002}
{Fenner}, Y., {Gibson}, B.~K., \& {Limongi}, M. 2002, \apss, 281, 537

\bibitem[{{Few} {et~al.}(2012){Few}, {Courty}, {Gibson}, {Kawata}, {Calura}, \&
  {Teyssier}}]{Few2012}
{Few}, C.~G., {Courty}, S., {Gibson}, B.~K., {et~al.} 2012, \mnras, 424, L11

\bibitem[{{Fisher} \& {Drory}(2012)}]{Fisher2012}
{Fisher}, D.~B. \& {Drory}, N. 2012, in American Astronomical Society Meeting
  Abstracts, Vol. 219, American Astronomical Society Meeting Abstracts \#219,
  417.01

\bibitem[{{Fran{\c c}ois} {et~al.}(2004){Fran{\c c}ois}, {Matteucci}, {Cayrel},
  {Spite}, {Spite}, \& {Chiappini}}]{Francois2004}
{Fran{\c c}ois}, P., {Matteucci}, F., {Cayrel}, R., {et~al.} 2004, \aap, 421,
  613

\bibitem[{{Fuhrmann} {et~al.}(2012){Fuhrmann}, {Chini}, {Hoffmeister}, \&
  {Bernkopf}}]{Fuhrmann2012}
{Fuhrmann}, K., {Chini}, R., {Hoffmeister}, V.~H., \& {Bernkopf}, J. 2012,
  \mnras, 420, 1423

\bibitem[{{Gibson} {et~al.}(2013){Gibson}, {Pilkington}, {Brook}, {Stinson}, \&
  {Bailin}}]{Gibson2013}
{Gibson}, B.~K., {Pilkington}, K., {Brook}, C.~B., {Stinson}, G.~S., \&
  {Bailin}, J. 2013, \aap, 554, A47

\bibitem[{{Gilmore} \& {Wyse}(1986)}]{Gilmore1986}
{Gilmore}, G. \& {Wyse}, R.~F.~G. 1986, \nat, 322, 806

\bibitem[{{Gonzalez} \& {Laws}(2000)}]{Gonzalez2000}
{Gonzalez}, G. \& {Laws}, C. 2000, \aj, 119, 390

\bibitem[{{Greggio}(2005)}]{Greggio2005}
{Greggio}, L. 2005, \aap, 441, 1055

\bibitem[{{Hartwick}(1976)}]{Hartwick1976}
{Hartwick}, F.~D.~A. 1976, \apj, 209, 418

\bibitem[{{Haywood}(2001)}]{Haywood2001}
{Haywood}, M. 2001, \mnras, 325, 1365

\bibitem[{{Haywood}(2006)}]{Haywood2006}
{Haywood}, M. 2006, \mnras, 371, 1760

\bibitem[{{Haywood}(2008)}]{Haywood2008}
{Haywood}, M. 2008, \mnras, 388, 1175

\bibitem[{{Haywood}(2014)}]{Haywood2014}
{Haywood}, M. 2014, \memsai, 85, 253

\bibitem[{{Haywood} {et~al.}(2013){Haywood}, {Di Matteo}, {Lehnert}, {Katz}, \&
  {G{\'o}mez}}]{Haywood2013}
{Haywood}, M., {Di Matteo}, P., {Lehnert}, M.~D., {Katz}, D., \& {G{\'o}mez},
  A. 2013, \aap, 560, A109

\bibitem[{{Hopkins} \& {Beacom}(2006)}]{Hopkins2006}
{Hopkins}, A.~M. \& {Beacom}, J.~F. 2006, \apj, 651, 142

\bibitem[{{Iwamoto} {et~al.}(1999){Iwamoto}, {Brachwitz}, {Nomoto},
  {Kishimoto}, {Umeda}, {Hix}, \& {Thielemann}}]{Iwamoto1999}
{Iwamoto}, K., {Brachwitz}, F., {Nomoto}, K., {et~al.} 1999, \apjs, 125, 439

\bibitem[{{J{\o}rgensen} \& {Lindegren}(2005)}]{Jorgensen2005}
{J{\o}rgensen}, B.~R. \& {Lindegren}, L. 2005, \aap, 436, 127

\bibitem[{{Juri{\'c}} {et~al.}(2008){Juri{\'c}}, {Ivezi{\'c}}, {Brooks},
  {Lupton}, {Schlegel}, {Finkbeiner}, {Padmanabhan}, {Bond}, {Sesar},
  {Rockosi}, {Knapp}, {Gunn}, {Sumi}, {Schneider}, {Barentine}, {Brewington},
  {Brinkmann}, {Fukugita}, {Harvanek}, {Kleinman}, {Krzesinski}, {Long},
  {Neilsen}, {Nitta}, {Snedden}, \& {York}}]{Juric2008}
{Juri{\'c}}, M., {Ivezi{\'c}}, {\v Z}., {Brooks}, A., {et~al.} 2008, \apj, 673,
  864

\bibitem[{{Karakas}(2010)}]{Karakas2010}
{Karakas}, A.~I. 2010, \mnras, 403, 1413

\bibitem[{{Kawata} \& {Gibson}(2003)}]{Kawata2003}
{Kawata}, D. \& {Gibson}, B.~K. 2003, \mnras, 340, 908

\bibitem[{{Kennicutt}(1983)}]{Kennicutt1983}
{Kennicutt}, Jr., R.~C. 1983, \apj, 272, 54

\bibitem[{{Kennicutt}(1998)}]{Kennicutt1998}
{Kennicutt}, Jr., R.~C. 1998, \apj, 498, 541

\bibitem[{{Kobayashi} {et~al.}(2011){Kobayashi}, {Karakas}, \&
  {Umeda}}]{Kobayashi2011}
{Kobayashi}, C., {Karakas}, A.~I., \& {Umeda}, H. 2011, \mnras, 414, 3231

\bibitem[{{Kobayashi} {et~al.}(2000){Kobayashi}, {Tsujimoto}, \&
  {Nomoto}}]{Kobayashi2000}
{Kobayashi}, C., {Tsujimoto}, T., \& {Nomoto}, K. 2000, \apj, 539, 26

\bibitem[{{Kobayashi} {et~al.}(1998){Kobayashi}, {Tsujimoto}, {Nomoto},
  {Hachisu}, \& {Kato}}]{Kobayashi1998}
{Kobayashi}, C., {Tsujimoto}, T., {Nomoto}, K., {Hachisu}, I., \& {Kato}, M.
  1998, \apjl, 503, L155

\bibitem[{{Kordopatis} {et~al.}(2013){Kordopatis}, {Gilmore}, {Wyse},
  {Steinmetz}, {Siebert}, {Bienaym{\'e}}, {McMillan}, {Minchev}, {Zwitter},
  {Gibson}, {Seabroke}, {Grebel}, {Bland-Hawthorn}, {Boeche}, {Freeman},
  {Munari}, {Navarro}, {Parker}, {Reid}, \& {Siviero}}]{Kordopatis2013}
{Kordopatis}, G., {Gilmore}, G., {Wyse}, R.~F.~G., {et~al.} 2013, \mnras, 436,
  3231

\bibitem[{{Kormendy} {et~al.}(2010){Kormendy}, {Drory}, {Bender}, \&
  {Cornell}}]{Kormendy2010}
{Kormendy}, J., {Drory}, N., {Bender}, R., \& {Cornell}, M.~E. 2010, \apj, 723,
  54

\bibitem[{{Kroupa}(2001)}]{Kroupa2001}
{Kroupa}, P. 2001, \mnras, 322, 231

\bibitem[{{Kroupa} {et~al.}(1993){Kroupa}, {Tout}, \& {Gilmore}}]{Kroupa1993}
{Kroupa}, P., {Tout}, C.~A., \& {Gilmore}, G. 1993, \mnras, 262, 545

\bibitem[{{Kunder} {et~al.}(2012){Kunder}, {Koch}, {Rich}, {de Propris},
  {Howard}, {Stubbs}, {Johnson}, {Shen}, {Wang}, {Robin}, {Kormendy}, {Soto},
  {Frinchaboy}, {Reitzel}, {Zhao}, \& {Origlia}}]{Kunder2012}
{Kunder}, A., {Koch}, A., {Rich}, R.~M., {et~al.} 2012, \aj, 143, 57

\bibitem[{{Lacey} \& {Fall}(1985)}]{Lacey1985}
{Lacey}, C.~G. \& {Fall}, S.~M. 1985, \apj, 290, 154

\bibitem[{{Larson}(1976)}]{Larson1976}
{Larson}, R.~B. 1976, \mnras, 176, 31

\bibitem[{{Lilly} {et~al.}(1996){Lilly}, {Le Fevre}, {Hammer}, \&
  {Crampton}}]{Lilly1996}
{Lilly}, S.~J., {Le Fevre}, O., {Hammer}, F., \& {Crampton}, D. 1996, \apjl,
  460, L1

\bibitem[{{Limongi} \& {Chieffi}(2002)}]{Limongi2002}
{Limongi}, M. \& {Chieffi}, A. 2002, \pasa, 19, 246

\bibitem[{{Limongi} {et~al.}(2000){Limongi}, {Straniero}, \&
  {Chieffi}}]{Limongi2000}
{Limongi}, M., {Straniero}, O., \& {Chieffi}, A. 2000, \apjs, 129, 625

\bibitem[{{Madau} \& {Dickinson}(2014)}]{Madau2014}
{Madau}, P. \& {Dickinson}, M. 2014, \araa, 52, 415

\bibitem[{{Madau} {et~al.}(1996){Madau}, {Ferguson}, {Dickinson}, {Giavalisco},
  {Steidel}, \& {Fruchter}}]{Madau1996}
{Madau}, P., {Ferguson}, H.~C., {Dickinson}, M.~E., {et~al.} 1996, \mnras, 283,
  1388

\bibitem[{{Madau} {et~al.}(1998){Madau}, {Pozzetti}, \&
  {Dickinson}}]{Madau1998}
{Madau}, P., {Pozzetti}, L., \& {Dickinson}, M. 1998, \apj, 498, 106

\bibitem[{{Mannucci} {et~al.}(2006){Mannucci}, {Della Valle}, \&
  {Panagia}}]{Mannucci2006}
{Mannucci}, F., {Della Valle}, M., \& {Panagia}, N. 2006, \mnras, 370, 773

\bibitem[{{Mannucci} {et~al.}(2005){Mannucci}, {Della Valle}, {Panagia},
  {Cappellaro}, {Cresci}, {Maiolino}, {Petrosian}, \& {Turatto}}]{Mannucci2005}
{Mannucci}, F., {Della Valle}, M., {Panagia}, N., {et~al.} 2005, \aap, 433, 807

\bibitem[{{Matteucci}(2009)}]{Matteucci2008}
{Matteucci}, F. 2009, {Chemical evolution}, ed. J.~{Cepa} (Cambridge University
  Press), 183

\bibitem[{{Matteucci} \& {Chiosi}(1983)}]{MatteucciC1983}
{Matteucci}, F. \& {Chiosi}, C. 1983, \aap, 123, 121

\bibitem[{{Matteucci} \& {Francois}(1989)}]{Matteucci1989}
{Matteucci}, F. \& {Francois}, P. 1989, \mnras, 239, 885

\bibitem[{{Matteucci} \& {Greggio}(1986)}]{Matteucci1986}
{Matteucci}, F. \& {Greggio}, L. 1986, \aap, 154, 279

\bibitem[{{Matteucci} {et~al.}(2009){Matteucci}, {Spitoni}, {Recchi}, \&
  {Valiante}}]{Matteucci2009}
{Matteucci}, F., {Spitoni}, E., {Recchi}, S., \& {Valiante}, R. 2009, \aap,
  501, 531

\bibitem[{{McMillan}(2011)}]{McMillan2011}
{McMillan}, P.~J. 2011, \mnras, 414, 2446

\bibitem[{{Micali} {et~al.}(2013){Micali}, {Matteucci}, \&
  {Romano}}]{Micali2013}
{Micali}, A., {Matteucci}, F., \& {Romano}, D. 2013, \mnras, 436, 1648

\bibitem[{{Morrison} {et~al.}(1990){Morrison}, {Flynn}, \&
  {Freeman}}]{Morrison1990}
{Morrison}, H.~L., {Flynn}, C., \& {Freeman}, K.~C. 1990, \aj, 100, 1191

\bibitem[{{Naab} \& {Ostriker}(2006)}]{Naab2006}
{Naab}, T. \& {Ostriker}, J.~P. 2006, \mnras, 366, 899

\bibitem[{Nelder \& Mead(1965)}]{Nelder1965}
Nelder, J.~A. \& Mead, R. 1965, The Computer Journal, 7, 308

\bibitem[{{Ness} {et~al.}(2012){Ness}, {Freeman}, {Athanassoula},
  {Wylie-De-Boer}, {Bland-Hawthorn}, {Lewis}, {Yong}, {Asplund}, {Lane},
  {Kiss}, \& {Ibata}}]{Ness2012}
{Ness}, M., {Freeman}, K., {Athanassoula}, E., {et~al.} 2012, \apj, 756, 22

\bibitem[{{Nidever} {et~al.}(2014){Nidever}, {Bovy}, {Bird}, {Andrews},
  {Hayden}, {Holtzman}, {Majewski}, {Smith}, \& et~al.}]{Nidever2014}
{Nidever}, D.~L., {Bovy}, J., {Bird}, J.~C., {et~al.} 2014, ArXiv
  e-prints:1409.3566

\bibitem[{{Nittler}(2005)}]{Nittler2005}
{Nittler}, L.~R. 2005, \apj, 618, 281

\bibitem[{{Nomoto} {et~al.}(2006){Nomoto}, {Tominaga}, {Umeda}, {Kobayashi}, \&
  {Maeda}}]{Nomoto2006}
{Nomoto}, K., {Tominaga}, N., {Umeda}, H., {Kobayashi}, C., \& {Maeda}, K.
  2006, Nuclear Physics A, 777, 424

\bibitem[{{Pagel} \& {Patchett}(1975)}]{Pagel1975}
{Pagel}, B.~E.~J. \& {Patchett}, B.~E. 1975, \mnras, 172, 13

\bibitem[{{Pagel} \& {Tautvaisiene}(1997)}]{Pagel1997}
{Pagel}, B.~E.~J. \& {Tautvaisiene}, G. 1997, \mnras, 288, 108

\bibitem[{Powell(1964)}]{Powell1964}
Powell, M. J.~D. 1964, The Computer Journal, 7, 155

\bibitem[{{Prantzos}(2003)}]{Prantzos2003}
{Prantzos}, N. 2003, \aap, 404, 211

\bibitem[{{Raiteri} {et~al.}(1996){Raiteri}, {Villata}, \&
  {Navarro}}]{Raiteri1996}
{Raiteri}, C.~M., {Villata}, M., \& {Navarro}, J.~F. 1996, \aap, 315, 105

\bibitem[{{Ram{\'{\i}}rez} {et~al.}(2013){Ram{\'{\i}}rez}, {Allende Prieto}, \&
  {Lambert}}]{Ramirez2013}
{Ram{\'{\i}}rez}, I., {Allende Prieto}, C., \& {Lambert}, D.~L. 2013, \apj,
  764, 78

\bibitem[{{Reddy} {et~al.}(2006){Reddy}, {Lambert}, \& {Allende
  Prieto}}]{Reddy2006}
{Reddy}, B.~E., {Lambert}, D.~L., \& {Allende Prieto}, C. 2006, \mnras, 367,
  1329

\bibitem[{{Robitaille} \& {Whitney}(2010)}]{Robitaille2010}
{Robitaille}, T.~P. \& {Whitney}, B.~A. 2010, \apjl, 710, L11

\bibitem[{{Romano} {et~al.}(2005){Romano}, {Chiappini}, {Matteucci}, \&
  {Tosi}}]{Romano2005}
{Romano}, D., {Chiappini}, C., {Matteucci}, F., \& {Tosi}, M. 2005, \aap, 430,
  491

\bibitem[{{Romano} {et~al.}(2010){Romano}, {Karakas}, {Tosi}, \&
  {Matteucci}}]{Romano2010}
{Romano}, D., {Karakas}, A.~I., {Tosi}, M., \& {Matteucci}, F. 2010, \aap, 522,
  A32

\bibitem[{{Salpeter}(1955)}]{Salpeter1955}
{Salpeter}, E.~E. 1955, \apj, 121, 161

\bibitem[{{Scalo}(1998)}]{Scalo1998}
{Scalo}, J. 1998, in Astronomical Society of the Pacific Conference Series,
  Vol. 142, The Stellar Initial Mass Function (38th Herstmonceux Conference),
  ed. G.~{Gilmore} \& D.~{Howell}, 201

\bibitem[{{Scalo}(1986)}]{Scalo1986}
{Scalo}, J.~M. 1986, \fcp, 11, 1

\bibitem[{{Schmidt}(1959)}]{Schmidt1959}
{Schmidt}, M. 1959, \apj, 129, 243

\bibitem[{{Schmidt}(1963)}]{Schmidt1963}
{Schmidt}, M. 1963, \apj, 137, 758

\bibitem[{{Shen} {et~al.}(2010){Shen}, {Rich}, {Kormendy}, {Howard}, {De
  Propris}, \& {Kunder}}]{Shen2010}
{Shen}, J., {Rich}, R.~M., {Kormendy}, J., {et~al.} 2010, \apjl, 720, L72

\bibitem[{{Snaith} {et~al.}(2014){Snaith}, {Haywood}, {Di Matteo}, {Lehnert},
  {Combes}, {Katz}, \& {G{\'o}mez}}]{Snaith2014}
{Snaith}, O.~N., {Haywood}, M., {Di Matteo}, P., {et~al.} 2014, \apjl, 781, L31

\bibitem[{{Sommer-Larsen}(2006)}]{Sommer2006}
{Sommer-Larsen}, J. 2006, \apjl, 644, L1

\bibitem[{{Spitoni} {et~al.}(2014){Spitoni}, {Matteucci}, \&
  {Sozzetti}}]{Spitoni2014}
{Spitoni}, E., {Matteucci}, F., \& {Sozzetti}, A. 2014, \mnras, 440, 2588

\bibitem[{{Stinson} {et~al.}(2013){Stinson}, {Brook}, {Macci{\`o}}, {Wadsley},
  {Quinn}, \& {Couchman}}]{Stinson2013}
{Stinson}, G.~S., {Brook}, C., {Macci{\`o}}, A.~V., {et~al.} 2013, \mnras, 428,
  129

\bibitem[{{Tacconi} {et~al.}(2010){Tacconi}, {Genzel}, {Neri}, {Cox}, {Cooper},
  {Shapiro}, {Bolatto}, {Bouch{\'e}}, {Bournaud}, {Burkert}, {Combes},
  {Comerford}, {Davis}, {Schreiber}, {Garcia-Burillo}, {Gracia-Carpio}, {Lutz},
  {Naab}, {Omont}, {Shapley}, {Sternberg}, \& {Weiner}}]{Tacconi2010}
{Tacconi}, L.~J., {Genzel}, R., {Neri}, R., {et~al.} 2010, \nat, 463, 781

\bibitem[{{Timmes} {et~al.}(1995){Timmes}, {Woosley}, \& {Weaver}}]{Timmes1995}
{Timmes}, F.~X., {Woosley}, S.~E., \& {Weaver}, T.~A. 1995, \apjs, 98, 617

\bibitem[{{Tinsley}(1974)}]{Tinsley1974}
{Tinsley}, B.~M. 1974, \apj, 192, 629

\bibitem[{{Travaglio} {et~al.}(2004){Travaglio}, {Hillebrandt}, {Reinecke}, \&
  {Thielemann}}]{Travaglio2004}
{Travaglio}, C., {Hillebrandt}, W., {Reinecke}, M., \& {Thielemann}, F.-K.
  2004, \aap, 425, 1029

\bibitem[{{van den Bergh}(1962)}]{vandenBergh1962}
{van den Bergh}, S. 1962, \aj, 67, 486

\bibitem[{van Dokkum {et~al.}(2013)van Dokkum, Leja, Nelson, Patel, Skelton,
  Momcheva, Brammer, Whitaker, Lundgren, Fumagalli, Conroy, Schreiber, Franx,
  Kriek, Labbé, Marchesini, Rix, van~der Wel, \& Wuyts}]{vanDokkum2013}
van Dokkum, P.~G., Leja, J., Nelson, E.~J., {et~al.} 2013, The Astrophysical
  Journal Letters, 771, L35

\bibitem[{{Woosley} \& {Weaver}(1995)}]{WW1995}
{Woosley}, S.~E. \& {Weaver}, T.~A. 1995, \apjs, 101, 181

\bibitem[{{Wyse}(2001)}]{Wyse2001}
{Wyse}, R.~F.~G. 2001, in Astronomical Society of the Pacific Conference
  Series, Vol. 230, Galaxy Disks and Disk Galaxies, ed. J.~G. {Funes} \& E.~M.
  {Corsini}, 71--80

\bibitem[{{Yamada} {et~al.}(2013){Yamada}, {Suda}, {Komiya}, {Aoki}, \&
  {Fujimoto}}]{Yamada2013}
{Yamada}, S., {Suda}, T., {Komiya}, Y., {Aoki}, W., \& {Fujimoto}, M.~Y. 2013,
  \mnras, 436, 1362

\end{thebibliography}

\end{document}